\pdfoutput=1
\documentclass[letterpaper,twocolumn,english,aps,showpacs,showkeys,preprintnumbers,prd,floatfix,nofootinbib,superscriptaddress]{revtex4-1}

\usepackage{amsmath,amssymb,bbm,color}
\usepackage{epsfig}
\usepackage{graphics}
\usepackage{euscript}
\usepackage{pslatex}
\usepackage{hyperref}
\usepackage{xcolor}
\usepackage{fancybox}
\usepackage{enumerate}

\newcommand{\Ecal}{{\cal E}}

\newcommand{\Peu}{\EuScript{P}}

\newcommand{\Ibbm}{\mathbbm{1}}
\newcommand{\Pbbm}{\mathbbm{P}}
\newcommand{\Rbbm}{\mathbbm{R}}

\newcommand{\PPopp}{\overleftrightarrow{\rm PP}}

\newcommand{\veps}{\varepsilon}

\newcommand{\alfb}{{\bar{\alpha}}}
\newcommand{\betb}{{\bar{\beta}}}

\newcommand{\bk}{{\bf{k}}}
\newcommand{\ba}{{\bf{a}}}

\newcommand{\tbet}{{\tilde{\beta}}}
\newcommand{\bsl}{{\!\!\!\!\slash}}

\begin{document}

\preprint{
\vbox{
\null \vspace{0.3in}
\hbox{IFJPAN-IV-2011-2}
\hbox{CERN-PH-TH/2011-034}
\hbox{SMU-HEP-12-12}
}
}

\title{\null \vspace{0.3in}
Inclusion of the QCD next-to-leading order corrections
in the quark-gluon Monte Carlo shower}

\author{S.\ Jadach}
\affiliation{Institute of Nuclear Physics, Polish Academy of Sciences,\\
  ul.\ Radzikowskiego 152, 31-342 Cracow, Poland}
\affiliation{CERN, PH Division, TH Unit, CH-1211 Geneva 23, Switzerland}

\author{A.~Kusina}
\affiliation{Southern Methodist University, Dallas, TX 75275, USA}

\author{W.\ P\l{}aczek}
\affiliation{Marian Smoluchowski Institute of Physics, Jagiellonian University,\\
 ul. Reymonta 4, 30-059 Cracow, Poland.}

\author{M.\ Skrzypek}
\affiliation{Institute of Nuclear Physics, Polish Academy of Sciences,\\
  ul.\ Radzikowskiego 152, 31-342 Cracow, Poland}

\author{M.\ Slawinska}
\affiliation{Institute of Nuclear Physics, Polish Academy of Sciences,\\
  ul.\ Radzikowskiego 152, 31-342 Cracow, Poland}


\begin{abstract}
Methodology of including QCD NLO corrections
in the quark--gluon Monte Carlo shower is outlined.
The work concentrates on two issues:
(i) constructing leading order (LO) parton
shower Monte Carlo from scratch, such that it 
rigorously extends collinear factorization into the exclusive
(fully unintegrated) one which we call the Monte Carlo factorization scheme;
(ii) introducing next-to-leading-order (NLO) corrections to 
the hard process in this new environment.
The presented solution is designed to be extended
to the full NLO level Monte Carlo, including NLO corrections
not only in the hard process but in the whole shower.
The issue of the difference between
the factorization scheme implemented in the Monte Carlo (MC)
solution and the standard $\overline{MS}$ scheme is addressed.
The principal MC implementation is designed for the electroweak boson
production process at the LHC, but in order to discuss
universality -- process independence, the deep inelastic lepton--hadron
scaterring is also brought into the MC framework.
\end{abstract}

\maketitle
\tableofcontents{}

\section{Introduction}

The excellent performance and fast experimental data accumulation
of the Large Hadron Collider (LHC) at CERN makes the
precise evaluation of the strong interactions effects
within perturbative Quantum Chromodynamics
(QCD)~\cite{GWP,Gross:1974cs,Georgi:1951sr}
a more and more important task.
The  principal role of QCD in hadron colliders data analyses
(LHC and Tevatron) is to provide precise predictions for
distributions and luminosities of quarks and gluons
accompanying production of heavy particles.

One of the most important theoretical tools of perturbative QCD (pQCD)
are factorization theorems~\cite{Ellis:1978ty,Collins:1984kg,Bodwin:1984hc},
which reformulate any scattering process in QCD
in terms of the on-shell hard process part
convoluted in the lightcone variable with the {\em ladder parts},
provided a single large scale $Q^2$ is involved
(short distance interaction).
The hard process is usually treated at a fixed perturbative order
and the ladder parts are resummed to infinite order,
for each coloured energetic ingoing/outgoing parton.
The initial state ladders give rise to PDFs,
the {\em inclusive} parton distribution functions.

The initial state ladder,
instead of being source of the inclusive PDF,
can be also modelled using Monte Carlo simulation
(including hadronization), as initiated in
refs.~\cite{Sjostrand:1985xi,Webber:1984if}.
Such an implementation of the QCD ladder
is referred to as the parton shower Monte Carlo (PSMC) program.
This kind of programs play an enormous practical role
in all collider experiments.
In today PSMCs the initial ladders are restricted to
the leading order (LO).
With growing requirements on the quality and precision
of the pQCD predictions for the LHC experiments
it becomes urgent to upgrade PSMC to the same next-to-leading-order (NLO) level,
which was reached for the inclusive PDFs two decades ago.
This is not easy, mainly because factorization
theorems of QCD~\cite{Ellis:1978ty,Collins:1984kg,Bodwin:1984hc}
were never meant for the MC implementation.
They are well suited for the simpler case of the hard process
upgraded with the finite order calculations
convoluted with the collinear inclusive PDFs.

There has been, however, a significant progress in implementing
pQCD in the framework of PSMC, which started with the work
of MC@NLO team~\cite{Frixione:2002ik,Frixione:2010ra}, and followed by development
of POWHEG method~\cite{Nason:2004rx,Frixione:2007vw}.
In these works hard process in PSMC is upgraded to
the NLO,
while the ladder part stays at the LO level,
essentially older solutions and software
for the LO PSMC are not modified.
This, of course, saves a lot of work but because of that
the methodology of combining the initial ladder parts
and the NLO-corrected matrix element (ME) for the hard process
is quite complicated.
The solution for this problem is to redesign the basic LO PSMC.
This would be too big an investment, if simplification of the
NLO corrections to the hard process ME were the only aim.
However, this effort is mandatory if we are aiming to upgrade also
the ladder parts of PSMC to the complete NLO level.

In this paper we outline a redesigned LO parton
shower MC and simultaneously present
methodology of including QCD NLO corrections
to the hard process which takes advantage of it.
Hence we shall concentrate on two issues:
(i) constructing once again the LO parton
shower Monte Carlo from scratch, such that it is
based firmly on the rigorous extension of the collinear factorization
theorems, which, contrary to the original collinear factorization, is
fully exclusive (unintegrated),
(ii) introducing the NLO corrections to the hard
process in this new environment.
It is natural to expect that
the issues of the difference between the
factorization scheme implemented in the Monte Carlo (MC)
solution and the standard $\overline{MS}$ will have to be addressed.
The important point of universality (process independence)
will be also discussed extensively.
Although the principal aim will be
a new MC implementation with the LO ladder (upgradable to NLO)
and the NLO hard process for the production of the electroweak (EW)
boson in quark--antiquark annihilation,
in order to address the issue of universality,
and for other practical reasons,
the deep inelastic lepton--hadron process
will be also brought into the consideration.
The next step in the project, that is the upgrade of the ladder part
to the NLO level, will be treated in 
a separate publication~\cite{IFJPAN-IV-2012-7},
although the general method was already outlined
in refs.~\cite{Skrzypek:2011zw,Jadach:2010aa,Jadach:2010ew}.
Many technical details needed for the NLO ladder are provided
in ref.~\cite{Jadach:2011kc} and auxiliary discussions
on the soft limit and the choice of the factorization scale
in the MC can be found in 
refs.~\cite{Kusina:2011xh,Kusina:2010gp,Slawinska:2009gn}.
The first numerical tests of the discussed method, of including
the NLO hard process corrections, are presented in ref.~\cite{Jadach:2012vs}.

Let us note that the ongoing effort undertaken 
in refs.~\cite{Tanaka:2011ig,Tanaka:2007mi} 
is in some aspects similar to the present work,
in particular the parton shower
is also redesigned at the NLO level and the departure from the standard
$\overline{MS}$ factorization scheme is also advocated.
These works are extending/exploiting techniques of
refs.~\cite{Kalinowski:1980ju,Konishi:1979cb}.

The remaining part of the paper is organized as follows.
In sec.~\ref{sec:factor} we discuss collinear factorization
in the form suitable for the MC implementation.
Section~\ref{sec:DYMC} covers construction of the LO MC for
EW boson production and the new method of introducing
the NLO corrections to the hard part of this process.
In sec.~\ref{sec:DISMC} we present a similar MC solution for
the DIS process, with the new LO MC modelling of the initial
and final state ladders and the NLO corrections to the hard process.
The issue of the universality as well as factorization-scheme
dependence are addressed in various steps of this presentation,
with the final discussion in sec.~\ref{sec:summary}, where we
also summarize and give outlook of further work.

\section{Generalities -- collinear factorization}
\label{sec:factor}

\begin{figure}[!ht]
  \centering
  {\includegraphics[width=75mm]{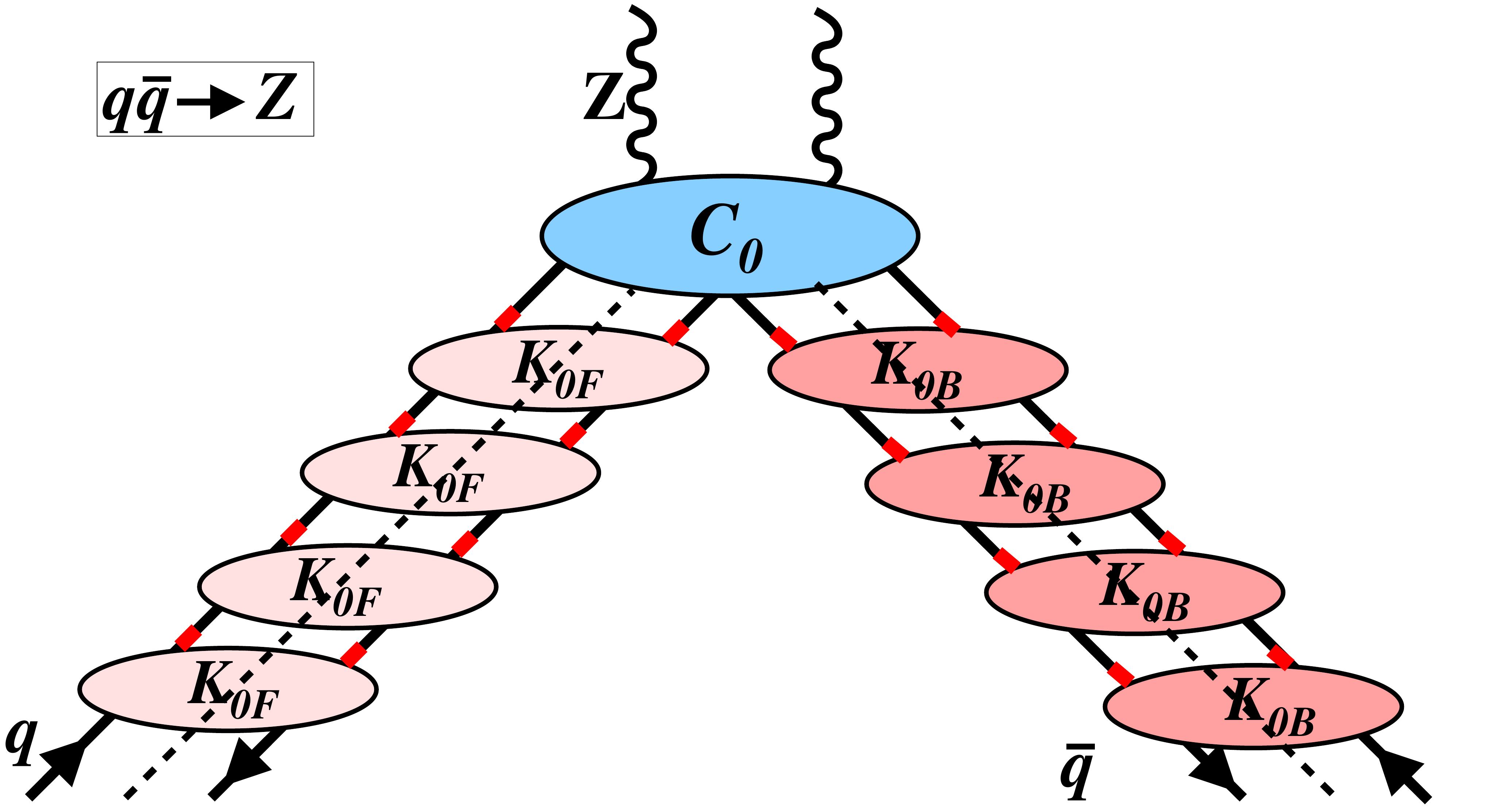}}\\
  {\includegraphics[width=75mm]{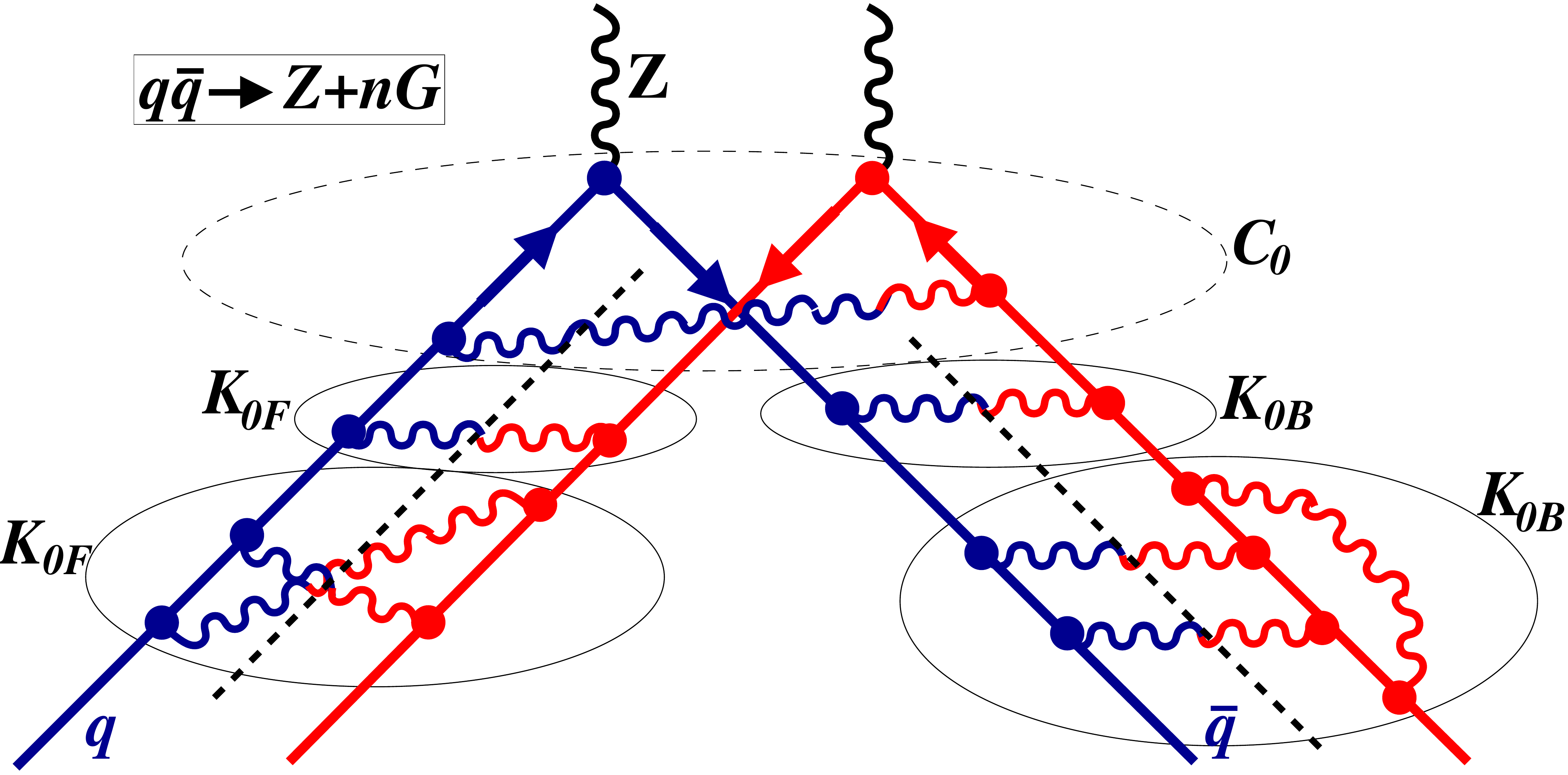}}
  \caption{
    The EGMPR~\cite{Ellis:1978ty} factorization for the EW boson production.
    Example 2PI kernels for a quark in the forward hemisphere $K_{0F}$,
    an antiquark in the backward hemisphere $K_{0B}$ and 
    the hard process part $C_0$
    are delimited by ellipses.
    The lower figure highlights the use of cut diagram notation by indicating
    the amplitude in blue and the complex conjugated amplitude in red.
    }
  \label{fig:EGMPR6}
\end{figure}

The precise definition of the LO approximation within
factorization of collinear singularities,
and the resulting distributions implemented in the Monte Carlo, 
is a necessary prelude of defining complete NLO distributions,
both in the ladder and in the hard process.
This is why in the following we shall define a new LO MC, 
``anchoring'' it in the collinear factorization
theorems~\cite{Ellis:1978ty,Bodwin:1984hc,Collins:1984kg}
as firmly as we can.

The production  process of electroweak (EW) bosons $W$, $Z$, $\gamma$, 
in the hadron--hadron scattering, or of other colour-neutral vector particle,
will be the object of the interest in the following.
We shall refer to it as Drell--Yan (DY) process for short.
We are going to describe a Monte Carlo algorithm
in which two initial state parton ladders
will be modelled up to the LO and the hard process to the NLO level,
both the hard and ladder parts in a completely exclusive way.
In the construction of the MC we will keep track of the precise relation
to the QCD factorization theorems, having in mind that the ladder
part (parton shower) will be upgraded to the NLO level in the next step.
Diagrammatically, we shall temporarily limit ourselves to the
$C_F^k$ part in the ladder, that is to gluon bremsstrahlung.
Because of copious soft gluon production,
this is the most difficult part in the MC construction of the LO
(later on of the NLO) MC implementation for the ladder.
Adding more diagrams (quark--gluon transitions, singlet diagrams)
will be discussed briefly,
but will be treated in a separate publication.

Let us start from the ``raw collinear factorization'' formula
of ref.~\cite{Ellis:1978ty} (in the axial gauge)
illustrated in Fig.~\ref{fig:EGMPR6} in a standard way (cut diagrams).
Following closely notation of ref.~\cite{Ellis:1978ty},
the standard Feynman amplitude
for the heavy boson production process is
$A^{r_F r_B}_{j_F j_B}(p_F,p_B,q_1,q_2,\Gamma)$,
where two incoming partons of the
type $j_F$ and $j_B$ have spin indices $r_F$ and $r_B$,
heavy boson decay lepton momenta are $q_i$ and any number of
the emitted on-shell gluons and quarks are
collectively denoted as $\Gamma$.
Denoting also collectively $(p_F,p_B,q_1,q_2)$ as $(p,q)$,
the partly integrated cut-diagram is defined as
\begin{equation}
\label{eq:AAstar}
\begin{split}
& {\cal M}^{r_F s_F r_B s_B}_{j_F j_B} (p,q) 
   \equiv 
   \sum_{\Gamma} \int d\Gamma\;
   \delta^{(4)}(p_F+p_B-q_1-q_2-p_\Gamma)\;
\\&~~~~~~~\times
   A^{r_F r_B}_{j_F j_B}(p,q,\Gamma)\;
   A^{s_F s_B}_{j_F j_B}(p,q,\Gamma)^*.
\end{split}
\end{equation}
It is related directly to differential cross section
$d\sigma_{j_F j_B}(p,q) = 
\frac{1}{flux} \sum_{r_F r_B}
{\cal M}^{r_F r_B r_F r_B}_{j_F j_B}(p,q)\; Z_F Z_B,
$
where $Z_{F,B}$ are wave renormalization factors.
The essence of the ``raw'' factorization theorem in ref.~\cite{Ellis:1978ty}
is that all collinear singularities are located in the ladders
with multiple 2-particle irreducible (2PI) kernels $K^{~r's'rs}_{0~j'~j}(k,p)$.
Suppressing for the moment ladder B and neglecting subscript F
the above statement reads
\[
\begin{split}
&{\cal M}^{r s}_{j} (p,q)=
{C_0}^{r s}_{j} (p,q)
+\sum_{n=1}^\infty
 \prod_{m=1}^n \int d^4 p_m\;
\\&~~~\times
\sum_{r_n,s_n,j_n} 
{C_0}^{r_n s_n}_{j_n} (p_n,q)
   K^{~~r_n s_n ~r_{n-1} s_{n-1} }_{0~j_{n}~j_{n-1}}(p_n,p_{n-1})
\\&~~~\times
\sum_{r_{n-1},s_{n-1},j_{n-1}} 
   K^{~~r_{n-1} s_{n-1} ~r_{n-2} s_{n-2} }_{0~j_{n-1}~j_{n-2}}(p_{n-1},p_{n-2})
\dots
\\&~~~\dots
\sum_{r_{1},s_{1},j_{1}} 
   K^{~~r_{1} s_{1} ~r s }_{0~j_{1}~j}(p_{1},p).
\end{split}
\]
Using the compact matrix notation
of refs.~\cite{Ellis:1978ty,Curci:1980uw},
the above expression in case of two ladders reads
\begin{equation}
\label{eq:CFTraw0}
\sigma
=C_0\;  \frac{1}{1- \cdot K_{0F}} \;
      \frac{1}{1- \cdot K_{0B}}
=\sum_{n_1,n_2=0}^\infty
 C_0\; (\cdot\; K_{0F})^{n_1} (\cdot\; K_{0B})^{n_2},
\end{equation}
see also the upper part of Fig.~\ref{fig:EGMPR6} for 
an equivalent graphical representation.
In the lower part of Fig.~\ref{fig:EGMPR6},
the above formula is illustrated diagrammatically
using the lowest order bremsstrahlung matrix element
where we explicitly indicate the 2PI kernels, with the
red part of the diagram representing the conjugate part $A^*$ 
of eq.~(\ref{eq:AAstar}).
Note that in the above expressions phase space of the emitted on-shell
partons (cut lines) is integrated over and treated inclusively.
In the following discussion it will be explicit and implemented
in the Monte Carlo parton shower.

According to ref.~\cite{Ellis:1978ty} all collinear singularities
in $\sigma$ of eq.~\eqref{eq:CFTraw0}
are coming from (dressed) propagators between kernels $K_{0F}$
($K_{0B}$) along the ladders.
The 2PI kernels for the initial quark ladder, $K_{0F}$,
and the antiquark ladder, $K_{0B}$, are expanded to the infinite order,
see ref.~\cite{Ellis:1978ty}.
In the following practical example
we shall truncate them to the (lowest) first order (LO) or
to the second order (NLO) $K_{0F}^{(2)}= K_{0F}^{[1]}+K_{0F}^{[2]}$,
taking into account the following $\sim C_F^2$ diagrams:
\[
\begin{split}
&{\Large {K_{0F}^{(1)} = {K_{0F}^{[1]}} =}}
             \raisebox{-10pt}{\includegraphics[width=9mm]{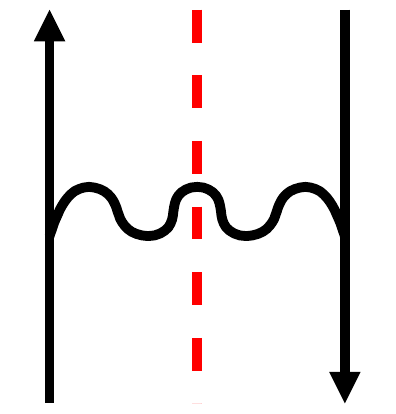}},\\
&{\Large {K_{0F}^{[2]} = }}
  \; \raisebox{-10pt}{\includegraphics[width=10mm]{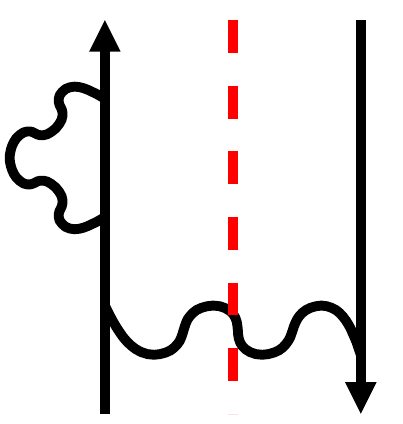}}
+ \; \raisebox{-10pt}{\includegraphics[width=10mm]{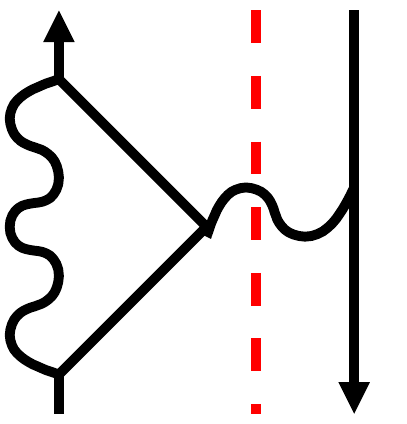}}
+ \; \raisebox{-10pt}{\includegraphics[width=10mm]{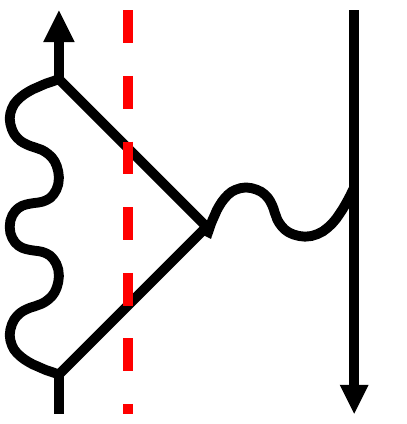}}
+ \; \raisebox{-10pt}{\includegraphics[width=10mm]{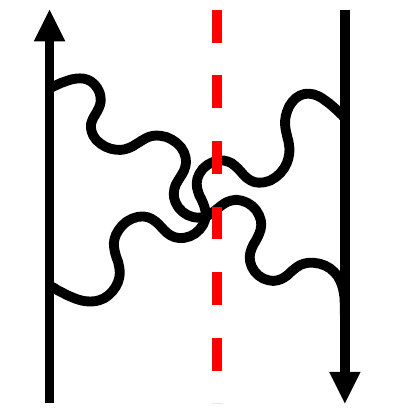}}.
\end{split}
\]
A similar expansion up to the 1st order (NLO) is done for the hard process part
$C_0^{(1)}= C_0^{[0]} +C_0^{[1]}$.
For simplicity, we are omitting the initial quark and antiquark distributions
in the beam hadrons and the flux factor is included in $C_0$.
The dot ``$\cdot$'' in the product $A \cdot B$ means full phase space
integration $\int d^4 q$ over the lines joining two subgraphs
in one ladder.

The next step in
the classic works of refs.~\cite{Ellis:1978ty,Curci:1980uw}
is the introduction of the projection operator $\Pbbm$.
Its role is to decouple kinematically not only $C_0$ and the ladder parts,
but the consecutive kernels $K_0$ along the ladders,
such that the integration over lightcone variables and collinear
logs become manifest and ready for analytical calculations.
Formally eq.~(\ref{eq:CFTraw0}) gets transformed 
(at the infinite order) into
\begin{widetext}
\begin{equation}
\label{eq:CFT1}
\begin{split}
&
\sigma
= C \otimes \Gamma_F \otimes \Gamma_B
= C \;  \frac{1}{1-\otimes K_F} \;
        \frac{1}{1-\otimes K_B}
=\sum_{n_1,n_2=0}^\infty
 C\; (\otimes K_F)^{n_1} (\otimes K_B)^{n_2},
\\&
C= C_0\; \frac{1}{1- \Rbbm K_{0F}}
      \; \frac{1}{1- \Rbbm K_{0B}},\quad
K_F = \frac{1}{1- \otimes \Pbbm K_{0F} \frac{1}{1- \Rbbm K_{0F}}},
\quad
K_B = \frac{1}{1- \otimes \Pbbm K_{0B} \frac{1}{1- \Rbbm K_{0B}}},
\\&
\Rbbm     =(\cdot 1-\otimes\Pbbm),
\end{split}
\end{equation}
\end{widetext}
where $A \otimes B$ means convolution in the lightcone
variable $\int dz_1 dz_2 \delta(x-z_1z_2) A(z_1) B(z_2) $,
while integration over transverse momenta is traded into
$\frac{1}{\veps^k}$ poles of dimensional regularization,
extracted  (upon integration) by $\Pbbm$,
see ref.~\cite{Curci:1980uw} for details.
The projection operator $\Pbbm$ used in ~\cite{Ellis:1978ty}
is slightly different.
However, both approaches are incompatible with any MC implementation,
as can be seen from the explicit expansion up to NLO%
\footnote{Omitting for simplicity quark wave function renormalization.}:
\begin{equation}
\begin{split}
C&=C_0^{(1)} \big(1 - \otimes \Pbbm K_{0F}^{[1]} 
                    - \otimes \Pbbm K_{0B}^{[1]} \big),
\\
\Gamma_F &= \Ibbm
 +\Pbbm K_{0F}^{[1]} + \Pbbm K_{0F}^{[2]}
 +\Pbbm  (K_{0F}^{[1]} \cdot K_{0F}^{[1]})
\\&\quad\;\;\;\;
 -\Pbbm  (K_{0F}^{[1]}\otimes \Pbbm K_{0F}^{[1]})
 +(\Pbbm K_{0F}^{[1]})\otimes (\Pbbm K_{0F}^{[1]}).
\end{split}
\end{equation}
Why? As we see, the above is a mixture of the original phase space
integrals like $(K_{0F} \cdot K_{0F}) $ and of partly integrated
integrals like in $(\Pbbm K_{0F})\otimes (\Pbbm K_{0F})$.
Even if we managed somehow to undo the transverse momentum integrations
implicit in $\Pbbm$ operator, we would still face huge
(double logarithmic) oversubtraction in
$\Pbbm K_{0F}^{[1]} ((\cdot\Ibbm-\otimes\Pbbm)K_{0F}^{[1]})$
compensated by
$(\Pbbm K_{0F}^{[1]})\otimes (\Pbbm K_{0F}^{[1]})$,
which would be deadly for any MC implementation.
For an explicit demonstration of this problem see also
the toy model considerations in ref.~\cite{Collins:1998rz},%
\footnote{In this work another example of the operator $\Pbbm$
  is presented and also the order in factorizing collinear singularities
  in eqs.~(\ref{eq:CFT1}) is reversed 
  -- it starts from the hard process.
  Nevertheless, it features the same oversubtraction problems 
  that inhibits MC implementation.}
or LO analysis to infinite order in ref.~\cite{Jadach:2010ew}.

\begin{figure}[!ht]
  \centering
  {\includegraphics[width=80mm]{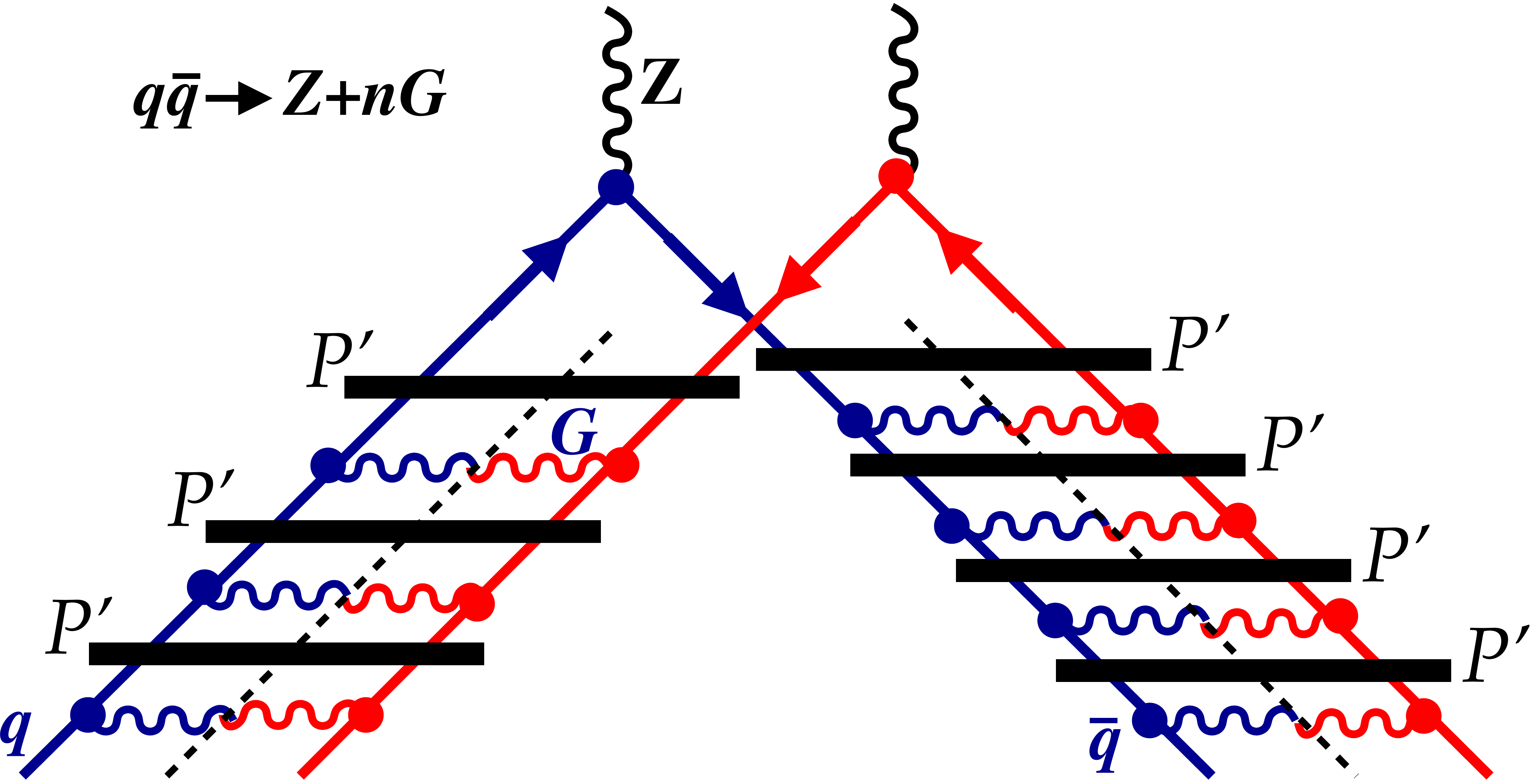}}
  \caption{
    The EGMPR factorization for EW boson production. The LO ladder with
    the projection operators $\Pbbm'$ inserted between the LO kernels.
    }
  \label{fig:EGMPR1a}
\end{figure}

The solution of the above over-subtraction problem is well known
and already employed in the existing LO parton shower MC 
since long~\cite{Sjostrand:1985xi,Webber:1984if}
-- in short, one has to introduce the time-ordered (T.O.) exponent,
see also ref.~\cite{Nagy:2007ty}.
Beyond the LO, a collinear factorization formula with T.O. exponent
was outlined in ref.~\cite{Jadach:2010ew} and we shall adopt it here.
We are going to use it for the LO MC ladders 
combined with the NLO hard process ME%
\footnote{
  Considerations concerning the NLO MC ladder can be found in
  refs.~\cite{Jadach:2010ew,Jadach:2010aa,Jadach:2011kc}.}.
According to ref.~\cite{Jadach:2010ew} eq.~(\ref{eq:CFT1})
is replaced by:
\begin{equation}
\begin{split}
& \sigma_{LO}=
\sum_{n_1=1}^\infty
\sum_{n_2=1}^\infty
  C_0^{(0)} 
\big\{
  (\Pbbm'  K_{0F}^{(1)})^{n_1}
\big\}_{T.O.}
\big\{
  (\Pbbm'  K_{0B}^{(1)})^{n_2}
\big\}_{T.O.},
\end{split}
\end{equation}
where $K_{0F}^{(1)}$ is the lowest order (LO) 2PI kernel,
the same as in eq.~(\ref{eq:CFT1}), but at the NLO and beyond
it is different, see the definition in ref.~\cite{Jadach:2010ew}.
The above LO process is depicted in Fig.~\ref{fig:EGMPR1a}.
As shown in ref.~\cite{Jadach:2010ew}, the complete and rigorous
definition of the new projection operator $\Pbbm'$ is not simple.
For the present purpose of the two LO ladders 
and the NLO hard process in the EW boson production
we shall define it step by step,
starting from the simple cases
of zero, one and two gluon ME,
$n_1+n_2=0,1,2$, rather than defining it immediately in the full form.

Let us denote the Born level differential cross section
for the EW vector boson production and decay process, 
$q\bar{q}\to l\bar{l}$, as follows:
\[
\frac{d\sigma_B}{d\Omega}(s,\theta).
\]
It may also include non-photonic EW radiative corrections.
The above differential cross section is so well known that we
may avoid defining its details explicitly.
From (two) Feynman diagrams we obtain
the following (exact) LO single-gluon emission differential distribution,
\begin{equation}
\label{eq:DYsig1}
\begin{split}
d\sigma_1 &= 
\frac{C_F \alpha_s}{\pi}\;
\frac{d\alpha d\beta}{\alpha\beta}\;
\frac{d\varphi}{2\pi}
\\&\times
\Bigg[
 \frac{d\sigma_B(\hat{s},\theta_{F})}{d\Omega}
 \frac{(1-\beta)^2}{2}
+\frac{d\sigma_B(\hat{s},\theta_{B})}{d\Omega}
 \frac{(1-\alpha)^2}{2}
\Bigg]
d\Omega,
\end{split}
\end{equation}
where the Sudakov variables $\alpha$ and $\beta$ are
defined in Appendix~\ref{app:A}.
This elegant formula is valid for any on/off-shell
vector particle production, $B=\gamma, W, Z$.
The polar angles $\theta_{F,B}$ are defined~\cite{Berends:1982ie}
with respect to $-\vec{p}_{0B}$ and $\vec{p}_{0F}$ respectively
in the rest frame of the $B$ boson
(the rest frame of $p_{0F}+p_{0B}-k$).

Two collinear limits $\beta\to 0$, $1-\alpha=z=const$
or  $\alpha\to 0$, $1-\beta=z=const$
in eq.~(\ref{eq:DYsig1}) are manifest.
For instance, in the 1st case we have:
\begin{equation}
\label{eq:ColF1g}
d\sigma_1 = 
\frac{C_F \alpha_s}{\pi}\;
\frac{dz d\beta}{(1-z)\beta}\;
\frac{d\varphi}{2\pi}\;
 \frac{1+z^2}{2}\;
 \frac{d\sigma_B(zs,\theta_{F})}{d\Omega}
d\Omega,
\end{equation}
that is the LO kernel
$P_{qq}(z)= \frac{2C_F \alpha_s}{\pi}\; \frac{1+z^2}{2(1-z)}$
shows up as expected.
Introducing the $\Pbbm'$ projector in this context
may look like an overkill,
but it will be instructive to explain how it
works in this simple case
before going to not so obvious case
of multiple use of $\Pbbm'$ in the following.

\begin{figure*}[!ht]
\centering
{\includegraphics[width=55mm]{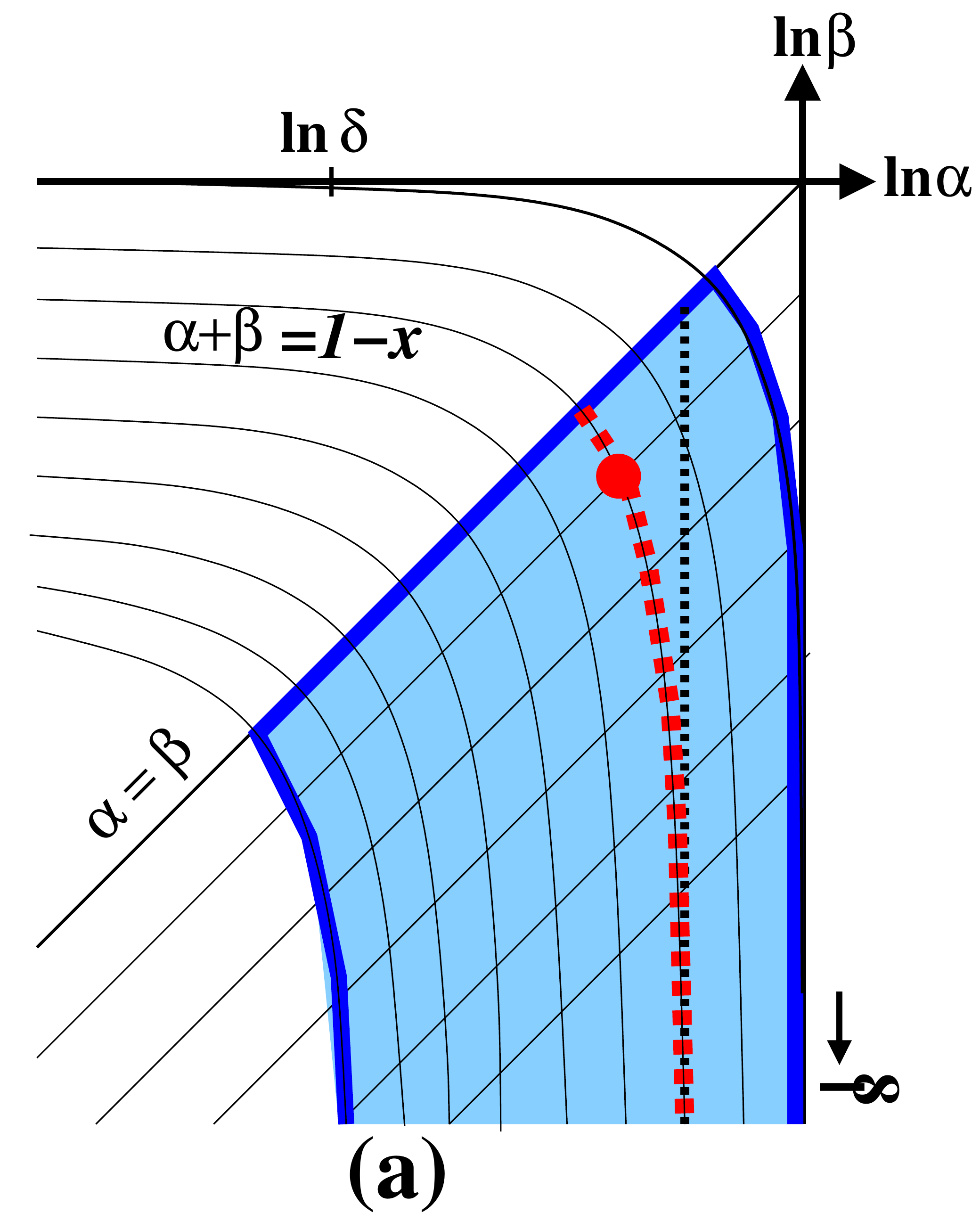}}
~~~~~~
{\includegraphics[width=55mm]{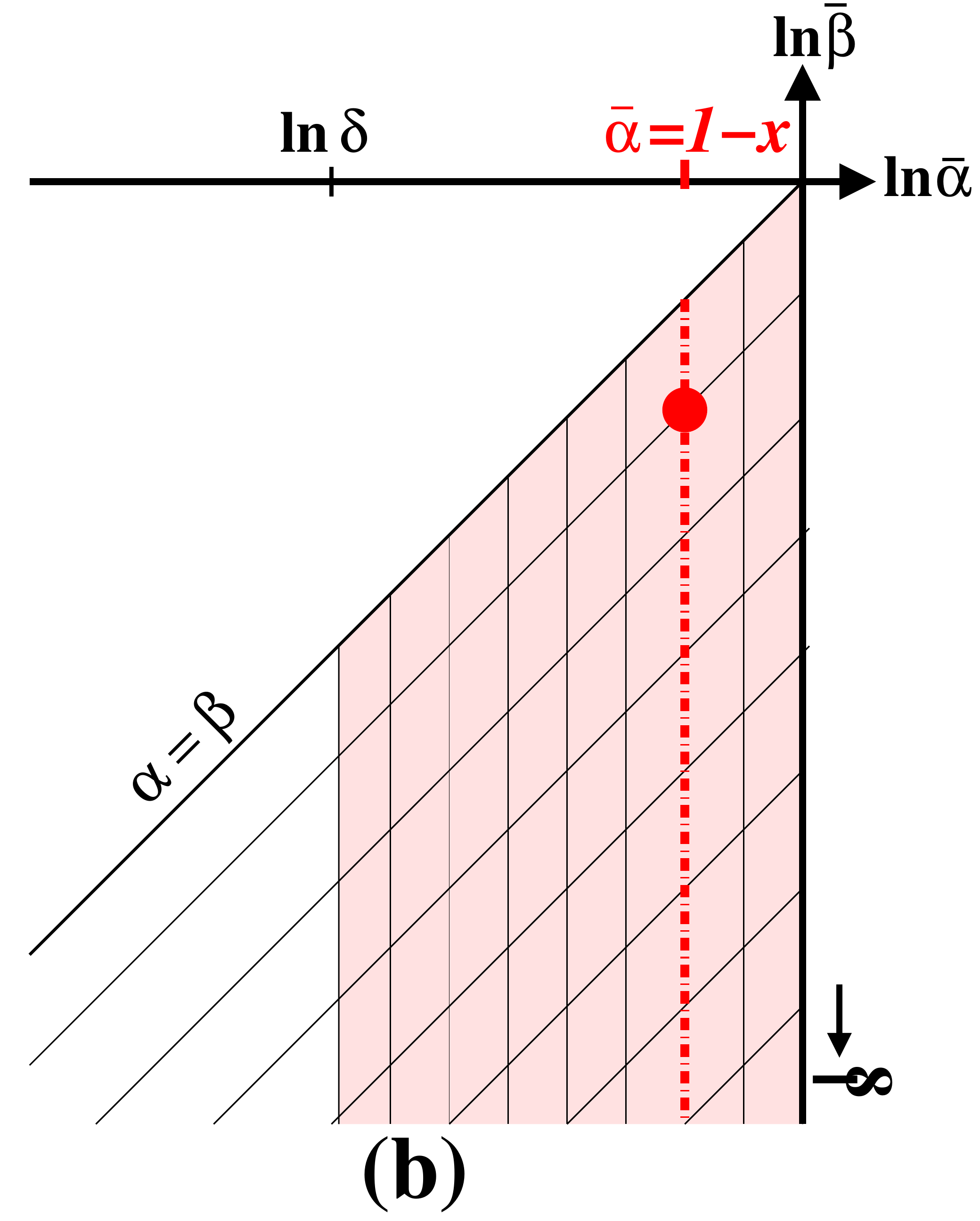}}
\caption{The Sudakov plane of single gluon emission in DY
  for angular ordering.}
\label{fig:SudPlanDY5}
\end{figure*}

The necessary ingredient is a spin projection operator
$P_{spin}$, which we define a little bit more rigorously
as compared to the Curci--Furmanski--Petronzio work (CFP)~\cite{Curci:1980uw}.
Our $P_{spin}$ acts definitely before
the phase space integration%
\footnote{
 In CFP spin projection is part of
 $H\;\Pbbm\; K=H{q\bsl}_i \big] P_\veps \big[\frac{n\bsl}{4n\cdot q_i}\; K$,
 where $P_\veps$ extracts the pole part and simultaneously sets $q^2\to 0$ in H.
 We have to be  more specific
 about the choice of $\hat{q}$ in the substitution $q\to\hat{q}$, $\hat{q}^2=0$.
}:
\begin{equation}
\label{eq:PspinPrim}
 H P_{spin}\; K 
  = H \hat{q\bsl}_i \Big] \Big[ \frac{n\bsl}{4n\cdot q_i}\; K,
\end{equation}
where $\hat{q}_i$ is the on-shell momentum entering the $H$ part,
such that it conserves the longitudinal (lightcone)
component $n\cdot\hat{q}_i = n\cdot q_i$,
and for the axial gauge vector $n$
we may take $n= p_{0B}$ or any other light-like vector,
the same for all rungs in a given ladder.
The same $n$ is defining transverse polarizations 
of gluons in the axial gauge
and enters a definition of lightcone variables $x_i = (nq_i)/(np_{0F})$.

In the present case of a single gluon emission, the action of
the $\Pbbm'$ projector, inserted in the squared, spin summed
Feynman diagram%
\footnote{Just one Feynman diagram for gluon emission from the initial quark.}
between the gluon emission vertex and the $q\bar{q} B$ vertex 
($B$ is the EW vector boson) can be summarized as follows:
\begin{enumerate}[(a)]
\item
Apply $P_{spin}$ to decouple a spinor $\gamma$-trace into two parts,
the hard ME part and the ladder part.
\item
Apply the explicit upper limit of the phase space for an emitted gluon
in the transverse momentum,
for example $a<M$, with $a^2\equiv s\beta/\alpha$.
\item
Take the expression for the hard part ME%
\footnote{The one-gluon ladder can remain unchanged, but
 in the case of two gluons in $K_0$
 taking a limiting expression is also done in the ladder.}
in the collinear limit,
$a\to 0$ ($\beta\to 0$)
keeping $\alpha=1-z=const$ (or $\alpha+\beta=1-z=const$)
and extrapolate it all over the phase space.
\item
Keep unchanged the phase space integration element and its limits.
\end{enumerate}
Point (c) of the above recipe is the most
important and requires more discussion.
Finding out the limiting collinear
expression is trivial, see eq.~(\ref{eq:ColF1g}).
What is non-trivial is the {\em off-collinear extrapolation} (OCEX)
of this formula,
out of $\beta=0$ point to all non-collinear phase space.
The simplest recipe: go back along $z=1-\alpha$ line and
use eq.~(\ref{eq:ColF1g})
\begin{equation}
\begin{split}
\label{eq:ColF1gAlf}
d\sigma_{LO} &= 
\frac{C_F \alpha_s}{\pi}\;
\frac{d\alpha d\beta}{\alpha\beta}\;
\frac{d\varphi}{2\pi}\;\\&\times
 \frac{1+(1-\alpha)^2 }{2}\;
 \frac{d\sigma_B((1-\alpha)s,\theta_{F})}{d\Omega}
d\Omega,
\end{split}
\end{equation}
would be acceptable, provided the Born cross section is flat.
In the presence of a narrow resonance in $d\sigma_B(\hat{s})$,
in the $\hat{s}=s(1-\alpha-\beta)=s z$ variable,
this would lead to a disastrous NLO correction $ d\sigma_1-d\sigma_{LO}$,
wildly varying over the phase space.
This kind of OCEX follows a vertical dashed line in 
Fig.~\ref{fig:SudPlanDY5}(a).

However, there is a freedom in the off-collinear extrapolation away
from $\beta=0$ point --
we may do it also along the line $x=1-\alpha-\beta=const$:
\begin{equation}
\begin{split}
\label{eq:ColF1bZ}
d\sigma_{LO} &= 
\frac{C_F \alpha_s}{\pi}\;
\frac{d\alpha d\beta}{\alpha\beta}\;
\frac{d\varphi}{2\pi}\;
\\&\times
 \frac{1+(1-\alpha-\beta)^2 }{2}\;
 \frac{d\sigma_B((1-\alpha-\beta)s,\hat\theta)}{d\Omega}
d\Omega.
\end{split}
\end{equation}
In Fig.~\ref{fig:SudPlanDY5}(a) this kind of OCEX goes along
the curved dashed line there.
The angle $\hat\theta$ has to be also defined within OCEX,
in any reasonable way which coincides
with the correct value  $\theta_{F}$ at the $\beta=0$ point%
\footnote{For instance, as an angle between
 $\vec{q}_2-\vec{q}_1$ and $\vec{p}_{0F}-\vec{p}_{0B}$
 in the rest frame of the EW boson.}.
In a sense, the above is fully compatible with
the methodology of calculating the NLO corrections to the EW
production process in ref.~\cite{Altarelli:1979ub}
where a lightcone variable of the collinear factorization
is also mapped into $x=1-\alpha-\beta$.
The essential difference is that with the help of OCEX
we are replacing the traditional collinear PDF
of ref.~\cite{Altarelli:1979ub}
with the exclusive distribution of eq.~(\ref{eq:ColF1bZ}).
A number of consequences of this replacement will unfold
gradually in the following.

For easier generalization to two and more gluons,
let us formalize slightly the above as:
\[
 \Pbbm'_M=  \PPopp_a \theta_{{M}>a}\; P_{spin},
\]
where $\PPopp_a$ takes collinear limit and implements
off-collinear extrapolation on both sides of its location
in the Feynman diagrams, without affecting the phase-space integration element:
\begin{widetext}
\begin{equation}
\begin{split}
&C_0^{(0)} \Pbbm'_{{M}} K_{0F}^{(1)}
=\int \frac{d^3 k}{2k^0}\;
  K_{0F}^{(1)}(k)\; \Pbbm'_{{M}}  \;
  \int d\tau_2(P-k; q_1,q_2)\; C_0^{(0)}(q_1,q_2,k)
\\&~~~
=\int
  \frac{2C_F \alpha_s}{\pi}
  \theta_{a<{M}}
  \frac{d a}{a}
  \frac{d\varphi}{2\pi}
  \frac{d\bar\alpha}{\bar\alpha}\;
  \frac{1+(1-\bar\alpha)^2}{2}\;\;
  d\Omega_q\;
  \frac{d\sigma_{B}(s(1-\bar\alpha),\hat\theta)}{d\Omega_q}
\theta_{s(1-\bar\alpha)>0}.
\end{split}
\end{equation}
\end{widetext}

Note that $z=1-\bar\alpha=\hat{s}/s$ is within 
the proper limits $0<z<1$.
The above can be formalized by means of introducing
the rescaled 4-momentum $\bar{k}^\mu=\lambda^{-1} k^\mu$, 
\[
\lambda=\frac{\alpha}{\alpha+\beta}=\frac{\alfb}{\alfb+\betb}\leq 1,
\]
where $\alfb=\alpha/\lambda$ and $\betb=\beta/\lambda$.
The dilatation transformation preserves angles,
hence in the MC one may generate $a$ and $\bar{\alpha}=1-z$
according to
\begin{equation}
\begin{split}
&C_0^{(0)} \Pbbm'_{{M}} K_{0F}^{(1)}
=\int \frac{d^3 \bar{k}}{2\bar{k}^0}\;
  \lambda^{2}\;
  K_{0F}^{(1)}(\bar{k})\; \Pbbm'_M  \;
\\&\qquad\qquad\quad\times
  \int d\tau_2(P-k; q_1,q_2)\; C_0^{(0)}(q_1,q_2,k)
\\&
=\int
  \frac{2C_F \alpha_s}{\pi}
  \theta_{a<{M}}
  \frac{d a}{a}
  \frac{d\varphi}{2\pi}
  \frac{d z}{1-z}\;
  \frac{1+z^2}{2}\;\;
  d\Omega_q\;
  \frac{d\sigma_{B}(s z,\hat\theta)}{d\Omega_q}
\theta_{s z>0},
\end{split}
\end{equation}
then construct $\bar{k}^\mu$ and finally rescale it $k=\lambda\bar{k}$.
The ``barred space'' of $\bar{k}^\mu$
depicted in Fig.~\ref{fig:SudPlanDY5}(b)
is merely a re-parametrization of the true phase space
in Fig.~\ref{fig:SudPlanDY5}(a).
For a single gluon the above parametrization
of the phase space may look trivial,
but for many gluons it will be useful.

As already said, the main role of the $\Pbbm$ 
operator in refs.~\cite{Ellis:1978ty,Curci:1980uw}
is to decouple kinematically the hard process and the ladder.
The above $\Pbbm'$ does it also, but more gently,
protecting the 4-momentum conservation.
The kinematic decoupling is seen from the 
phase-space integration
\begin{equation}
\begin{split}
&C_0^{(0)} \Pbbm'_{{M}} K_{0F}^{(1)}
=  \ln\frac{{M}}{q_0}\;\;
   \int_0^1 dz\;
   \frac{2C_F \alpha_s}{\pi}
   \frac{1+z^2}{2(1-z)}\;\;
   \sigma_B(sx),
\end{split}
\end{equation}
which provides exactly the same result
as the collinear factorization in these classic works
where the 4-momentum conservation is broken.
An additional cut-off $a>q_0$ (phase space boundary in the LO MC)
was used in the above.

One may also easily define, within the above scheme,
a prototype of an universal inclusive collinear PDF:
\begin{equation}
\begin{split}
D({M},x) &=
\Pbbm'_{{M}} K_{0F}^{(1)} \Big|_x
=\int \frac{d^3 \bar{k}}{2\bar{k}^0}\;
  K_{0F}^{(1)}(\bar{k})\; \Pbbm'_M  \;
  \delta_{x=1-\alfb}
\\&
=  \ln\frac{{M}}{q_0}\;\;
   \frac{2C_F \alpha_s}{\pi}
   \frac{1+x^2}{2(1-x)},
\end{split}
\end{equation}
thanks to $\Pbbm'_{{M}}$ closing the phase space
from the above by means of the factorization scale ${M}$.
As we shall see later on, in the analogous construction
for the deep inelastic electron--proton scattering,
the role of $\Pbbm'_{{M}}$ is to map the phase space of the ladder
into an idealized phase space of $\bar{k}^\mu$,
decoupled kinematically from the hard process,
thus removing process dependence%
\footnote{A phase space shape (upper limits)
 usually depends on the process type.}
and gaining universality of the ladder part!

\subsection{NLO correction to hard process -- one real gluon}

Having defined the single-gluon ladder parts
$C_0^{(0)} \Pbbm' K_{0F}^{(1)}$ 
and $C_0^{(0)} \Pbbm' K_{0B}^{(1)}$
in the {\em exclusive} way,
within the same exact phase space where the
complete exact 1-gluon distribution%
\footnote{ Here $C_0^{[1]}$ is the interference term, while
 the other two terms are amplitude squares.
} of eq.~(\ref{eq:DYsig1}),
\begin{equation}
\label{eq:DYD1}
D^{[1]}
=C_0^{[1]} + C_0^{[0]} \cdot K_{0F}^{(1)}
           + C_0^{[0]} \cdot K_{0B}^{(1)}
\end{equation}
is defined, it is straightforward to define
the LO $+$ NLO factorized hard process part
in the exclusive (unintegrated) manner:
\begin{equation}
\begin{split}
&
(C_0^{[0]} +C_0^{[1]} )\Big[1+(1- \Pbbm') K_{0F}^{(1)} 
                             +(1- \Pbbm') K_{0B}^{(1)}\Big] \simeq
\\&~~~~~
\simeq
 C_0^{[0]} +C_0^{[1]} +C_0^{[0]}(1-\Pbbm') K_{0F}^{(1)}
                      +C_0^{[0]}(1-\Pbbm') K_{0B}^{(1)}
\\&~~~~~
=  C_0^{[0]} +C_0^{[1]}
  +C_0^{[0]} K_{0F}^{(1)}
  +C_0^{[0]} K_{0B}^{(1)}
  -C_0^{[0]}\Pbbm' K_{0F}^{(1)}
  -C_0^{[0]}\Pbbm' K_{0B}^{(1)}
\\&~~~~~
=  C_0^{[0]} + D^{[1]}
  -C_0^{[0]}\Pbbm' K_{0F}^{(1)}
  -C_0^{[0]}\Pbbm' K_{0B}^{(1)}.
\end{split}
\end{equation}
The above difference of the exact and approximate ME
at the level of the integrand reads
\begin{widetext}
\begin{equation}
\label{eq:DYbeta1F}
\begin{split}
&C_0^{[1]} +C_0^{[0]}(1-\Pbbm') K_{0F}^{(1)}
           +C_0^{[0]}(1-\Pbbm') K_{0B}^{(1)}
=D^{[1]} -C_0^{[0]}\Pbbm' K_{0F}^{(1)}
         -C_0^{[0]}\Pbbm' K_{0B}^{(1)}=
\\&~~
= \int 
  \frac{d\alpha d\beta}{\alpha\beta}
  \frac{d\varphi}{2\pi}\;
  d\Omega_q
  \frac{2C_F \alpha_s}{\pi}\;
\Bigg\{
\Big[
  \frac{(1-\beta)^2}{2}
  \frac{d\sigma_{B}}{d\Omega_q}(\hat{s},\theta_{F})
 +\frac{(1-\alpha)^2}{2}
  \frac{d\sigma_{B}}{d\Omega_q}(\hat{s},\theta_{B})
\Big]
\\&~~~~~~~~~~~~~~~~~~~~
-\theta_{\alpha>\beta}
 \frac{1+(1-\alpha-\beta)^2}{2}
 \frac{d\sigma_{B}}{d\Omega_q}(\hat{s},\hat\theta)
-\theta_{\alpha<\beta}
 \frac{1+(1-\alpha-\beta)^2}{2}
 \frac{d\sigma_{B}}{d\Omega_q}(\hat{s},\hat\theta)
\Bigg\}
= \int dx\; C_{2r}(x)\; \sigma_{B}(sx),
\end{split}
\end{equation}
\end{widetext}
where the $ C_{2r}(x)$ function is calculated in
the Appendix \ref{app:B}.

Note that in the above integral the phase space is
covered by the LO distribution completely,
without any gap or overlap,
provided the factorization scale variable ${M}$ in both
hemispheres is adjusted conveniently.
Similarly, the entire integrand of the NLO correction
is also defined all over the 1-gluon phase space.

\subsection{LO two-gluon $C_0^{(0)} \Pbbm' K_{0F}^{(0)}\Pbbm' K_{0F}^{(0)}$
               -- a prelude for the LO ladder MC}
\label{sec:LO2Gprelude}

Let us take the exact integrated distribution
for the ladder diagram (no projections) with two gluons emission
\begin{equation}
\begin{split}
& \sigma_2=
  C_0^{(0)}\cdot K_{0F}^{(1)} \cdot K_{0F}^{(1)}
=\int\!\! dx_1
 \frac{d^3k_1}{2k_1^0}
 \frac{d^3k_2}{2k_2^0}
 d\tau_2(P-k_1-k_2;q_1,q_2)\;
\\&~~~~~~~~~~~~~~\times
 \rho_B(p_{0F},p_{0B},k_1,k_2,q_1,q_2)\;
 s\delta_{sx_1=(P-k_1)^2}
\end{split}
\end{equation}
in which we have introduced explicitly
a variable being the effective mass squared
\[
\hat{s}_1=sx_1
=(q_1+q_2+k_2)^2=(P-k_1)^2
\]
of the final state system after emitting the gluon $k_1$.

Let us start with the same operation of parametrization
of the phase space in terms of
$k_1=\lambda_1 \bar{k}_1$ as in the previous case of the single gluon,
for a gluon at the end of the ladder:
\begin{equation}
\label{eq:C0K0K0}
\begin{split}
& C_0^{(0)} K_{0F}^{(1)} K_{0F}^{(1)}
=\int\!\! dx_1
 \frac{d^3 \bar{k}_1}{2\bar{k}_1^0}
 \frac{d^3 k_2}{2k_2^0}
 d\tau_2(P-k_1-k_2;q_1,q_2)\;
\\&~~~~~~~~~~~~~~\times
 \lambda_1^2
 \rho_B(p_{0F},p_{0B},\lambda_1 \bar{k}_1,k_2,q_1,q_2)\;
 \delta_{x_1=1-\alfb_1},
\\&
\lambda_1=\frac{s(1-x_1)}{2P \bar{k}_1}
         =\frac{\alfb_1}{\alfb_1+\betb_1}.
\end{split}
\end{equation}
The factor $\lambda_1^2$ from the phase space is compensated by a similar factor
in the 1-gluon distribution,
as it is for the 1-gluon case.
No approximations nor projections are present yet.

Insertion of the first $\Pbbm'$ requires 
examination of the collinear limit
$a_1\to 0$ ($\betb_1\to 0$) keeping $\alfb_1=const$;
in this limit we have also $\beta_1\to 0$,
$\alfb_1\to\alpha$ and $\lambda_1\to 1$.
The spin projection operator of eq.~(\ref{eq:PspinPrim}) is also used.
The collinear limit is well known
\begin{equation}
\begin{split}
\lim_{a_1\to 0}
  a_1^2\; \rho_B(\lambda \bar{k}_1,k_2,...)
&=\frac{1}{\alfb_1^2}
 \frac{2C_F\alpha_s}{\pi^2}
 \frac{1+(1-\alfb_1)^2}{2}\;
\\&\times
 \rho_1\big(x_1p_{0F},p_{0B};k_2,q_1,q_2\big),
\end{split}
\end{equation}
where $\rho_1$ is the already discussed distribution of the 
single-gluon emission $\rho$ of eq.~(\ref{eq:DYsig1}),
in the reduced center-of-mass system of
$sx_1 =(x_1p_{0F}+p_{0B})^2$,
provided we rename $k\to k_2$.

The formula obtained above,
valid originally at the collinear point $a_1=0$, 
is now extrapolated to the off-collinear phase space
using $\bar{k}_1$:
\begin{widetext}
\begin{equation}
\begin{split}
\sigma_{2F} &=
C_0^{(0)} K_{0F}^{(1)} \Pbbm' K_{0F}^{(1)}
\\&
=\int\!\! dx_1
 \frac{d^3 \bar{k}_1}{2\bar{k}_1^0} \lambda_1^2\;
 \frac{d^3 k_2}{2k_2^0}
 d\tau_2(P-\lambda_1 \bar{k}_1-k_2;q_1,q_2)\;
 \frac{2C_F\alpha_s}{\pi^2}
 \theta_{a_2>a_1}
 \frac{1+(1-\alfb_1)^2}{2a_1^2\alfb_1^2}\;
 \rho_1\big(x_1p_{0F},p_{0B};k_2,q_1,q_2\big)
 \delta_{x_1=1-\alfb_1}
\\&
=\int\!\! dx_1
 \delta_{x_1=z_1}
 \frac{C_F\alpha_s}{\pi}
 \frac{\bar{P}(z_1)}{1-z_1}
 \frac{da_1}{a_1}
 \frac{d\varphi_1}{2\pi}
 \theta_{a_2>a_1}
 d\tau_3(P-\lambda_1 \bar{k}_1;q_1,q_2,k_2)\;
 \rho_1\big(z_1p_{0F},p_{0B};k_2,q_1,q_2\big),
\end{split}
\end{equation}
\end{widetext}
where the LO splitting kernel for the first emission of $k_1$
\[
P^{(0)}_{qq}(z)=\frac{\bar{P}(z)}{1-z}
            =\frac{1+z^2}{2(1-z)}
\]
is factorized off explicitly
and the factorization scale for $\Pbbm'_{M_1}$ is just $M_1=a_2$
of the gluon in the next $K_{0F}(k_2)$.

This fact that the factorization scale for the first emission
is defined to be $a_2$ of the second emission is the
essential difference with the standard EGMPR/CFP scheme~\cite{Ellis:1978ty,Curci:1980uw},
where $a_i<\mu$ for both emissions and
therefore $a_2\to 0$ is not blocked by $a_1$ like here.
The EGMPR arrangement has advantage of being similar to
the system of UV subtractions~\cite{Curci:1980uw} but causes
oversubtractions, unfriendly for the MC implementation.
We assume implicitly a cutoff regularizing $a_1\to 0$ limit,
for instance $a_1>a_0$.
Note that, although the distribution of $a_1$ seems to be simple,
we cannot perform $\int \frac{d a_1}{a_1}$ to get a pure log,
because the upper limit $a_2>a_1$ is still nontrivial;
and we have to wait until the next simplifications
due to insertion of the second $\Pbbm'$, 
before getting the pure log from the integration.

Let us now insert second $ \Pbbm'$
into $C_0^{(0)}  \Pbbm' K_{0F}^{(1)} \Pbbm' K_{0F}^{(1)}$.
Again, we examine the limit $a_2\to 0$, keeping $a_2/a_1=const$.
While taking this limit we keep $\alfb_2=x_1-x_2=const$, such that
$\hat{s}=(P-k_1-k_2)^2=const$,
in addition to previous $\hat{s}_1=const$.
More precisely, we start by introducing
\[
\hat{s}=\hat{s}_2=sx=sx_2
=(P-k_1-k_2)^2 =\hat{s}_1 -2(P-k_1)\cdot k_2
\]
as an integration variable:
\begin{equation}
\begin{split}
\sigma_{2F} &=
 \int\limits_{a_2>a_1}\!\!
 dx_1 dx_2
 \frac{d^3 \bar{k}_1}{2\bar{k}_1^0}
 \frac{d^3 k_2}{2k_2^0}
 d\tau_2(P-\lambda_1 \bar{k}_1-k_2;q_1,q_2)\;
\\&\times
 \frac{2C_F\alpha_s}{\pi^2}
 \frac{\bar{P}(1-\alfb_1)}{a_1^2 \alfb_1^2}\;
 \rho_1\big(x_1p_{0F},p_{0B};k_2,q_1,q_2\big)\;
\\&\times
 \delta_{x_1=1-\alfb_1}\;
 s\delta(sx_2 -sx_1 +2(P-k_1)\cdot k_2).
\end{split}
\end{equation}
Next, we perform the same transformations on $\delta$-functions
accompanied by the rescaling $k_2=\lambda_2 \bar{k}_2$:
\begin{equation}
\begin{split}
& s\delta(sx_2 -sx_1 +2(P-k_1)\cdot k_2)
\\&
=\int dY
 s\delta(sx_2 -sx_1 +2(P-k_1)\cdot k_2)\;\;
 Y^{-2} 2p_{0B} \cdot k_2\;
\\&\qquad\times
  \delta(sx_2 -sx_1 +2p_{0B} \cdot k_2 Y^{-1})
\\&
=\int dY
 s\delta(sx_2 -sx_1 +2(P-k_1)\cdot Y \bar{k}_2)\;\;
 Y^{-1} 2p_{0B} \cdot \bar{k}_2\;
\\&\qquad\times
 \delta(sx_2 -sx_1 +2p_{0B} \cdot \bar{k}_2)
\\&
= \frac{\lambda_2^{-1} 2p_{0B} \cdot \bar{k}_2}%
       { 2(P-k_1)\cdot \bar{k}_2 }
 \delta\Big(x_2 -x_1 +\frac{2p_{0B} \cdot \bar{k}_2}{s} \Big)
\\&
=\delta\Big(x_2 -x_1 +\frac{2p_{0B} \cdot \bar{k}_2}{s} \Big)
=\delta_{x_1 -x_2 =\alfb_2},
\end{split}
\end{equation}
where
\[
\lambda_2(\bar{k}_1,\bar{k}_2)
=\frac{s(x_1-x_2)}{2(P-k_1)\cdot \bar{k}_2}
=\frac{s(x_1-x_2)}{2(P-\lambda_1(\bar{k}_1)\; \bar{k}_1)\cdot \bar{k}_2}.
\]
Note that the scaling factor $\lambda_2\to 1$ in the collinear limit
$\betb_2\to 0$.

Let us stress that the integral under consideration 
is now transformed into a new equivalent form,
but the limit $a_2\to 0$ is yet to be taken!
In the transformed variables the integral reads
\begin{equation}
\label{eq:C0K0PK0}
\begin{split}
\sigma_{2F} &=
\int\limits_{a_2>a_1>a_0}\!\!
 dx_1 dx_2
 \frac{d^3 \bar{k}_1}{2\bar{k}_1^0}
 \frac{d^3 \bar{k}_2}{2\bar{k}_2^0}
 d\tau_2(P-\lambda_1 \bar{k}_1-\lambda_2 \bar{k}_2;q_1,q_2)\;
\\&\times
 \frac{2C_F\alpha_s}{\pi^2}
 \frac{\bar{P}(1-\alfb_1)}{a_1^2 \alfb_1^2}\;
 \lambda_2^{2}\;
 \rho_1\big(x_1p_{0F},p_{0B};\lambda_2 \bar{k}_2,q_1,q_2\big)\;
\\&\times
 \delta_{x_1=1-\alfb_1}\;
 \delta_{x_1 -x_2 =\alfb_2}.
\end{split}
\end{equation}
Now we are ready to take the limit $a_2\to 0$ keeping $a_1/a_2=const$
(also $a_0\to 0$) and $\alfb_i=const$:
\begin{equation}
\label{eq:LOLO}
\begin{split}
& 
C_0^{(0)}  \Pbbm' K_{0F}^{(1)} \Pbbm' K_{0F}^{(1)}=
\\&
=\int\limits_{{M}>a_2>a_1}\!\!
 dx_1 dx_2
 \frac{d^3 \bar{k}_1}{2\bar{k}_1^0}
 \frac{d^3 \bar{k}_2}{2\bar{k}_2^0}
 d\tau_2(P-\lambda_1 \bar{k}_1-\lambda_2 \bar{k}_2;q_1,q_2)\;
\\&~~~~~~~~~~~~~~\times
 \frac{2C_F\alpha_s}{\pi^2}
 \frac{\bar{P}(x_1)}{\alfb_1^2 a_1^2}\;
 \frac{2C_F\alpha_s}{\pi^2}
 \frac{\bar{P}(x_2/x_1)}{\alfb_2^2 a_2^2}\;
 \frac{d\sigma_B}{d\Omega}(sx_2,\hat\theta)\;
\\&~~~~~~~~~~~~~~\times
 \delta_{x_1=1-\alfb_1}\;
 \delta_{x_2 =1-\alfb_1-\alfb_2}.
\end{split}
\end{equation}
Note that the $\lambda_2^2$ factor from the phase space and
the matrix element cancels out as before.
The above formula is the principal result
of this subsection. It defines the double use of $\Pbbm'$,
the transformation $k_i(\bar{k}_j)$ and its inverse
$\bar{k}_j(k_i)$, $i,j=1,2$.
Note that in the above formulae we could use
variables $z_1=1-\alfb_1=x_1$ and 
$z_2=(1-\alfb_1-\alfb_1)/(1-\alfb_1)=x_2/x_1$
instead of $\alfb_i,\; i=1,2$.
In the following we may find it useful to switch to the $z_i,\; i=1,2,$ variables.

With the global factorization scale ${M}$ inserted at the end
of the LO ladder the transverse-plane integration
$
\!\! \int\limits_{{M}>a_2>a_1>a_0}\!\!
\frac{d a_2}{a_2} \frac{d a_1}{a_1}
=\frac{1}{2!} \ln^2\frac{{M}}{a_0}
$
now decouples and provides a pure double log%
\footnote{
  With our phase space parametrization in terms of $\bar{k}_i$
  the above mechanism of producing pure logs is a general phenomenon,
  because ${M}$ is always at the end of the ladder,
  and the ladder has a built-in time-ordered exponential.
}:
\begin{equation}
\begin{split}
\label{eq:PureLog2}
C_0^{(0)}  \Pbbm' K_{0F}^{(1)} \Pbbm' K_{0F}^{(1)}
&= \frac{1}{2!} \ln^2\frac{{M}}{a_0}\;
  \Big( \frac{2C_F\alpha_S}{\pi} \Big)^2
\\&~~~\times
  \int\limits_0^1 dx
  \big[P_{qq}\otimes P_{qq}\big]_{2R}(x)\;
  \sigma_B(sx),
\end{split}
\end{equation}
where
\[
\begin{split}
4[P^{(0)}_{qq}\otimes P^{(0)}_{qq}]_{2R}(z) &=
 \frac{1+z^2}{1-z} \Big[ 4\ln\frac{1}{\delta} +4\ln(1-z)\Big]
\\&~~~
 +(1+z)\ln z -2(1-z)
\end{split}
\]
is just a double convolution of the LO kernel
with the IR regularization $\alpha_i>\delta$.

The distribution of eq.~(\ref{eq:LOLO}) is easy to generate
in the Monte Carlo.
First, one generates $a_i$ and $\alfb_i$
paying attention to the constraint $x=x_2=1-\alfb_1-\alfb_2$,
and $\bar{k}_i$ are constructed in the laboratory frame.
Then  $\lambda_i$ are calculated and
the rescaling $\bar{k}_i^\mu\to k_i^\mu$ is done (in two steps!).
Finally, in the frame $P-k_1-k_2$ one generates $q_i$ according
to the Born differential distribution.
The phase space boundary $\hat{s}\geq 0$ is obeyed automatically.
In the MC the soft IR regulator
$z_i=x_i/x_{i-1}<1-\delta$ will be introduced
and the overall virtual Sudakov form-factor
$\exp\big(-\frac{2C_F\alpha_s}{\pi} 
 \ln\frac{{M}}{q_0}\ln\frac{1}{\delta}\big)$
will also be supplemented.

The above example demonstrates the most important features of the $\Pbbm'$
projector 
(see ref.~\cite{Jadach:2010ew} for more details).
In particular the following lessons are to be learnt:
\begin{enumerate}[(a)]
\item
The phase space parametrization $k_i \to \bar{k}_i $ plays an important role
in $\Pbbm'$,
as it is instrumental in implementing off-collinear extrapolation,
and also helps to take the collinear limit in the first place.
\item
The rescaled 4-momenta $\bar{k}_i $ violate the 4-momentum conservation,
similarly like 4-momenta after action of the kinematical projector
of refs.~\cite{Ellis:1978ty,Collins:1998rz}.
In our case, however, the off-collinear extrapolation 
is {\em effectively undoing} this kinematical projection%
\footnote{The undoing is ``effective'',
  because the kinematical projection
  of refs.~\cite{Ellis:1978ty,Collins:1998rz} 
  in our methodology is never done.
  A clever parametrization of the phase space is the only 
  thing really done.}
and allows to operate in the original phase space,
with the 4-momentum conservation untouched.
\item
A nice accident of the Jacobian
$|\partial(\bar{k}_1,\bar{k}_2) /\partial(k_1,k_2)|$
being compensated by the matrix element is generally not guaranteed. 
However, if it were not true, we would
have to impose this by hand,
such that a pure logarithm results from the phase space integration,
like in eq.~(\ref{eq:PureLog2}),
similarly as in the definition of the collinear counterterm in
refs.~\cite{Frixione:2002ik,GehrmannDeRidder:2007hr}, for example.
\item
The role of the parametrization $k_i \to \bar{k}_i $
in assuring universality (process-independence)
of the ladder parts will be clarified once the MC
for the DY and DIS processes with the NLO corrections
to the hard part are defined, see below.
\item
The phase space parametrization in terms of $\bar{k}_i$
will be used also inside $K_0$ to parametrize
the two-gluon phase space,
for instace for the 2-gluon crossed diagram in the NLO ladder.
\item
The soft eikonal limit is protected by $\Pbbm'$,
because rescaling $\bar{k}_i \to k_i$ preserves it.
\end{enumerate}
For a better (complete) understanding of the construction of $\Pbbm'$ one needs
to examine in a fine detail the case of two gluons in the middle
of the ladder, for instance%
\footnote{ This task is pursued
   in separate works~\cite{Jadach:2011kc,IFJPAN-IV-2012-7}.}
in $C_0 \Pbbm' K_{0F} (1-\Pbbm') K_{0F}$,
which provides the NLO correction to the evolution kernel.

\section{Monte Carlo for EW boson production}
\label{sec:DYMC}

The insertion of the $\Pbbm'$ operator into the LO gluonstrahlung
ladder with any number of gluons can be done similarly 
as in Sec.~\ref{sec:LO2Gprelude} obtaining
the distribution ready for the LO MC modelling
of the production process of the EW boson
with multiple gluons emitted from the incoming quarks.

\subsection{Simplified single ladder case}
We start with the gluonstrahlung
ladder just in one hemisphere in order to avoid algebraic
complications of the two-ladder case
(to be dealt with in the next subsection):
\begin{equation}
\label{eq:LOMCF}
\begin{split}
& 
C_0^{(0)}  \cdot\Gamma_F=
\sum_{n=1}^\infty
\left\{C_0^{(0)} (\Pbbm' K_{0F}^{(1)})^n\right\}_{T.O.}=
\\&
=e^{-S_F}
\sum_{n=0}^\infty
\int dx\;\;
\bigg(
\prod_{i=1}^n
 d^3\Ecal(\bar{k}_i)\;
 \theta_{\eta_i<\eta_{i-1}}
 \frac{2C_F\alpha_s}{\pi^2} \bar{P}(z_i)
\bigg)\;
\\&~~~~~~~~~~~~~~\times
 d\tau_2(P-\sum_{j=1}^n k_j;q_1,q_2)\;
\theta_{\Xi<\eta_n}
\delta_{x =\prod_{j=1}^n z_j}\;
\frac{d\sigma_B}{d\Omega}(sx,\hat\theta),
\end{split}
\end{equation}
where
\[
\begin{split}
&k_i=\lambda_i \bar{k}_i,\quad
\lambda_{i}= \frac{s(x_{i-1}-x_i)}%
                {2(P-\sum_{j=1}^{{i-1}} k_j )\cdot \bar{k}_i},
\\&
x_i=1- \sum_{j=1}^{i}\alfb_i=\prod_{j=1}^{i} z_j, \quad
z_i= \frac{x_i}{x_{i-1}},
\end{split}
\]
and $\bar{P}(z)=\frac{1}{2}(1+z^2)$.
The $a$-ordering, ${M}>a_i>a_{i-1}\;\; i=1,...,n$,
is rephrased into an equivalent rapidity 
ordering $\Xi<\eta_i<\eta_{i-1}<\eta_0$,
where ${M}= \sqrt{s} e^{-\Xi}$,
see the definitions of the phase space--integration element $d^3\Ecal(k)$
and of other kinematic notations in Appendix \ref{app:A}.
The $T.O.$ subscript stands for the time-ordering exponential
structure in the factorization scale,
see ref.~\cite{Jadach:2010ew} for a general definition.
The $S_F$ function is the usual MC Sudakov 
form-factor depending on the shape
of the IR-boundary $\alfb_i\to 0$.
Factorization scale is now defined as
the minimum rapidity $\Xi$ for gluons in the F hemisphere
(maximum rapidity for gluons in the B hemisphere).
For the moment $\Xi$ is a free parameter, to be defined more
precisely later on.

Note that the definition of $\lambda_i$ 
is {\em recursive}, that is to define $\lambda_i$
one must know $\lambda_{i-1}$.
In a typical MC event first $\lambda$'s,
corresponding to very collinear gluons 
will be very close to one, $\lambda_i\simeq 1$;
only the last ones corresponding to {\em non-collinear non-soft} gluons,
i.e. close to the hard process,
will be rescaled by a significant $\lambda_i \neq 1$ factor.

Similarly as in the case of two gluons in eq.~(\ref{eq:PureLog2}),
the transverse integration decouples and is feasible analytically:
\begin{equation}
\begin{split}
& 
C_0^{(0)}  \cdot\Gamma_F
=\int_0^1 dx\; G_F({M},x)\; \sigma_B(sx),
\\&
G_F({M},x)=
 e^{ -\frac{2C_F\alpha_S}{\pi} \ln\frac{1}{\Delta} \ln\frac{{M}}{q_0}}
\\&\qquad\qquad\times
 \bigg\{
 \delta_{x=1}
+\sum_{n=1}^\infty 
 \frac{1}{n!} 
 \Big( \frac{2C_F\alpha_S}{\pi} \Big)^n
 \ln^n\frac{{M}}{q_0}\;\;
 [P^{(0)}_{qq}]^{\otimes n}(x)
 \bigg\},
\end{split}
\end{equation}
\newline
where the IR regularization $(1-z_i)<\Delta$
is used in the $n$-times convolution of the LO kernel 
$[P^{(0)}_{qq}]^{\otimes n}$.
It should be stressed that the above LO formula
represents the LO MC without any approximation.

The inclusive (bare) PDF $G_F({M},x)$ of the MC obeys
by construction
the LO evolution equation:
\begin{equation}
\frac{\partial}{\partial \ln {M}} 
 G_F({M},x)
= [\Peu^{(0)}_{qq} \otimes G_F({M})](x),
\end{equation}
where 
$
\Peu^{(0)}_{qq}(z) = 
 \frac{2C_F\alpha_S}{\pi}
 \left[\frac{1+z^2}{2(1-z)}\right]_+
$.
This is essentially due to the use of the $T.O.$ exponent 
in eq.~(\ref{eq:LOMCF}).

\subsection{Two ladder LO case}
Let us now consider the case of two ladders.
In the backward (B) hemisphere $x_i$ are related to $\betb_i$
(instead of $\alfb_i$),
the evolution runs towards larger rapidity,
otherwise all algebraic structure is the same.
Again, we first express the LO MC master formula in terms
of rescaled 4-momenta $\bar{k}_i=\lambda_i k_i$,
but we postpone the definition of $\lambda_i$,
as it will be a little bit special.
We propose the following multigluon distribution
to be implemented in the LO approximation:%
\footnote{
 The reader should keep in mind that the above is for the 
``primordial'' quark and antiquark initial beams,
 and their distributions in hadron will be added in the MC program.}
\begin{widetext}
\begin{equation}
\label{eq:LOMCFBmaster}
\begin{split}
& 
C_0^{(0)}  \cdot \Gamma_F^{(1)} \cdot \Gamma_B^{(1)}=
\sum_{n_1=1}^\infty
\sum_{n_2=1}^\infty
\big[
  C_0^{(0)} 
  (\cdot \Pbbm'  K_{0F}^{(1)})^{n_1}
  (\cdot \Pbbm'  K_{0B}^{(1)})^{n_2}
\big]_{T.O.}=
\\&
=\sum_{n_1=0}^\infty\;
 \sum_{n_2=0}^\infty
 \int dx _F\; dx_B\;
e^{-S_{_F}}
\int_{\Xi<\eta_{n_1}}
\bigg(
\prod_{i=1}^{n_1}
 d^3\Ecal(\bar{k}_i)
 \theta_{\eta_i<\eta_{i-1}}
 \frac{2C_F\alpha_s}{\pi^2} \bar{P}(z_{Fi})
\bigg)\;
\delta_{x_F =\prod_{i=1}^{n_1} z_{Fi}}\;
\\&\qquad\qquad\qquad~~~~~~~~~\times
e^{-S_{_B}}
\int_{\Xi>\eta_{n_2}}
\bigg(
\prod_{j=1}^{n_2}
 d^3\Ecal(\bar{k}_j)
 \theta_{\eta_j>\eta_{j-1}}
 \frac{2C_F\alpha_s}{\pi^2} \bar{P}(z_{Bj})
\bigg)\;
\delta_{x_B =\prod_{j=1}^{n_2} z_{Bj}}\;
 d\tau_2(P-\sum_{j=1}^{n_1+n_2} k_j ;q_1,q_2)\;
\frac{d\sigma_B}{d\Omega}(sx_Fx_B,\hat\theta).
\end{split}
\end{equation}
\end{widetext}
In the above we understand that the F part
is the forward part of the phase space
$\eta_{0F}>\eta_i>\Xi$ and
the B part is the backward part
$\Xi>\eta_i>\eta_{0B}$.
The rapidity boundary $\Xi$ between the F and B parts
is kept as a free parameter as long as possible,
to be adjusted later on.
In particular $\Xi=0$ is perfectly legal
and may serve as the first choice, before something
better is found%
\footnote{In practice the choice of $\Xi$ may be
 quite complicated, for instance it can be 
 correlated with the rapidity position of the EW boson.
 In the NLO corrections calculation,
 for a single gluon, $\Xi=0$ should be set.}.
Variables $z_{Fi}$ and $z_{Bj}$ could be defined similarly
as in the single hemisphere case, provided we perform
kinematical mappings asymmetrically,
first in one hemisphere and then in the other one.
We shall do it below, however, in a more sophisticated symmetric way.
The definition of $z_{Fi}$ and $z_{Bj}$ will result from that.

The angle $\hat\theta$ can be defined with respect to any reasonable
$z$-axis in the rest frame of $P-\sum k_j$, for instance
along $\vec{p}_{0F}-\vec{p}_{0B}$ in this frame.
The boundary between the F and B phase space is at the rapidity $\Xi$.
It is also understood that the differential cross section
of eq.~(\ref{eq:LOMCFBmaster}) is implicitly convoluted with
some initial quark distributions at $\eta_{0F}$ and $\eta_{0B}$
and the appropriate boost is done from the reference frame
of the quark--antiquark frame $p_{0F}+p_{0B}$
to the laboratory frame.

Again, in eq.~(\ref{eq:LOMCFBmaster}),
phase space can be integrated analytically over transverse momenta,
providing the classical factorization formula
\begin{equation}
\label{eq:LOfactDY2LO}
C_0^{(0)}  \Gamma_F^{(1)} \Gamma_B^{(1)} =
\int_0^1 dx_F\;dx_B\;
G_F(\Xi, x_F)\; G_B(\Xi, x_B)\;
\sigma_B(sx_Fx_B),
\end{equation}
where $G_B(\Xi, x)=G_F(\Xi, x)$.
The remarkable feature is that the above LO formula
represents the exact LO MC without any approximations!

We could define the dilatation parameters $\lambda_i$
(recursively) first for one ladder and then for the other,
but this solution would be asymmetric.
Instead, we define the dilatation parameters
in $k_i=\lambda \bar{k}_i$ in a more sophisticated way,
for both hemispheres simultaneously.
For that purpose we introduce a new ordering (indexing) of gluons 
according to the distance  $|\eta_i-\Xi|$ from the 
rapidity $\Xi$ of the EW boson%
\footnote{%
    Which is also the phase space boundary between the F and B ladders.}.
Formally, we define a permutation
\[
 \pi=\{\pi_1,\pi_2,...,\pi_{n_1+n_2}\} 
\]
of all gluons in the F+B phase space, such that 
\[
 |\eta_{\pi_{i}}-\Xi|>|\eta_{\pi_{i-1}}-\Xi|,\quad i=1,...,n_1+n_2,
\]
With the help of the above simultaneous ``double ordering''
in the F and B hemispheres,
we define in a {\em recursive} way:
\begin{equation}
\label{eq:lambdaFB}
k_{\pi_i}=\lambda_{i} \bar{k}_{\pi_i},\quad
\lambda_{i}= 
\frac{s(x_{i-1}-x_{i})}%
     {2(P-\sum_{j=1}^{i-1} k_{\pi_j} )\cdot \bar{k}_{\pi_i}},\quad
i=1,2,...,n_1+n_2.
\end{equation}
where $x_{i}$ now is
\[
 x_i= \prod_{j=1}^i z_{(F,B)\pi_j},
\]
where $(F,B)$ means that we insert in the above product
either $z_{Fj}$ or $z_{Bj}$,
depending whether $\pi_j$ points to the F or B region.
The parameter $\lambda_i$ is defined in
eq.~(\ref{eq:lambdaFB}) recursively by means of solving
the following equation
\[
 \bar{s}_i= s x_i= (P-\sum_{j=1}^{i} k_{\pi_j})^2
          = (P-\sum_{j=1}^{i} \lambda_j \bar{k}_{\pi_j})^2.
\]

The LO Monte Carlo algorithm using the above algebraic
framework can be described step by step as follows:
\begin{enumerate}
\item
Variables $x_F$ and $x_B$ 
are generated with the help of the {\tt FOAM} program~\cite{foam:2002},
next parton (gluon) multiplicities $n_{1,2}$ and
variables $z_{Fi}$ and $z_{Bj}$ are generated
using the constrained MC (CMC) algorithm~\cite{Jadach:2005bf}.%
\footnote{In the CMC algorithm parton multiplicity is
  generated as the first variable,
  contrary to the Markovian and backward
  evolution algorithms where parton multiplicity is generated
  at the end of the MC algorithm.}
The $\Xi$ variable is set.
\item
The four-momenta $\bar{k}_i^\mu$ are defined separately
in the F and B parts of the phase space using the CMC module,
with the corresponding constraints:
$1-\sum\limits_{j\in F} \alfb_j= \prod\limits_{j\in F} z_{Fj}=x_F$
and $1-\sum\limits_{j\in B} \betb_j= \prod\limits_{j\in B} z_{Bj}=x_B$.
\item
Permutation $\pi$ with the simultaneous ordering in $F+B$ space is established.
\item
Using $P$ and $\bar{k}_{\pi_1}$ the rescaling parameter $\lambda_1$
is calculated, then $k_{\pi_1}=\lambda_1 \bar{k}_{\pi_1}$ is set.
At this stage $(P-k_{\pi_1})^2=sx_1$,
where $x_1=z_{\pi_1}=1-\alfb_{\pi_1}$ 
or $x_1=z_{\pi_1}=1-\betb_{\pi_1}$,
depending whether $k_{\pi_1}$ was in the F or B part of
the phase space.
\item
Using $P-k_{\pi_1}$ and $\bar{k}_{\pi_2 }$
the parameter $\lambda_2$ is found and
$k_{\pi_2}=\lambda_2 \bar{k}_{\pi_2}$ is set,
enforcing $(P-k_{\pi_1}-k_{\pi_2})^2=sx_2=sz_{\pi_1}z_{\pi_2}$.
The recursive procedure continues until the last gluon.
\item
In the rest frame of $\hat{P}= P-\sum_j k_{\pi_j}$,
lepton 4-momenta $q_1^\mu$ and $q_2^\mu$ 
are generated according to the Born angular distribution.
\end{enumerate}
The definition of $z_{Fj}$ and $z_{Bj}$ 
in terms of $\alfb_j,\; j\in F$ and $\betb_j,\; j\in B,$ follows
from the above algorithm and 
is more complicated that in the case of one hemisphere.
The main advantage of the above scenario is that
this way the kinematics of the two hemispheres is interrelated
very gently, starting from very collinear gluons 
(for which $\lambda_i \simeq 1$)
and finishing with the least collinear ones, 
next to the hard process EW boson
(the angular ordering is the same for $k_i$ and $\bar{k}_i$).

As compared to ref.~\cite{Jadach:2007zz}, where a similar algorithm
based on CMC~\cite{Jadach:2005bf} and rescaling 
of 4-momenta was proposed,
the present algorithm does the ``rescaling''%
\footnote{The ``rescaling'' in our case
 is merely a synonym for parametrization of
 exact phase space in terms of the rescaled 4-momenta.}
in a more sophisticated way.
The rescaling affects mainly the hard non-collinear gluons,
not the soft and collinear ones,
while the rescaling in ref.~\cite{Jadach:2007zz} was global,
similarly as global manipulations on the 4-momenta 
(boosts and rescaling) used in other
parton shower MCs~\cite{Sjostrand:1985xi,Webber:1984if}
in order to impose the 4-momentum conservation.
Moreover, kinematic parametrization of the phase space in the present
MC is based on the projector $\Pbbm'$
of the collinear factorization (instead of being {\it ad hoc}),
which is essential for completing the NLO.

\subsection{Real NLO correction to hard process}

The NLO correction in the EW boson production hard process
(non-singlet) will be implemented
using a single ``monolithic'' MC weight, which reweighs the LO distributions
defined in the previous subsection.
The real emission part of the NLO correction in this weight
comes from the integrand of eq.~(\ref{eq:DYbeta1F})
which we rewrite as follows:
\begin{widetext}
\begin{equation}
\label{eq:DYbeta1FB}
\begin{split}
&C^{[1]}
=C_0^{[1]} +C_0^{[0]}(1-\Pbbm') K_{0F}^{(1)}
           +C_0^{[0]}(1-\Pbbm') K_{0B}^{(1)}
= \int d^3\Ecal(k)\;
  d\Omega_q
  \frac{2C_F \alpha_s}{\pi^2}\;
  \tbet_1(\hat{p}_F,\hat{p}_B;q_1,q_2,k);
\\&
\tbet_1(\hat{p}_F,\hat{p}_B;q_1,q_2,k)=
\Big[
  \frac{(1-\beta)^2}{2}
  \frac{d\sigma_{B}}{d\Omega_q}(\hat{s},\theta_{F})
 +\frac{(1-\alpha)^2}{2}
  \frac{d\sigma_{B}}{d\Omega_q}(\hat{s},\theta_{B})
\Big]
\\&\qquad\quad~~~~~~~~~~~~~~~~~~~~
-\theta_{\alpha>\beta}
 \frac{1+(1-\alpha-\beta)^2}{2}
 \frac{d\sigma_{B}}{d\Omega_q}(\hat{s},\hat\theta)
-\theta_{\alpha<\beta}
 \frac{1+(1-\alpha-\beta)^2}{2}
 \frac{d\sigma_{B}}{d\Omega_q}(\hat{s},\hat\theta).
\end{split}
\end{equation}
\end{widetext}
In the following use of function
$\tbet_1(\hat{p}_F,\hat{p}_B,...)$, defined in eq.~(\ref{eq:DYbeta1FB}),
vectors $\hat{p}_F$ and $\hat{p}_B$ ($\hat{p}_{F,B}^2=0$)
result from the last insertion of $\Pbbm'$ before the hard process
and they are defined in the rest frame of $\hat{P}=q_1+q_2$
to determine 
$\hat\theta=\angle(\vec{q}_1,\vec{p}_{0F})$
in the LO part of the Born cross section.
On the other hand, angles $\theta_F$ and $\theta_B$
are defined with respect to the original
$-\vec{p}_{0B}$ and $\vec{p}_{0F}$ in this frame.
They will all coincide when all gluons get collinear.

The distribution of the MC with the LO+NLO hard process
is now defined as follows:
\begin{equation}
\label{eq:NLODYMCmaster}
\begin{split}
& C^{(1)}  \Gamma_F^{(1)} \Gamma_B^{(1)}
=\sum_{n_1=0}^\infty\;
 \sum_{n_2=0}^\infty
 \int dx_F\; dx_B\;
\\&\times
e^{-S_{_F}}
\int_{\Xi<\eta_{n_1}}
\bigg(
\prod_{i=1}^{n_1}
 d^3\Ecal(\bar{k}_i)
 \theta_{\eta_i<\eta_{i-1}}
 \frac{2C_F\alpha_s}{\pi^2} \bar{P}(z_{Fi})
\bigg)\;
\delta_{x_F =\prod_i z_{Fi}}\;
\\&\times
e^{-S_{_B}}
\int_{\Xi>\eta_{n_2}}
\bigg(
\prod_{j=1}^{n_2}
 d^3\Ecal(\bar{k}_j)
 \theta_{\eta_j>\eta_{j-1}}
 \frac{2C_F\alpha_s}{\pi^2} \bar{P}(z_{Bj})
\bigg)
\delta_{x_B =\prod_j z_{Bj}}\;
\\&\times
 d\tau_2\Big(P-\sum_{j=1}^{n_1+n_2} k_j ;\; q_1,q_2\Big)\;
\frac{d\sigma_B(sx_Fx_B,\hat\theta)}{d\Omega}\;
W^{NLO}_{MC},
\end{split}
\end{equation}
where the MC weight is
\begin{equation}
\label{eq:NLODYMCwt}
\begin{split}
W^{NLO}_{MC}=
1+\Delta_{S+V}
&+\sum_{j\in F} 
 \frac{\tbet_1(\hat{s},\hat{p}_F,\hat{p}_B;a_j, z_{Fj})}%
      {\bar{P}(z_{Fj})\;d\sigma_B(\hat{s},\hat\theta)/d\Omega}
\\&
+\sum_{j\in B} 
 \frac{\tbet_1(\hat{s},\hat{p}_F,\hat{p}_B;a_j, z_{Bj})}%
      {\bar{P}(z_{Bj})\;d\sigma_B(\hat{s},\hat\theta)/d\Omega},
\end{split}
\end{equation}
with the NLO soft+virtual correction $\Delta_{V+S}$
to be defined separately in the following subsection.

Our construction of the above MC weight of eq.~(\ref{eq:NLODYMCwt})
is quite similar to that proposed in ref.~\cite{Jadach:2009gm}
for the NLO corrections in the middle of the ladder%
\footnote{
 The MC weight of ref.~\cite{Jadach:2009gm},
 when applied to the hard process,
 would also render complete NLO in the hard process and
 provide the same inclusive LO+NLO cross section.
}.
The difference is that the proposal of ref.~\cite{Jadach:2009gm}
was based entirely on the Bose--Einstein symmetrization
of the multigluon emission in the LO MC and the resulting
weight was more complicated,
while the MC weight of eq.~(\ref{eq:NLODYMCwt})
is significantly simpler and algebraically similar to that of 
ref.~\cite{yfs:1961}
(albeit for the collinear rather 
than  soft gluon resummation).

The meaning of the arguments in $\tbet_1$
is such that  in eq.~(\ref{eq:DYbeta1FB})
$\hat{s}=sx_Fx_B$, and three angles $\hat\theta$,
$\theta_F$ and $\theta_B$ have already been explained above.
What remains to be specified
is the definition of $\alpha_j$ and $\beta_j$
in terms of the variables $a_j, z_{j}$
in $j\in F,B$ parts of the phase space as follows:
\[
\begin{split}
&\alpha_j=1-z_{Fj},\quad
\beta_j=  \alpha_j\; a^2_j/a^2_{\Xi},\quad 
{\rm for}~~~ j\in F,
\\&
\beta_j=1-z_{Bj},\quad
\alpha_j=\beta_j\;  a^2_{\Xi}/a^2_j ,\quad
{\rm for}~~~ j\in B,
\end{split}
\]
where the rapidity $\eta_j$ is translated properly
into $a_j$ and 
the rapidity $\Xi$ corresponds to $a_\Xi$.

As compared to earlier attempts to consistently 
implement the NLO corrections to the hard process in the parton shower MC,
the proposals of refs.~\cite{Stephens:2007zz,Jadach:2007zz}
were going in a similar direction.
However, the present work differs from these works
in three important points:
(i) the virtual corrections $\Delta_{S+V}$
are added here,
(ii) the method of combining
the NLO correction with the LO MC proposed here 
is systematic and NLO-complete,
(iii) the treatment of the kinematics is
compatible with the principles of the collinear factorization.

Comparison of our methodology with the well established methods of
MC@NLO~\cite{Frixione:2002ik} and POWHEG~\cite{Nason:2004rx}
is done in section~\ref{sec:MCatNLO}.

\subsection{NLO analytical factorization formula}

For the LO MC defined in eq.~(\ref{eq:LOMCFBmaster})
we have seen that without compromising the exact
phase space for multiple gluons (keeping the 4-momentum conservation)
we could get the contributions of ladder parts factorizing off,
exactly as in the traditional collinear factorization
(in which the 4-momentum is not conserved).
The above seems to be almost miraculous in view of the complicated
nature of the exact complete phase space
and the fact that no approximation was done.

What is even more amazing is that the same nice exact factorization
is also true for the MC with the NLO-corrected hard process
included according to eq.~(\ref{eq:NLODYMCmaster}).
The key point is that analytical integration of the phase space,
again without any approximation,
in eq.~(\ref{eq:NLODYMCmaster}) is feasible (!)
and leads to a simple familiar result
\begin{equation}
\begin{split}
\label{eq:DYanxch}
C^{(1)}  \Gamma_F^{(1)} \Gamma_B^{(1)} &=
\int\limits_0^1 dx_F\;dx_B\; dz\;
G_F(\Xi, x_F)\;G_B(\Xi, x_B)\;
\\&\times
\sigma_B(szx_Fx_B)
\bigg\{
\delta_{z=1}(1+\Delta_{S+V})
+C_{2r}(z)
\bigg\},
\end{split}
\end{equation}
where
$C_{2r}(z) 
=\frac{2C_F \alpha_s}{\pi}\; \left[ -\frac{1}{2}(1-z) \right]
$, see Appendix B.
Two LO PDFs, $G_F(\Xi, x_F)$ and $G_B(\Xi, x_B)$
are these of eq.~(\ref{eq:LOfactDY2LO}).
The  algebraic proof of the above formula can
be found in Appendix C%
\footnote{
 A quite similar formula for introducing the NLO correction to the LO
 kernel in the middle of the ladder was already tested numerically
 in the MC exercise with 4-digit precision, see ref.~\cite{Jadach:2010ew}.
}.

Note that the LO PDFs, $G_{F,B}(\Xi, x)$,
are by construction in the collinear factorization scheme
specific for the MC, 
it will be the same in the MC for the DIS process --
the scheme dependence of physical observables will cancel as usual,
for instance while transferring experimental knowledge on the parton
distributions from the DIS process to the DY process (or vice versa).

Let us comment on the $z$-dependent $C_{2r}(z)$ term
in eq.~(\ref{eq:DYanxch}), since it is different from
what we see in ref.~\cite{Altarelli:1979ub}%
\footnote{
 It is also absent in QED, in the 
 well known formula of Bonneau-Martin~\cite{Bonneau:1971mk}.},
where it is simply absent.
It is not present there
because it is compensated by the twin terms
$\sim -\frac{1}{2}\veps(1-z) \frac{1}{\veps}\delta(1-y^2)$
originating from the $\gamma$-traces and located at $y=\pm 1$,
that is exactly at the collinear poles
(for gluon strictly parallel with the quark).
This cancellation
is not disturbed by the $\overline{MS}$ 
collinear counterterm subtraction.
In our MC factorization scheme this term is included in the counterterm,
hence the net contribution from one of these terms,
contrary to $\overline{MS}$,
stays uncancelled in eq.~(\ref{eq:DYanxch}).

\subsection{NLO soft+virtual corrections}
Let us consider the unsubtracted results
of Altarelli--Ellis--Martinelli (AEM) work~\cite{Altarelli:1979ub}%
  \footnote{ See eq.~(89) there.},
in which real and virtual 1-gluon emission diagrams are combined
and integrated over the phase space,
keeping the variable $z=x=\hat{s}/s=1-\alpha-\beta$ under control:
\begin{equation}
\label{eq:AEM79}
\begin{split}
&\sigma^{AEM}_{DY}
= \int_0^1 dz\; f^{\rm AEM}_{DY}(z)\; \sigma_B(zs),
\\
&f^{\rm AEM}_{DY}(z)
 =\delta(1-z)\;
 +\delta(1-z)
  \frac{C_F \alpha_s}{\pi}\;\left( \frac{2}{3}\pi^2 -\frac{7}{4} \right)
\\&\qquad\quad~~
 +2\frac{C_F \alpha_s}{\pi}
  \left( \frac{\hat{s}}{\mu^2}\right)^\veps
  \left( \frac{\bar{P}(z)}{1-z} \right)_+
  \left(
    \frac{1}{\veps} +\omega_2
  \right)
\\&\qquad\quad~~
 +\frac{C_F \alpha_s}{\pi}\;
 \left\{ 
      4\bar{P}(z)\frac{\ln(1-z)}{1-z}
     -2\bar{P}(z)\frac{\ln z}{1-z}
 \right\}_+,
\end{split}
\end{equation}
where $\bar{P}(z)=\frac{1}{2}(1+z^2)$, $\omega_2=\gamma_E -\ln 4\pi$
and the incoming quark and antiquark are massless and on-shell.
The NLO real 1-gluon emission distribution
entering in the above AEM result is identical to 
$D^{[1]}$ of eq.~(\ref{eq:DYD1})
and $\sigma^{AEM}_{DY}$ includes the complete all (gluonstrahlung) virtual corrections,
including the quark wave--function renormalization constant $Z_F$.
Since $\sigma^{AEM}_{DY}$ is gauge invariant,
the calculation of eq.~(\ref{eq:AEM79})
is done in ref.~\cite{Altarelli:1979ub} in the convenient Landau gauge%
\footnote{ In Landau gauge $Z_F=1$ up to the NLO level.}.

In the formal standard $MS$ methodology one subtracts
from $f^{\rm AEM}_{DY}(z)$ two LO collinear counterterms%
\begin{equation}
\label{eq:KMS}
K_F^{MS}(z)+K_B^{MS}(z)=
2\; K_F^{MS}(z)=
2\; \frac{C_F \alpha_s}{\pi}\;
  \left( \frac{\bar{P}(z)}{1-z} \right)_+
  \frac{1}{\veps},
\end{equation}
in order to avoid double counting with the ladder and/or experimental PDF,
thus obtaining $C_0(1-\Pbbm K_{0F} -\Pbbm K_{0B})$.
This gives rise to the standard subtracted
DY analog of the DIS coefficient function
in the $MS$ scheme
\begin{equation}
\label{eq:fDYMS}
\begin{split}
f^{MS}_{DY}(z) &=
f^{\rm AEM}_{DY}(z) - K_F^{MS}(z) - K_B^{MS}(z)
\\&
 =\delta(1-z)
 +\delta(1-z)
  \frac{C_F \alpha_s}{\pi}\;
   \left( \frac{2}{3}\pi^2 -\frac{7}{4} \right)
\\&
 +2\frac{C_F \alpha_s}{\pi}\;
  \left( \frac{\bar{P}(z)}{1-z} \right)_+
  \left(
    \ln\frac{\hat{s}}{\mu^2}
   +\omega_2
  \right)
\\&
 +\frac{C_F \alpha_s}{\pi}\;
 \left\{ 
      4\bar{P}(z) \frac{\ln(1-z)}{1-z}
     -2\bar{P}(z)\frac{\ln z}{1-z}
 \right\}_+.
\end{split}
\end{equation}
Note that the $\ln\frac{\hat{s}}{\mu^2} +\omega_2$ 
term in the above will be absent if the relation
$\hat{s}=\mu^2e^{-\omega_2}$ is adopted%
\footnote{ This corresponds
 to the use of $\frac{1}{\veps} +\omega_2$ instead of the pure pole
 in the counterterm and setting $\hat{s}=\mu^2$.
},
as in refs.~\cite{Mele:1990bq,Brock:1993sz}.
It is well known~\cite{Altarelli:1979ub}
that the numerically dominant term
$\frac{\ln(1-z)}{(1-z)_+}$  in the above function
is correcting for the misrepresenting
of the soft gluon behaviour and incorrect phase space limits
of the $\overline{MS}$ dimensional regularization (subtraction) scheme.
In our MC scheme this term will not be present (gets transferred to the ladder).

The plus regularization $(\dots)_+$ of the IR singularity
in eq.~(\ref{eq:KMS}), in the
diagrammatic approach of the CFP~\cite{Curci:1980uw},
comes from $Z_F$ in the axial gauge.
It is also shown by CFP that there is
a diagram-per-diagram correspondence between Feynman
diagrams in the axial gauge and the diagrams
of the operator product expansion
(OPE)~\cite{Floratos:1977au,Floratos:1978ny}%
\footnote{
 In this sense the axial gauge is implicitly present in eq.~(\ref{eq:KMS}),
 while the unsubtracted~eq.~(\ref{eq:AEM79}) is gauge invariant.
}.

In the context of the subtraction/factorization scheme aligned
with the MC, we should apply eq.~(\ref{eq:DYbeta1F}), 
that is in order to avoid double counting with the ladder (PDF),
we should subtract from eq.~(\ref{eq:DYsig1}) the LO contribution
of the MC:
\[
\rho_{1Fc}(\alpha,\beta)=
\frac{1}{\beta}
\frac{C_F \alpha_s}{\pi}
\frac{\bar{P}(1-\alpha-\beta)}{\alpha}
\theta_{\beta<\alpha}.
\]
However, in order to combine it properly with the virtual corrections,
the above distribution
should be extrapolated to $n=4+2\veps$ and integrated over the phase space.
This is done for the F hemisphere in the following (keeping in mind
that $1-\alpha-\beta=1-\bar\alpha$):
\begin{widetext}
\begin{equation}
\label{eq:KFz}
\begin{split}
&
K_F^{MC}(z,\veps)=
\frac{C_F \alpha_s}{\pi}
\int \frac{d\alfb d\betb }{\alfb\betb}
\int\!\! d\Omega_{2+2\veps}
\bigg( \frac{\hat{s}\alfb\betb}{z\mu^2} \bigg)^{\veps}\;
\bar{P}(1-\alfb,\veps)
\theta_{\betb<\alfb}
\delta_{\alfb=1-z}
\theta_{\alfb>\delta}
-\delta(1-z)\; S_{MC}(s,\veps)
\\&
=\frac{C_F \alpha_s}{\pi}\;
\bigg( \frac{\hat{s}}{z\mu^2} \bigg)^{\veps}\;
\frac{\Omega_{2+2\veps}}{\veps}
\frac{\bar{P}(z,\veps)}{(1-z)^{1-2\veps}}
\theta_{1-z>\delta}
-\delta(1-z)\; S_{MC}(s,\veps)
=\frac{C_F \alpha_s}{\pi}\;
  \Bigg( \frac{\bar{P}'(z,\veps)}{1-z} 
  \left[
    \frac{1}{\veps} +\omega_2
   +\ln\frac{\hat{s}}{z\mu^2}
  \right]
  \Bigg)_+,
\end{split}
\end{equation}
\end{widetext}
where the ${\cal O}(\veps)$ contribution 
from $\gamma$-trace is added to the LO kernel
\[
\begin{split}
\bar{P}'(z,\veps)&=\bar{P}(z)\big(1+2\veps\ln(1-z)\big)
  +\frac{1}{2}\veps(1-z)^2,
\\
\bar{P}(z,\veps) &= \bar{P}(z) + \frac{1}{2}\veps(1-z)^2.
\end{split}
\]
The overall plus-prescription is coming from
the 1st order expansion of the Sudakov form-factor in the MC
in $n=4+2\veps$ dimensions
and from the usual sum rule 
$\int dz\; K_F^{MC}(z,\veps) =0$
(treating $\hat{s}$ as $z$-independent):
\begin{equation}
S_{MC}(\hat{s},\veps)
=\frac{C_F \alpha_s}{\pi}\;
\frac{\Omega_{2+2\veps}}{\veps}
\int_0^{1-\delta} dz\; \frac{\bar{P}(z,\veps)}{(1-z)^{1-2\veps}}\;
\bigg( \frac{\hat{s}}{z\mu^2} \bigg)^{\veps},
\end{equation}
with both ${\cal O}(\veps)$ terms
in $\bar{P}'(z,\veps)$ necessarily participating in the $(...)_+$
prescription%
\footnote{In the MC practice regularization of $1/(1-z)$
  is done with some cut-off
  $1-z>\delta$ rather than with $\veps$ of the dimensional regularization.
}.

The expression of eq.~(\ref{eq:AEM79})
is the result of the following phase space integration
\begin{equation}
\begin{split}
f^{\rm AEM}_{DY}(z,\veps)
&=\frac{C_F \alpha_s}{\pi}
\int \frac{d\alpha d\beta }{\alpha\beta}
\int d\Omega_{2+2\veps}
\bigg( \frac{s\alpha\beta}{\mu^2} \bigg)^{\veps}\;
\\&\times
\rho_1(\alpha,\beta)
\delta_{1-z=\alpha+\beta}
\theta_{\alpha+\beta>\delta}
-\delta(1-z)\; U_{S+V},
\end{split}
\end{equation}
where $\rho_1(\alpha,\beta)$ is the one real gluon distribution
and $U_{S+V}$ sums up the real soft gluon, $\alpha+\beta>\delta$,
and vertex virtual contribution.
The complete 1-gluon contribution 
(including the expanded Sudakov form-factor)
of the LO MC reads:
\begin{equation}
\begin{split}
&K_F^{MC}(z,\veps)+K_B^{MC}(z,\veps)=
\frac{C_F \alpha_s}{\pi}
\int \frac{d\alfb d\betb }{\alfb\betb}
d\Omega_{2+2\veps}
\bigg( \frac{\hat{s}\alfb\betb}{z\mu^2} \bigg)^{\veps}\;
\\&~~~~~~\times
[
\bar{P}(1-\alfb,\veps)
\theta_{\betb<\alfb}
\delta_{1-z=\alfb}\;
+
\bar{P}(1-\betb,\veps)
\theta_{\betb>\alfb}
\delta_{1-z=\betb}\;
]
\\&~~~~~~\times
\theta_{1-z>\delta}
-\delta(1-z)\; 2S_{MC}(s,\veps).
\end{split}
\end{equation}
The difference between the complete NLO of 
eq.~(\ref{eq:AEM79}) and the above LO MC contribution,
after partial phase space integration, reads:
\begin{equation}
\label{eq:DYfinalVS}
\begin{split}
f^{MC}_{DY}(z)
&=f^{\rm AEM}_{DY}(z,\veps) - K^{MC}_F(z,\veps)-K^{MC}_B(z,\veps)
\\&
= -\frac{C_F \alpha_s}{\pi}\; (1-z)
+\delta(1-z)\Delta_{V+S},
\\
\Delta_{V+S}&=
 \frac{C_F \alpha_s}{\pi}\left( \frac{2}{3}\pi^2 -\frac{7}{4} \right)
+\frac{C_F \alpha_s}{\pi} \frac{1}{2}
=\frac{C_F \alpha_s}{\pi} \left( \frac{2}{3}\pi^2 -\frac{5}{4} \right).
\end{split}
\end{equation}
From the above we are able to determine the $z$-independent
soft+virtual correction $\Delta_{V+S}$ in the NLO MC weight%
\footnote{ The last term in $\Delta_{V+S}$ 
 is due to the plus-prescription in the $(1-z)_+$ part
 of the MC counterterm of eq.~(\ref{eq:KFz}).}.
The above does not include any singular
terms like $\ln(1-z)/(1-z)_+$, as advertised earlier.

The difference between the standard $MS$ function of
eq.~(\ref{eq:fDYMS})
and that of eq.~(\ref{eq:DYfinalVS})
is entirely due to the difference between
the $MS$ counterterm of eq.~(\ref{eq:KMS})
and the MC counterterm of eq.~(\ref{eq:KFz}):
\begin{equation}
\begin{split}
& f^{MS}_{DY}(z)-f^{MC}_{DY}(z)=
 -2K_F^{MS}(z,\veps)+2K_F^{MC}(z,\veps)
\\&~~~
=\frac{C_F \alpha_s}{\pi}\; (1-z)_+
+2\frac{C_F \alpha_s}{\pi}\;
  \left( \frac{\bar{P}(z)}{1-z} \right)_+
  \left(
    \ln\frac{\hat{s}}{\mu^2}
   +\omega_2
  \right)
\\&~~~~~~~~
 +\frac{C_F \alpha_s}{\pi}\;
 \left\{ 
      4\bar{P}(z)\frac{\ln(1-z)}{1-z}
     -2\bar{P}(z)\frac{\ln z}{1-z}
 \right\}_+
\end{split}
\end{equation}
and it represents clearly the difference between the $MS$
and the MC factorization schemes.

One may ask how to interpret this change from
$MS$ factorization scheme to MC factorization scheme, in
particular, 
how unique the modified MC counterterms of eq.~(\ref{eq:KFz}) are.
One may answer this question in two complementary ways.
One way is that the new MC counterterm of eq.~(\ref{eq:KFz})
represents just the collinear limit of the {\em exact}
matrix element in $n=4+2\epsilon$ dimensions
(keeping higher order terms in $\veps$)
in a sense of the $\Pbbm'$ projection operator. 
This definition has to be supplemented with the plus prescription
in the soft limit or, alternatively, by saying
that $Z_F$, which in CFP ($MS$) provides for plus prescription,
is replaced by the Sudakov form-factor.
This approach represents an effort in combining the best from
the two, the collinear and soft resummation.
Another way of addressing this question is to say that
the real backbone in the collinear factorization is OPE
with CFP providing a solid bridge to OPE, and the only thing
that has to be explained and kept track of is the difference
between CFP and MC (in a similar way like finite UV renormalization).
This approach was already advocated in 
refs.~\cite{Ellis:1978ty,Furmanski:1981cw} and in other
papers~\cite{Mele:1990bq}, where factorization scheme dependence
was discussed.
In our approach, we are 
using both ways of addressing the above question.

The related question is whether the counterterm of eq.~(\ref{eq:KFz})
is universal?
Basically the answer is that it is universal  thanks to the fact
that it is defined in terms of the $\bar{k}^\mu= \lambda k^\mu$
variables.
In other words, the kinematic mapping, inherent
in the new $\Pbbm'$ operator, should remove the hard process dependence
on the side of the ladder, in the same way as the pole-part operation
in CFP~\cite{Curci:1980uw} or $P_{kin}$ of ref.~\cite{Ellis:1978ty}.
To be completely certain that the above aim of the universality of
the new MC factorization scheme is achieved, in next section we shall define
a similar MC scheme for the DIS process, define and use the collinear
counterterm of this MC scheme, and in Sect.~\ref{sec:FSindep-relation}
we shall check validity of the factorization scheme independent relation (DY$-2\times$DIS)
of ref.~\cite{Altarelli:1979ub} between
the coefficient functions of DY and DIS,
both taken in the MC factorization scheme.

\subsection{Differences compared with POWHEG and MC@NLO methods}
\label{sec:MCatNLO}

In this subsection we outline main differences of our method compare
to the well established approaches of POWHEG~\cite{Nason:2004rx} and
MC@NLO~\cite{Frixione:2002ik} used today to combine NLO-corrected hard
process with the LO parton shower.

First and most obvious difference between our method and those of
POWHEG and MC@NLO is use of different factorization schemes.
In our approach we use factorization scheme~\cite{Jadach:2010ew,Jadach:2011kc}
designed especially for MC simulations whereas POWHEG and MC@NLO
use the standard $\overline{MS}$ scheme.
This allows them to use directly the standard $\overline{MS}$ collinear PDFs,
while we need additional work here%
  \footnote{One possibility is refitting PDFs, which shouldn't be
  to complicated as the difference between MC and $\overline{MS}$ scheme
  on the inclusive level is small.
  }.
Moreover, we build the LO parton shower MC from scratch whereas
POWHEG and MC@NLO profits from the well established (unmodified) LO MC programs.

At first it may seem that these general features result in unnecessary
complications in our approach, however, profits are more important,
especially if we have in mind construction
of the fully NLO MC parton shower (with NLO corrections, not only in the hard process,
but also in the ladder parts).
Our method features:
\begin{enumerate}[(i)]
\item
simple and positive MC weight implementing the NLO on top of the LO MC,
see ref.~\cite{Jadach:2012vs} (MC@NLO features negative weights);
\item
no need to correct for the difference in the collinear counterterm of
the LO MC and the standard $\overline{MS}$ scheme;
\item
the virtual+soft corrections $\Delta_{V+S}$ are completely kinematics independent
-- all annoying $d\Sigma^{c\pm}$ contributions of MC@NLO are gone;
\item
built-in resummation of the $\frac{\ln^n(1-x)}{1-x}$ terms is provided;
\item
direct relation to the collinear factorization procedures.
\end{enumerate}

Note also that,
in the presented method there is no need to define the hardest emission,
as in POWHEG, as it is automatically included into the sum over spectator
gluons in the formula for the MC weight in eq.~\eqref{eq:NLODYMCwt}.
In fact, we can explicitly
see that the dominant contribution is from the ``hardest'' (in $k_T$) gluon%
  \footnote{This is just a relabeling according to $k_T$, we do not need to change
  previously generated, angular-ordered gluons. It is only exploited here for
  the purpose of efficient evaluation of the NLO MC weight.},
for numerical illustration see ref.~\cite{Jadach:2012vs}.
This allows us to avoid truncated/vetoed gluons needed in POWHEG
methodology in case of angular ordering.

A detailed comparison of MC@NLO and POWHEG methods themselves can be found in
refs.~\cite{Nason:2012pr,Hoeche:2011fd}.

\section{Deep inelastic electron--proton scattering}
\label{sec:DISMC}

As already said, the process of deep inelastic electron--proton
scattering (DIS) is included in the scope because it is
is an important source of information on parton distributions
in a proton and, by comparing the DIS and DY processes,
the question of universality in the collinear factorization
implemented in the MC can be fully discussed.

In the following subsections we shall first introduce
kinematics, phase space and notation for one real gluon emission.
Next, we shall define the multigluon LO MC distribution
with initial-state radiation (ISR) and
final-state radiation (FSR) LO ladders and the LO matrix
element for the hard process for electron-hadron DIS.
Analytical integration of the MC distribution
will lead to the familiar formula for
structure function $F_2$ in the form of the convolution of PDF
with the Born cross section.
Then we shall give a close simple formula for the MC
weight implementing the NLO correction to the hard process.
The analytical integration will be again possible giving
the structure functions $F_2$ and $F_1$ in 
form of the convolution of PDF
with the NLO coefficient function.
Of course, the above NLO coefficient function will be in
the MC factorization scheme, 
but we shall see that universality is preserved,
by means of checking validity of the 
factorization scheme independent relation DY$-2\times$DIS
of ref.~\cite{Altarelli:1979ub}
between the coefficient functions of the DY and DIS processes.

\subsection{One-real gluon distribution and kinematics}

\begin{figure}[h]
\centering
{\includegraphics[width=80mm]{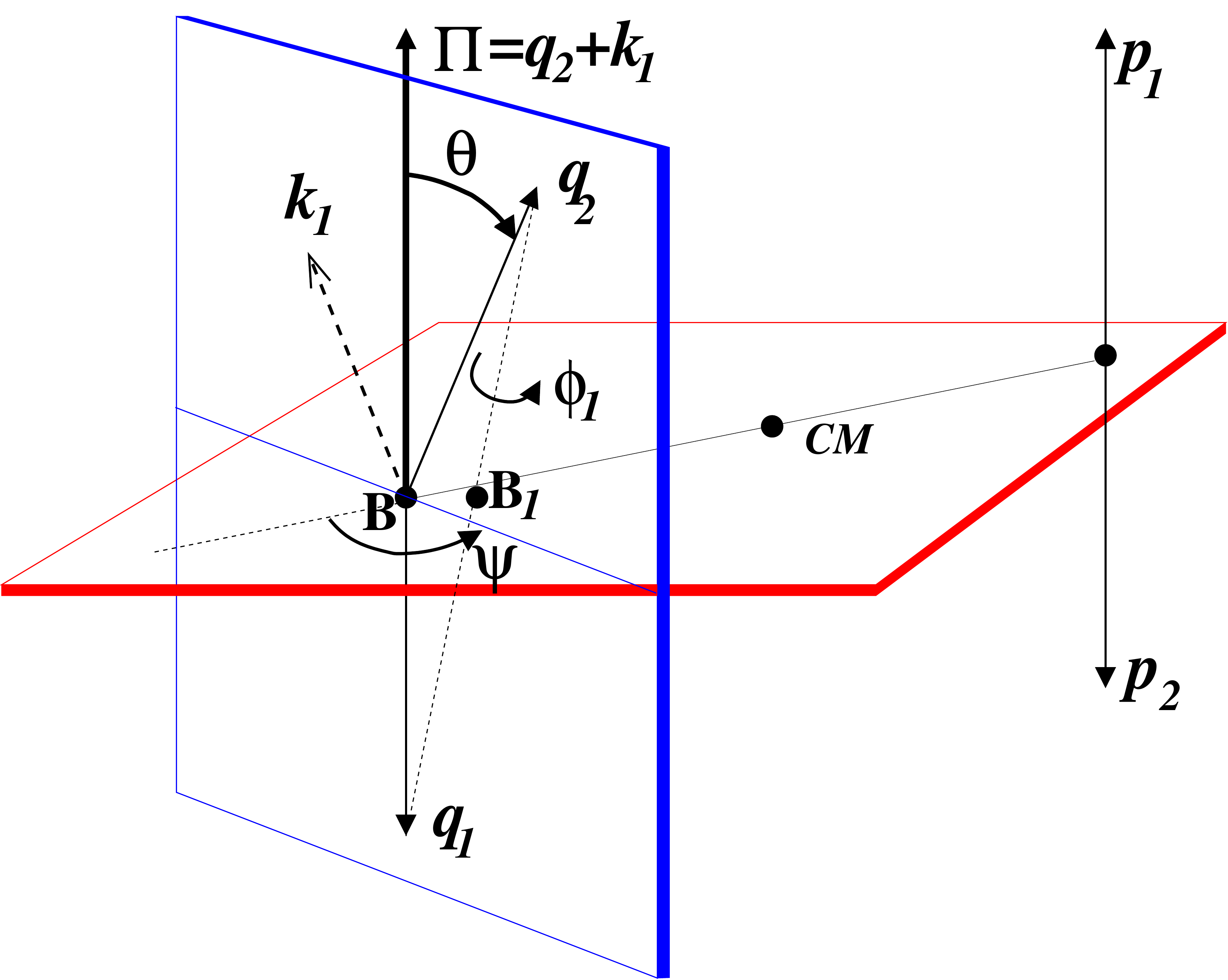}}
\caption{ Kinematics of one gluon emission in the Breit frame.}
\label{fig:Breit1G}
\end{figure}

The Born differential cross section of the electron-quark scatterning
$
 e(p_1)+q(q_1)\rightarrow e(p_2)+q(q_2)
$
in terms of the standard variables%
\footnote{ We omit the minus sign in the variables like $t$ and $u$
  with respect to the standard notation.}
$s=2p_1q_1$, $t=2p_1p_2$, $u=2p_1q_2$
reads
\begin{equation}
d\sigma_B = \frac{\alpha^2}{s} 
   d\left(\frac{t}{s}\right) d\varphi\;\;
   Q_q^2\; \frac{s^2+u^2}{t^2},
\end{equation}
where $Q_q$ is the quark charge.
Next, consider the process with emission of an additional gluon
from the quark line:
\begin{displaymath}
 e(p_1)+q(q_1)\rightarrow e(p_2)+q(q_2)+g(k).
\end{displaymath}
The differential distribution in this case reads
\begin{equation}
\begin{split}
& d\sigma_1 = 
  \frac{Q_q^2 \alpha_{QED}^2}{s} d\left(\frac{t}{s}\right) d\varphi\;\;
   \frac{s^2+u_1^2+s_1^2+u^2}{2 t t_1}\;
  \frac{d\psi}{2\pi}\;
\\&\qquad\qquad\qquad\times
  \frac{C_F \alpha_s}{\pi}
  \frac{d\alpha d\beta}{\alpha\beta}\;
  \frac{t_1}{t}.
\end{split}
\end{equation}
The additional invariants
$s_1=2p_2q_2$, $u_1=2q_1p_2$, $t_1=2q_1q_2$ are introduced in this case.
The factor $\frac{t_1}{t}$ is the Jacobian due to parametrization of
the phase space in terms of 
the rescaled Sudakov variables~\cite{bhlumi2:1992,Jadach:1991ty}:
\begin{equation}
\alpha=\frac{2kq_2}{t_1+2kq_1},\quad 
\beta =\frac{2kq_1}{t_1+2kq_1}.
\end{equation}
The angle $\psi$ is the azimuthal angle of $\vec{k}$ around $z$-axis
in the Breit frame of $Q=q_2+k-q_1$,
that is where $Q^0=0$,
with an additional requirement that $\vec{q}_1$ is parallel to the $z$ axis.
We call this the reference frame $B$, see Fig.~\ref{fig:Breit1G}.

Yet another Breit frame $B_1$ is marked
in Fig.~\ref{fig:Breit1G},
that of $Q_1=q_2-q_1$ with the $z$-axis along $\vec{q}_1$.
It will be used in the MC
and in the analytical calculations.
Note that the integration is over the angle $\psi$ of the $k,q_1,q_2$
plane as a whole around $\vec\Pi$, while another azimuthal angle
$\phi_1$ of $\vec{k}$ in the $B_1$ frame is frozen to zero.
Note that the standard Sudakov variables are:
\begin{equation}
\alpha'=\frac{kq_2}{q_1q_2},\quad 
\beta' =\frac{kq_1}{q_1q_2},\quad
t=t_1 (1-\alpha'+\beta').
\end{equation}
They are not convenient because $\beta'\in(0,\infty)$,
and the following transformation is mandatory%
\footnote{In the collinear limit $k\simeq q_2 \beta/(1-\beta)$,
  with $z=1-\beta$ being the lightcone variable in the LO splitting
  kernel.}:
\begin{equation}
\label{eq:alfbet}
\begin{split}
&\alpha= \frac{\alpha'}{1+\beta'},\quad
 \beta = \frac{\beta'}{1+\beta'},\quad
\alpha'= \frac{\alpha}{1-\beta},\quad
\beta' =  \frac{\beta}{1-\beta},\quad
\\&
0<\alpha\leq 1-\frac{t}{s},\quad
0<\beta\leq 1,\qquad
t_1= t\frac{1-\beta}{1-\alpha}.
\end{split}
\end{equation}

\subsection{Bjorken variables, structure functions, collinear limits}
\label{sect:DISkinema}
The standard Bjorken variables are
\begin{equation}
x_{B}\equiv \frac{t}{2q_1 Q}
  =\frac{|Q^2|}{2q_1 Q},\quad
1 \geq x_B >0,\quad
y_B \equiv \frac{Qq_1}{p_1q_1}=\frac{t}{sx_B}.
\end{equation}
In the case of a single gluon they are expressed as follows
\begin{equation}
x_{B}=\frac{t}{2q_1 Q}
  = \frac{1-\alpha'+\beta'}{1+\beta'}
  = 1-\alpha,\quad
y_B=\frac{2q_1 Q}{2q_1 p_1}
  =\frac{t}{s(1-\alpha)}.
\end{equation}
The reader should keep in mind that
for simplicity $x_B$ is the fraction of the parton momentum
in the initial quark%
\footnote{Returning to the normal definition in the MC
  (a fraction of the hadron momentum) is quite trivial.}.

Let us recall the definitions of the standard 
deep-inelastic structure functions
in terms of the above Bjorken variables:
\begin{equation}
\begin{split}
&\frac{d^2\sigma}{dt dx_B}
= \frac{2\pi \alpha_{QED}^2  Q_q^2}{t^2} \;x_B^{-1}
  \left\{ y_B^2\; 2x_B F_1(x_B) +2(1-y_B) F_2(x_B) \right\},
\\&~~~~
= \frac{2\pi \alpha_{QED}^2  Q_q^2}{t^2} \;x_B^{-1}
  \left\{ [1+(1-y_B)^2]\; F_2(x_B) -y_B^2\; x_B F_L(x_B) \right\},
\\&~~~~
= \frac{2\pi \alpha_{QED}^2  Q_q^2}{t^2} \;x_B^{-1}
  \big\{ [1+(1-y_B)^2]\; 2x_B F_1(x_B)
\\&~~~~~~~~~~~~~~~~~~~~~~~~~~~~~~~~~~~~~~~
 +2(1-y_B) x_B F_L(x_B) \big\},
\end{split}
\end{equation}
where we have employed the standard definition $ 2xF_1 \equiv F_2-xF_L $.
In the LO case the Callan-Gross relation $2xF_1=F_2$ is fulfilled
and the longitudinal structure function $F_L=0$
(it will receive a nonzero contribution at NLO).
The LO relation to the parton distribution function (luminosity) is
$2F_1(x) =F_2/x= {\rm PDF}(x)$.

It is instructive to investigate the collinear ISR and FSR  limits.
The slightly reorganized single-gluon emission distribution reads
\begin{equation}
\begin{split}
d\sigma_1 &= 
 \alpha_{QED}^2  Q_q^2\; 
 \frac{dt}{t^2} d\varphi\;\;
  \frac{d\psi}{2\pi}\;
  \frac{C_F \alpha_s}{\pi}
  \frac{d\alpha d\beta}{\alpha\beta}\; W,
\\
  W&=\frac{s^2+u_1^2+s_1^2+u^2}{2 s^2}.
\end{split}
\end{equation}
The soft limit is already manifest
in the eikonal phase space factor
$ \frac{d\alpha d\beta}{\alpha\beta}.$
The following explicit expressions for the invariants
in terms of our Sudakov variables are useful:
\begin{equation}
\label{eq:DISmandelst}
\begin{split}
&\frac{t_1}{s}= (1-\beta)y_B,\quad
 \frac{u_1}{s}= 1 - y_B,\quad
 \frac{u}{s}=\frac{s_1}{s}-(1-\alpha-\beta)y_B,
\\&
\frac{s_1}{s} \simeq (1-\alpha)(1-\beta) 
    +\alpha\beta(1 -y_B)
\\&~~~~~~~~~
 +2\cos\psi\sqrt{(1-\alpha)(1-\beta)\; \alpha\beta (1-y_B)}.
\end{split}
\end{equation}
In the FSR collinear limit, 
$\alpha\simeq 0$, 
$\beta \simeq 1-z $,
$k\simeq q_2 (1-z)/z$ and
$y_B \simeq y_0=t/s$, we have
\begin{equation}
\begin{split}
&s^2+u_1^2 \simeq s^2+(s-t)^2,\quad
s_1^2+u^2  \simeq (s^2+(s-t)^2)(1-\beta)^2,
\\& 
W\simeq \frac{1+(1-\beta)^2}{2} \frac{s^2+(s-t)^2}{s^2}
       =\frac{1+z^2}{2}\; 
        [1+ (1-y_0 )^2 ].
\end{split}
\end{equation}
In the ISR collinear limit 
$\beta\simeq 0$,
$\alpha \simeq 1-z$ and
$k\simeq (1-z) q_1$ we  have:
\begin{equation}
\begin{split}
&s^2+u_1^2\simeq s^2+\Big(s- \frac{t}{1-\alpha} \Big)^2,\quad
\\&
s_1^2+u^2 \simeq [s(1-\alpha)]^2+[ s(1-\alpha) -t ]^2,
\\&
W\simeq \frac{1+(1-\alpha)^2}{2} [ 1+(1-y_B )^2 ]
    =   \frac{1+z^2}{2}\; [1+ (1-y_B )^2 ].
\end{split}
\end{equation}

\subsection{Bare structure functions for 1-gluon emission}
\label{sect:exactC2}
Our immediate aim is now to reproduce 
the well known~\cite{Bardeen:1978yd,Curci:1980uw,Altarelli:1978id}
result for the NLO correction to the $F_2(x)$ structure function
by means of integration of the one-gluon phase space
(the NLO correction to $F_L(x)$ will also be found).
The aim is to test our Monte Carlo phase space parametrization,
prepare ground for determining the soft+virtual correction in the MC
and put FSR under control.

The unsubtracted (bare) contribution 
to $F_2(x)/x$ corrected due to
the real gluon emissions plus the vertex correction
(eq.~(59) in \cite{Altarelli:1979ub})
can be rewritten as:
\begin{equation}
\label{eq:Altarelli97DIS}
\begin{split}
C^{\rm AEM}_{2,bare}(z)&=
\delta(1-z)+
\frac{C_F \alpha_s}{\pi}
\bigg\{
  P_{qq}(z) \Big[ \frac{1}{\veps} +\omega_2 \Big]
\\&
 +P_{qq}(z) 
  \ln\frac{t(1-z)}{z\mu^2}
 -\frac{3}{4}\frac{1}{1-z}
 +\frac{1}{2}(3+2z)
\bigg\}_+,
\end{split}
\end{equation}
where
$P_{qq}(z)=\frac{1+z^2}{2(1-z)}$,
$\omega_2= \gamma_E-\ln(4\pi)$ comes from the
$\Omega_{2+2\veps} =2\pi(1+\veps\omega_2+\dots)$ expansion.
The baryon number conservation sum rule  
$\int_0^1 dz\; C^{\rm AEM}_{2,bare}(z)=1$
holds explicitly.

The standard NLO $MS$ correction $C_2^s$ 
to $z^{-1} F_2(z)$ form-factor
is obtained simply by means of subtracting the $MS$
collinear counterterm%
\footnote{We subtract a pure pole as in the original CFP work
 and not $(\frac{1}{\veps}+\omega_2) \{P_{qq}(z)\}_+ $,
 as it is a common practice nowadays.
}
$\frac{1}{\veps} \{P_{qq}(z)\}_+ $ (i.e. the pole part).
The formula of ref.~\cite{Bardeen:1978yd}
to be reproduced reads:
\begin{equation}
\label{eq:bardeen78}
\begin{split}
& \Delta F_2^{NLO}(x_B)=
C_2^s(x_B)=
\\&
=\frac{C_F \alpha_s}{\pi}
\bigg\{
 P_{qq}(x_B)
 \Big[ \ln\frac{t(1-x_B)}{\mu^2 x_B}+\omega_2\Big]
 -\frac{3}{4}\frac{1}{1-x_B}
 +\frac{3}{2}+x_B
\bigg\}_+
\\&
=\frac{C_F \alpha_s}{\pi}
\left\{
 P_{qq}(x_B)
 \Big[ \ln\frac{t(1-x_B)}{\mu^2 x_B} +\omega_2\Big]
 -\frac{3}{4}\frac{1}{1-x_B}
 +1+\frac{3}{2}x_B
\right\}_+
\\&~~~~~~
+\frac{C_F \alpha_s}{\pi}\;
\left\{
 \frac{1-x_B}{2}
\right\}_+.
\end{split}
\end{equation}
The last term $(1-x_B)/2$ in the non-singular part
is due to the $\veps$-term from the $\gamma$-trace for
the intital-state collinear singularity and $\omega_2$
from the $n$-dimensional phase space.

\begin{figure}
\centering
\fbox{\includegraphics[width=60mm]{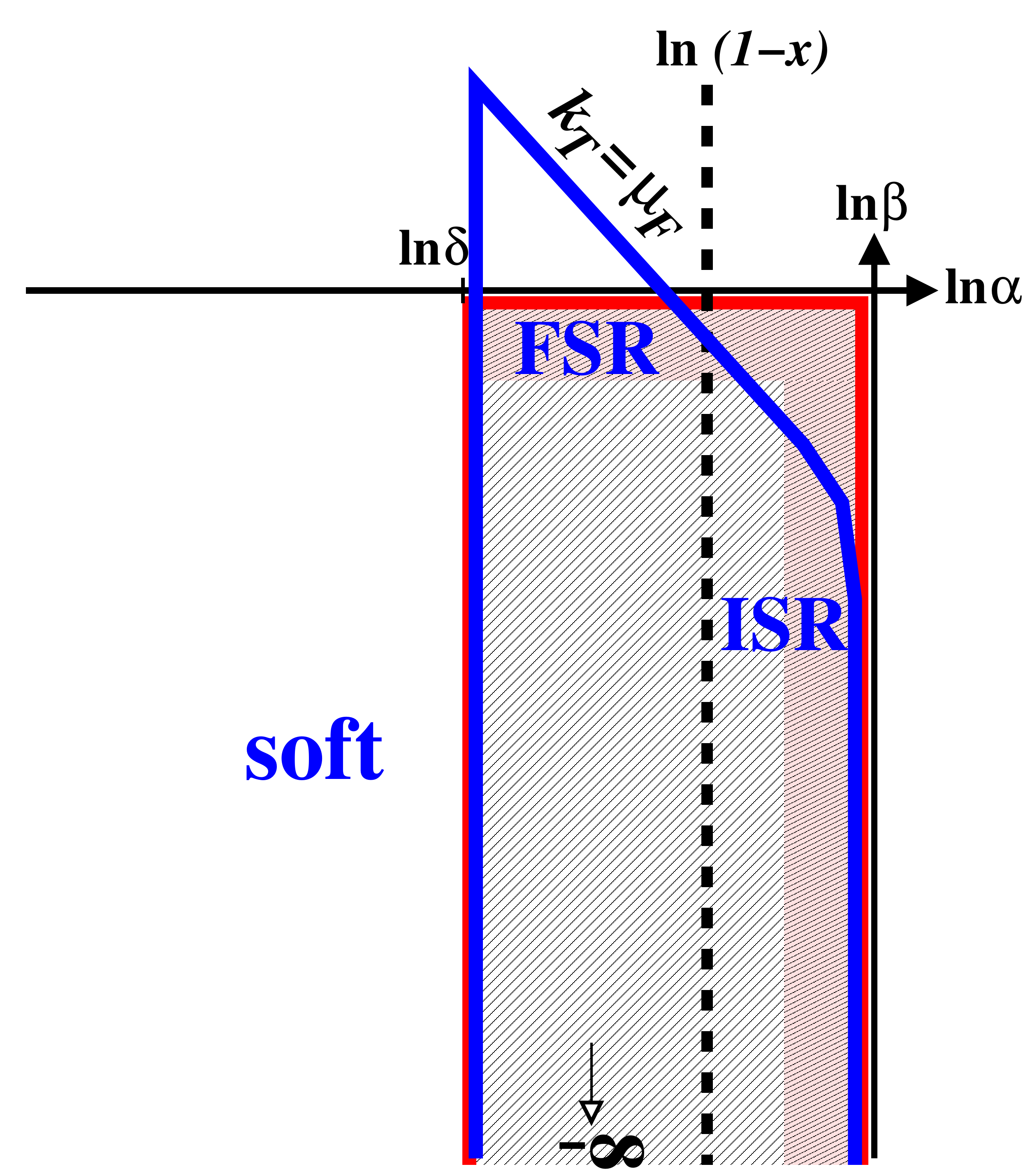}}
\caption{The logarithmic Sudakov plane for 1-gluon emission in the DIS process.}
\label{fig:DIStrapez}
\end{figure}

We start from the
unsubtracted (bare) DIS distribution coming from
two real-gluon emission diagrams from 
the quark line plus the vertex virtual correction
in $n=4+2\veps$ dimensions:
\begin{equation}
\label{eq:BareDIS}
\begin{split}
& \frac{d^2\sigma}{dt dx_B}
= \frac{2\pi \alpha_{QED}^2  Q_q^2}{t^2}
 \left[
 (1-U) \delta_{x_B=1}+\bar{G}_2(x_B,y_B)
 \right],
\\&
\bar{G}_2(x_B,y_B)=
  \int d\alpha d\beta\; 
  \int \frac{d\Omega^\psi_{2+2\veps}}{2\pi}\;\;
  \rho_2(\alpha,\beta)\; \delta_{x_B=1-\alpha},
\\&
\rho_2(\alpha,\beta)=
  \frac{C_F \alpha_s}{\pi}
  \frac{1}{\alpha\beta}
  W(\alpha,\beta,y_B,\veps)
  \left(\frac{t\alpha\beta}{\mu^2(1-\alpha)(1-\beta)} \right)^\veps
\\&~~~~~~~~~~~~~~\times
  \theta_{1>\alpha>\delta} \theta_{1>\beta}
\\&
W(\alpha,\beta,y_B,\veps)=\frac{s^2+u_1^2+s_1^2+u^2}{2s^2}
 +\veps\frac{s^2+u_1^2}{s^2}\frac{(t-t_1)^2}{t^2},
\end{split}
\end{equation}
where we have reinstalled in $W$ the $\veps$ term from the $\gamma$-trace%
\footnote{Only for the ISR collinear singularity;
 the one for FSR falls into the $U$ factor.
}.

The real emission phase space is explicitly integrated for $\alpha>\delta$,
where $\delta\ll 1$ is an IR cut-off.
The above phase space division is graphically 
shown in Fig.~\ref{fig:DIStrapez}.
Note that the $\alpha>\delta$ part of the phase space,
which we are going to integrate over includes not only
the hard collinear ISR but also the hard collinear FSR!
The costant $U$ must include the vertex correction 
summed up with the soft real
emission $\alpha<\delta$.
For determining $U$ it will be enough to
know~\cite{Bardeen:1978yd,Altarelli:1978id}
that the $F_2$ part of the distribution in eq.~(\ref{eq:BareDIS})
fulfills exactly the Adler sum rule in the dimensional regularization,
and in this way we may omit the details of its calculation.
The complicated phase space factor is simply due to the fact
that the transverse momentum of the gluon in the Breit frame is:
\[
   k_T^2=|\bk|^2 = t_1 \alpha'\beta'
  =\frac{t\alpha\beta}{(1-\alpha)(1-\beta)}.
\]
\newline
Finally, $\int d\Omega^\psi_{2+2\veps}$ is the $n$-dimensional
extension of $\int_0^{2\pi} d\psi$.

In the CFP scheme the ISR collinear singularity upon integration
gives rise to the LO pole part
\[
C_0^{[0]}\Pbbm K_{0I}=
C_0^{[0]}\Gamma_I^{[1]}(x_B)=
\frac{1}{\veps}
[1+(1-y_B)^2]
\frac{C_F \alpha_s}{\pi}\;
\bigg(\frac{\bar{P}(x_B)}{1-x_B} \bigg)_+,
\]
where $K_{0I}$ is the lowest-order 2PI kernel for the ISR ladder
and the $+$ prescription comes from $Z_F$, as usual.
The subtracted hard process matrix element in the CFP scheme
is $C_0^{[1]}-C_0^{[0]}\Pbbm K_{0I} $.
We shall calculate it with the help of the usual counterterm technique.
The ISR collinear/soft counterterm (SCC) we define as follows:
\begin{equation}
\begin{split}
\rho_{2c}(\alpha,\beta) &=
[1+(1-y_B)^2]\;
\frac{C_F \alpha_s}{\pi}\;
\frac{\bar{P}(1-\alpha)}{\alpha}\;
\\&\times
\beta^{\veps-1}\;
B^{-\veps}
\theta_{\beta<B(\alpha)}
\theta_{1>\alpha>\delta},
\\
B(\alpha) &=\frac{\mu^2 (1-\alpha)}{t\alpha}.
\end{split}
\end{equation}
It is defined such that it integrates to the pure pole
part exactly:
\[
C_0^{[0]}\Gamma_I^{[1]}(x_B)
\equiv
\int d\alpha d\beta\; \rho_{2c}(\alpha,\beta) \delta_{1-x_B=\alpha}.
\]
In Fig.~\ref{fig:DIStrapez} we have also 
marked the integration area for the above counterterm.
As we see, in this area the upper phase space integration limit
from the energy momentum conservation is replaced by the limit
on the gluon transverse momentum equal (approximately) to $\mu$.

With the help of the above ISR collinear counterterm
our task is reduced to calculate the subtracted DIS distribution
in $n=4$ dimensions:
\begin{equation}
\label{eq:G2subtr}
G_2(x_B,y_B)=
  \int d\alpha d\beta\; 
  \int_0^{2\pi} \frac{d\psi}{2\pi}\;\;
  [\rho_2(\alpha,\beta) - \rho_{2c}(\alpha,\beta)]
  \delta_{x_B=1-\alpha},
\end{equation}
except for the trivial $\veps$ term in $W$, which contributes
$\frac{C_F \alpha_s}{\pi}\; \big( \frac{1-x_B}{2}\big)_+$,
to be added at the end. 
The same with the similar $\sim\veps\omega_2$ term
from the phase space.

The integration can be summarized as follows:
\begin{widetext}
\begin{equation}
\label{eq:SfunMSprim}
\begin{split}
\frac{d^2\sigma^{NLO}_{subt.}}{dt dx_B}
&= \frac{2\pi \alpha_{QED}^2  Q_q^2}{t^2} \;
\frac{2C_F \alpha_s}{\pi^2}\int\! d^3\Ecal(k_1)\;
\Big[
  W(\alpha,\beta,y_{B}) 
 -W_0(y_{B}) \bar{P}(1-\alpha)\theta_{\beta_1<B(\alpha_1)}
\Big]
\delta_{1-x_B=\alpha_1}
\\&
= \frac{2\pi \alpha_{QED}^2  Q_q^2}{t^2} \;
  \left( 
     [1+(1-y_B)^2] C_2^s(x_B) -y_B^2\; C_L(x_B) 
  \right),
\\
C_2^s(x_B) &=
\frac{C_F \alpha_s}{\pi}
\left\{
 P_{qq}(x_B)\Big[
 \ln\frac{t(1-x_B)}{\mu^2 x_B}
 +\omega_2 \Big]
 -\frac{3}{4}\frac{1}{1-x_B}
 +1 +\frac{3}{2}x_B
\right\}_+
+\frac{C_F \alpha_s}{\pi} \left\{
   \frac{1-x_B}{2}
\right\}_+,
\\
C_L(x_B) &= \frac{C_F \alpha_s}{\pi}x_B,\qquad
W_0(y)\equiv 1+(1-y)^2,
\end{split}
\end{equation}
\end{widetext}
where the $+$ prescription is provided
by the virtual corrections.
We have also included
the $\veps$-contribution from the  $\gamma$-trace
and $\omega_2$ from the phase space.
As we see, $C_2^s(z)$ is equal to the finite part
of $C^{\rm AEM}_{2,bare}(z)$
of eq.~(\ref{eq:Altarelli97DIS}),
thus we have reproduced the classic 
result~\cite{Bardeen:1978yd}, as promised.

In the MC scheme the ISR counterterm $C_{0}^{(0)}\Pbbm' K_{0I}$
is defined as the single-gluon distribution
which is extrapolated to $n=4+2\veps$ dimensions and integrated
over the phase space:
\begin{equation}
\label{eq:KIdim}
\begin{split}
&
K_I(z,\veps)=
\frac{C_F \alpha_s}{\pi}
\int \frac{d\alpha d\beta }{\alpha\beta}
\int d\Omega_{2+2\veps}
\bigg( \frac{t\alpha\beta}{(1-\alpha)\mu^2} \bigg)^{\veps}\;
\\&~~~~~~~~~~~~~~~~~\times
\bar{P}(1-\alpha,\veps)\;
\theta_{\beta<\alpha}\;
\delta_{1-z=\alpha}\;
\theta_{\alpha>\delta}
-\delta_{z=1} S_{I}
\\&~~~~~
=\frac{C_F \alpha_s}{\pi}\;
\bigg( \frac{t}{z\mu^2} \bigg)^{\veps}\;
\frac{\Omega_{2+2\veps}}{\veps}
\frac{\bar{P}'(z,\veps)}{(1-z)^{1-2\veps}}
\theta_{1-z>\delta}
-\delta(1-z)\; S_{I}
\\&~~~~~
=\frac{C_F \alpha_s}{\pi}\;
  \Bigg( \frac{\bar{P}'(z,\veps)}{1-z}
  \left[
    \frac{1}{\veps}+\omega_2
   +\ln\frac{t}{z\mu^2}
  \right]
  \Bigg)_+,
\end{split}
\end{equation}
where $\bar{P}'(z,\veps)=\bar{P}(z)\big(1+2\veps\ln(1-z)\big)
  +\frac{1}{2}\veps(1-z)^2$.
The source of the $+$ prescription
in this case is the MC Sudakov form-factor
calculated in $n=4+2\veps$ in such a way that the sum rule 
$\int dz\; K_I(z,\epsilon)=0$ is preserved also in $n$-dimensions:
\[
S_{I}=\frac{C_F \alpha_s}{\pi}\;
\frac{\Omega_{2+2\veps}}{\veps}
\int_0^{1-\delta} dz\; \frac{\bar{P}'(z,\veps)}{(1-z)^{1-2\veps}}
\bigg( \frac{t}{z\mu^2} \bigg)^{\veps}\;,
\]
hence two ${\cal O}(\veps)$ terms
in $\bar{P}'(z,\veps)$ necessarily participate in the $(...)_+$
prescription.

Subtracting $K_I(z,\veps)$ of eq.~(\ref{eq:KIdim})
from the complete ${\cal O}(\alpha^1)$ 
result of eq.~(\ref{eq:Altarelli97DIS}) 
gives us the following coefficient function
\begin{equation}
\label{eq:C2angOrd}
 C_2^{MC}(z)=
 \frac{C_F \alpha_s}{\pi}
 \left\{
 -\frac{1+z^2}{2(1-z)} \ln(1-z)
  -\frac{3}{4}\frac{1}{1-z}
  +1+ \frac{3}{2}z
 \right\}_+
\end{equation}
in the MC factorization scheme, with the angular ordering.

The most important
part of the difference between the above MC structure functions
and the $MS$ variant is coming from
the different cut-off in the ISR counterterms:
\begin{equation}
\begin{split}
& \frac{2\pi \alpha_{QED}^2  Q_q^2}{t^2} \;
\frac{C_F \alpha_s}{\pi}\int\! d^3\Ecal(k_1)\;
W_0(y_{B}) 
\\&~~~~~~~\times
\Big[
 -\bar{P}(1-\alpha)\theta_{\beta_1<B(\alpha_1)}
 +\bar{P}(1-\alpha)\theta_{\beta_1<\alpha_1}
\Big]
\delta_{1-x_B=\alpha_1}
\\&
=\frac{2\pi \alpha_{QED}^2  Q_q^2}{t^2} \; W_0(y_{B})\;
 \frac{C_F \alpha_s}{\pi}\;
\\&~~~~~~~\times
 \Big[
 2P_{qq}(x_B)
 \ln(1-x_B)
+P_{qq}(x_B)
 \ln\frac{t}{\mu^2 x_B}
 \Big].
\end{split}
\end{equation}

A few comments are in order:
\begin{itemize}
\item
Why not $k^-$-ordering? 
In such a DIS-like factorization scheme%
\footnote{In the DIS factorization scheme $C_2=0$ exactly,
  while in the $k^-$ ordering it would be only less singular.}
term $\left(\frac{\ln(1-x)}{1-x}\right)_+$
would have been gone from eq.~(\ref{eq:C2angOrd}).
FSR could be treated in the DIS MC without the LO resummation,
with the unexponentiated FSR NLO corrections.
However, if the universality is to be maintained, and the same
$k^-$-ordering is applied to the W/Z production process,
that would either mean asymmetric treatment of the emission from
the quark and antiquark lines
or a large double logarithmic dead zone in the corresponding LO MC,
between the ISR and FSR phase spaces. 
Both options are unacceptable.
\item
Is there also a kinematic mapping involved in the above $\Pbbm'$,
like in the previous $W/Z$ production process?
Yes, it is implicitly included in the definition of $\alpha$ and $\beta$
variables in eq.~(\ref{eq:alfbet}), where dilatation using
factor $1/(1+\beta')$ factor is seen.
\item
From the point of view of the MC the above considerations
are incomplete, as they still keep FSR in 
the inclusive/integrated form. 
\end{itemize}

\subsection{DIS multigluon LO Monte Carlo}
Let us start with the raw distribution for $n$ gluons,
the $(\alpha C_F)^n$ part only, relevant for the LO MC.
\begin{displaymath}
 e(p_1)+q(q_1)\rightarrow e(p_2)+q(q_2)+g(k_1)+g(k_2)+...+g(k_n).
\end{displaymath}
The corresponding differential distribution reads
\begin{equation}
\begin{split}
d\sigma_n &= 
  Q_q^2 \alpha_{QED}^2 dt d\varphi\;\;
   \frac{W}{t t_1}\;
  \frac{d\psi}{2\pi}\;
  \left(\frac{C_F \alpha_s}{\pi}\right)^n
\\&\times
  \left(
  \prod_{i=1}^n
  \frac{d\alpha_i d\beta_i}{\alpha_i \beta_i}\;
  \frac{d\phi_i}{2\pi}
  \right)
  \delta_y\Big(\sum_j \vec{k}_j\Big)
  \frac{t_1}{t},
\end{split}
\end{equation}
where $W$ is a mild function to be defined later on.
The invariants
$s_1=2p_2q_2$, $u_1=2q_1p_2$, $t_1=2q_1q_2$ are the same as previously.
The factor $\frac{t_1}{t}$ is again the Jacobian due to the parametrization of
the phase space in terms of the Sudakov 
variables~\cite{Jadach:1988ec}, see bellow.
The angle $\psi$ is the azimuthal angle of $\vec{k}$ around the $z$-axis
in the Breit frame of $Q=q_2+k-q_1$ with $Q^0=0$,
with the additional requirement that $\vec{q}_1$ is parallel to the $z$ axis.
We call this reference frame $B$.
Another Breit frame $B_1$ is used in the MC,
that of $Q_1=q_2-q_1$ with $z$-axis also along $\vec{q}_1$ (and $\vec{q}_2$).
The illustration of the
kinematics in Fig.~\ref{fig:Breit1G} is still valid,
provided we replace $k_1$ by $\sum_j k_j$.

The integration is done over the angle $\psi$ of the $(\Pi,q_1,q_2)$
plane as a whole around 
$\vec\Pi=\vec{q}_2+\sum_j \vec{k}_j$,
while there is a single restriction on $n$ azimuthal angles
$\phi_i$ of $\vec{k}_i$ in the $B_1$ frame,
namely the vector
$\sum\vec{k}_j$ must be co--planar with $p_1$ and $p_2$.

The standard Sudakov variables are:
\begin{equation}
\begin{split}
\alpha'_i &= \frac{k_iq_2}{q_1q_2},\qquad 
\beta'_i =\frac{k_iq_1}{q_1q_2},\quad
\\
t &= t_1 \big(1-\sum_j\alpha'_j+\sum_j\beta'_j\big)-K^2,\qquad
K=\sum_j k_j.
\end{split}
\end{equation}
Next, we transform them as follows~\cite{Jadach:1988ec}
\begin{equation}
\begin{split}
&\alpha_i= \frac{\alpha'_i}{1+\sum_j\beta'_j},\quad
 \beta_i= \frac{\beta'_i}{ 1+\sum_j\beta'_j},\quad
\\&
\alpha'_i= \frac{\alpha_i}{1-\sum_j\beta_j},\quad
\beta'_i = \frac{\beta_i}{1-\sum_j\beta_j},\quad
\\&
0<\sum_j\alpha_j\leq 1-\frac{t}{s},\quad
0<\sum_j\beta_j\leq 1.
\end{split}
\end{equation}

The Bjorken variable $x_B$ (of parton in the initial quark)
can be expressed in terms of the Sudakov variables.
Using $Q=K+q_2-q_1$ and $K=\sum_j k_j$, we obtain
\begin{equation}
\begin{split}
x_{B}&
= \frac{2q_1q_2 +2q_1K -2q_2K-K^2}{2q_1q_2+2q_1K}
= \frac{1+\sum_j\beta'_j-\sum_j\alpha'_j}%
       {1+\sum_j\beta'_j}-\tilde{K}^2
\\&
= 1-\sum_j\alpha_j-\tilde{K}^2
= \frac{t}{t_1(1+\sum_j\beta'_j)}.
\end{split}
\end{equation}
In the NLO world the term
\[
 \tilde{K}^2 =\frac{K^2}{2q_1 Q}
   =\frac{2\sum_{i>j} k_i\cdot k_j}{2q_1 Q}
\]
can be either omitted
or taken care of in the collinear limit.
Note that $\tilde{K}^2$ is absent in the case of the single-gluon
calculation of the NLO coefficient function.

\begin{figure*}
\centering
{\includegraphics[width=70mm]{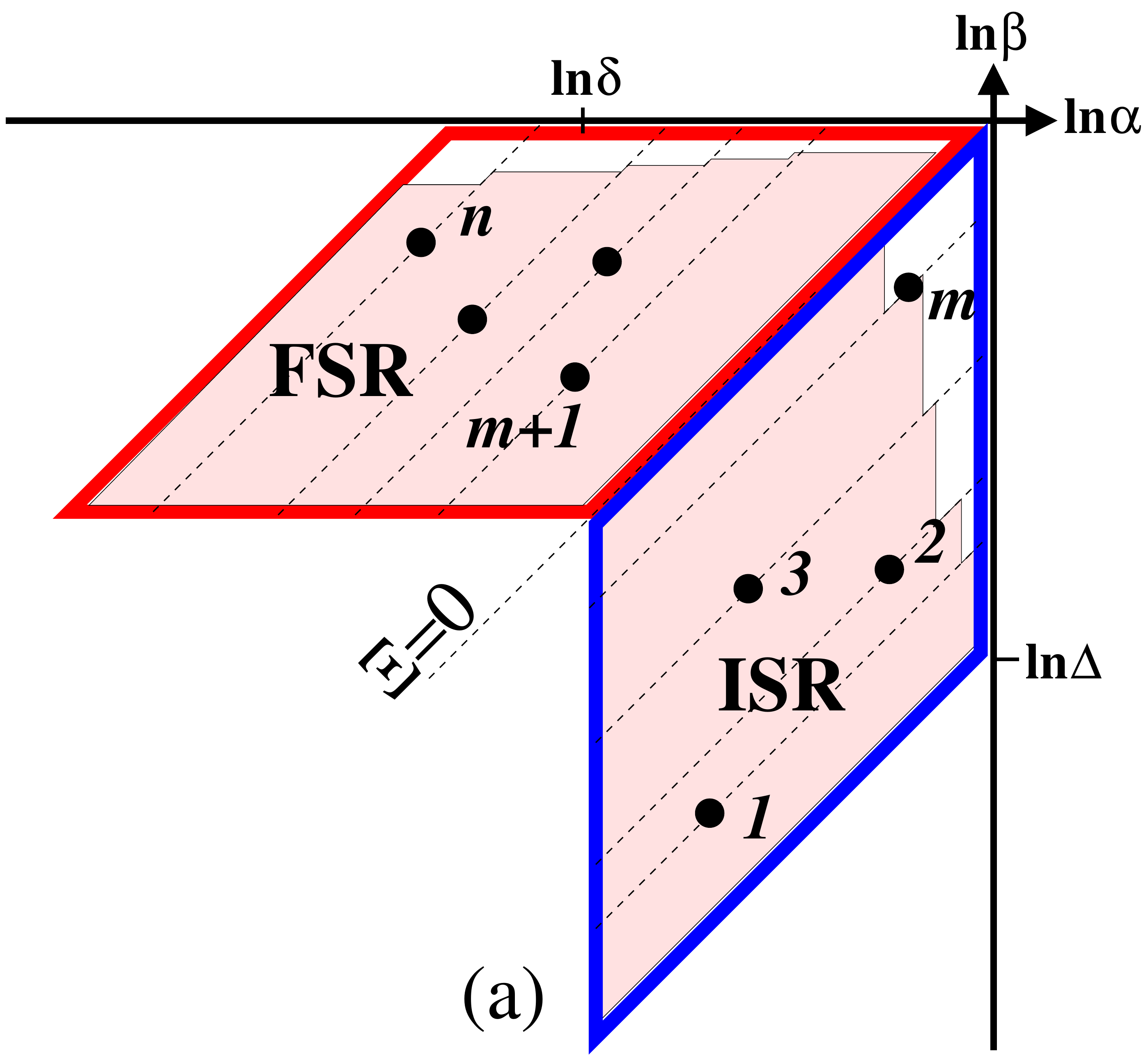}}~~~
{\includegraphics[width=70mm]{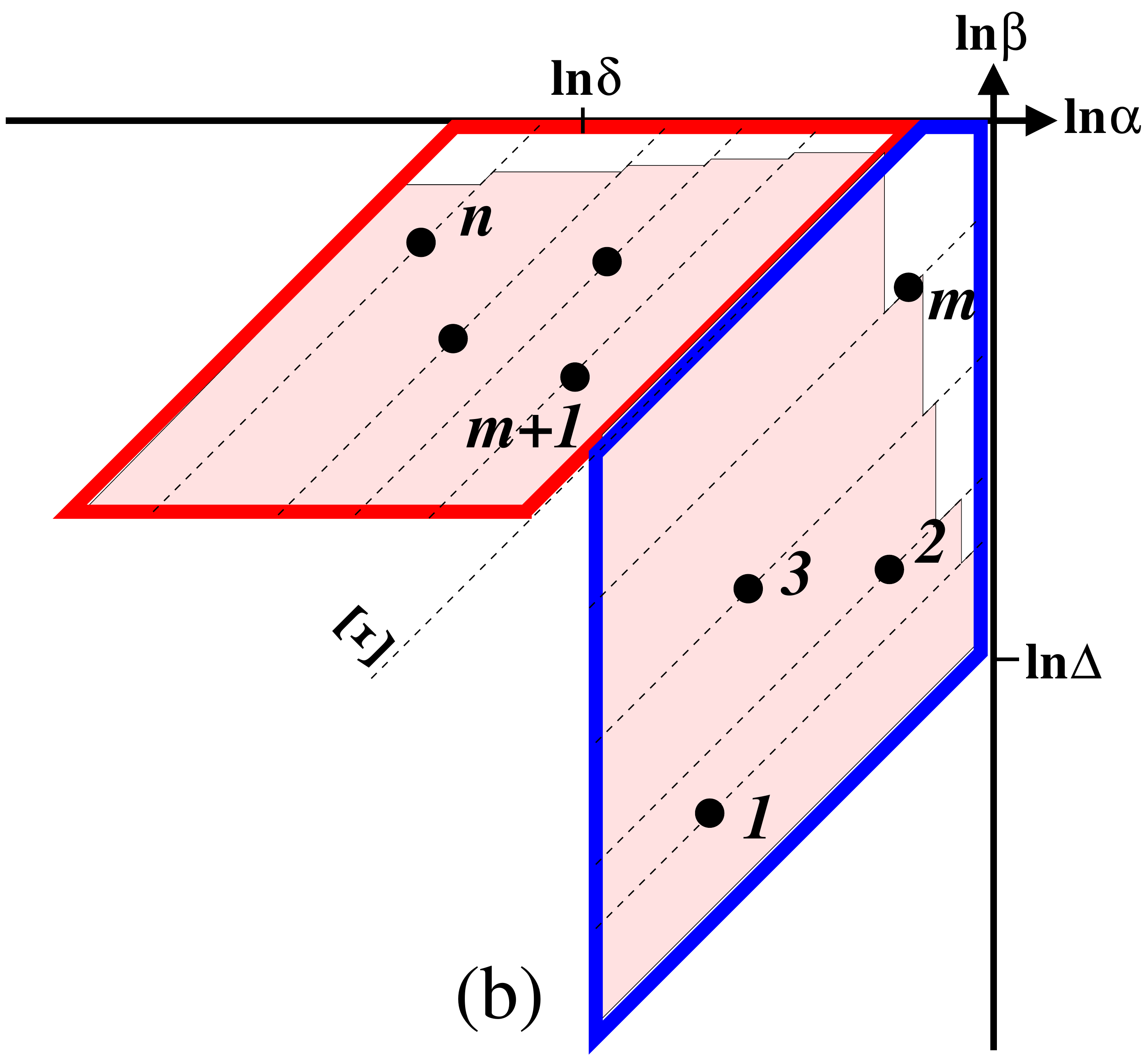}}
\caption{
  The Sudakov plane of the LO MC for DIS.
  The shaded area denotes the integration domain for the Sudakov 
  form-factor $S$ of the LO MC. 
  Rapidity equal to $\Xi$ marks the boundary between ISR and FSR.}
\label{eq:SudPlanDIS2}
\end{figure*}

The fully differential distribution for 
emitting $n$ gluons in the LO MC for DIS
we define as follows:
\begin{equation}
\label{eq:DISmasterLO}
\begin{split}
d\sigma_n &= 
  Q_q^2 \alpha_{QED}^2 dt d\varphi\;\;
   \frac{1}{t^2}\;
  e^{-S}\;
  \frac{d\psi}{2\pi}\;
  \delta_y\Big(\sum_j \vec{k}_j\Big)\;
\\&\times
  \left(
  \prod_{i=1}^n
  \frac{C_F \alpha_s}{\pi}
  \frac{d\alpha_i d\beta_i}{\alpha_i \beta_i}\;
  \frac{d\phi_i}{2\pi}
  \bar{P}(\hat{z}_i)
  \theta_{a_i>a_{i-1}}\!\!
  \right)
  \theta_{\sum \alpha_i<1}
  \theta_{\sum \beta_i<1}.
\end{split}
\end{equation}
The key objects to be defined are the variables $\hat{z}_j$ 
and the Sudakov form-factor $S$.
For this LO modeling of the gluonstrahlung we use ordering
according to the factorization scale (evolution) variable
\begin{equation}
a_i^2= t \frac{\beta_i}{\alpha_i}
   \in \bigg(t\Delta,\;\frac{t}{\Delta} \bigg),
\end{equation}
which is the variable of the angular ordering of the MC.

The ISR part of the Sudakov plane (the blue trapezoid in Fig.\
\ref{eq:SudPlanDIS2})  
contains the gluons ${\cal I}=(1,2,3,... m)$
which have  $\beta_i/\alpha_i < e^\Xi$
and the FSR part (the red trapezoid in Fig.\ \ref{eq:SudPlanDIS2}) 
hosts the gluons ${\cal F}=(m+1,m+2,...,n)$
which have  $\beta_i/\alpha_i > e^\Xi$.
We shall indicate that the gluon $j$ belongs to one of these two subsets
by $j\in {\cal I}$ or $j\in {\cal F}$.
The variable $\hat{z}_j$ of the ISR and FSR gluon is defined
in terms of either $\alpha$'s or $\beta$'s:
\begin{equation}
\begin{split}
&
\text{ for}~~ j\in {\cal I}:\quad
  \hat{z}_j 
  = z^{\cal I}_j = \frac{x^{\cal I}_j}{x^{\cal I}_{j-1}},\;\;
  x^{\cal I}_j \equiv 1-\sum_{i=1}^{j} \alpha_i,\;
\\&
\text{ for}~~ j\in {\cal F}:\quad
  \hat{z}_j = z^{\cal F}_j 
  = \frac{x^{\cal F}_j}{x^{\cal F}_{j-1}},\;\;
  x^{\cal F}_j \equiv 1-\sum_{i=m+1}^{j} \beta_i.
\end{split}
\end{equation}
The Sudakov form-factor $S$ is the integral over the area in 
the logarithmic Sudakov
plane available for the real emission 
in the step-by-step Markovian process.
This area is visualized in Fig.~\ref{eq:SudPlanDIS2}
as a shaded polygon.
The rapidity $\Xi$ defines the boundary between the ISR and FSR
emissions according to the corresponding LO distribution
and can be treated as an arbitrary parameter, 
for example $\Xi=0$ in Fig.~\ref{eq:SudPlanDIS2}a
is an acceptable LO choice.
In fact, another more clever choice of $\Xi$, like the one
indicated in Fig.~\ref{eq:SudPlanDIS2}b, can be made,
for instance, within the Markovian LO MC algorithm.
It is also possible to switch from one value of $\Xi$ to another
in the final stage of the LO MC by means of reweighting
MC events.
In view of the above flexibility, we leave out the exact definition
of $\Xi$ to the later stage of the MC code implementation.

\subsection{Structure function for LO MC}
The standard double-differential distribution of the DIS process,
as realized in our LO MC,
is obtained by inserting the $\delta$-function defining $x_B$:
\begin{equation}
\label{eq:DISMCxQ}
\begin{split}
\frac{d^2\sigma^{LO}}{dt dx_{B}} &= 
  \frac{2\pi Q_q^2 \alpha_{QED}^2}{t^2}
  W_0(y_B)
\\&\times
  \sum_{n=0}^\infty
  \left(
  \prod_{i=1}^n
  \frac{2C_F \alpha_s}{\pi^2}
  \int d^3\Ecal(k_i)\;
  \bar{P}(z^k_i)\;
  \theta_{a_i>a_{i-1}}\!\!
  \right)
\\&\times
  e^{-S}\;
  \delta_{\sum_j \vec{k}^y_j=0}\;
  \theta_{\sum \alpha_i<1}
  \theta_{\sum \beta_i <1}
  \delta_{x_{B}=1-\sum\alpha_j -\tilde{K}^2}.
\end{split}
\end{equation}

The above distribution is directly implementable
in the MC form, for instance using the Markovian algorithm,
and the above double
differential distribution of $x_B$ and $t$ is coming just
from histogramming, 
using MC events with all 4-momenta of all leptons, quarks and gluons
explicitly defined.

On the other hand, we may explicitly show analytically
that the above distribution is proportional to PDF
convoluted with the coefficient function.
This is done by means of inserting into the integrand
$
 1=\prod_{i=1}^n (\theta_{a_i>a_\Xi} 
                 +\theta_{a_i<a_\Xi})
$
and after expanding/reordering the sums of
the integrals. 
The distribution of eq.~(\ref{eq:DISMCxQ})
almost factorizes into the ISR and FSR parts:
\begin{widetext}
\begin{equation}
\begin{split}
\frac{d^2\sigma}{dt dx_{B}} &= 
  \frac{2\pi Q_q^2 \alpha_{QED}^2}{t^2}\;
 W_0(y_B)
 \Bigg\{
 \Bigg[
  \sum_{n=0}^\infty
  \left(
  \prod_{i=1}^n
  \frac{2C_F \alpha_s}{\pi^2}
  \int d^3\Ecal(k_i)\;
  \bar{P}(z^{\cal I}_i)
  \theta_{a_\Xi \geq a_i>a_{i-1} }\!\!
  \right)
  e^{-S_I}
 \Bigg]
 \\&\times
  \left[
  \sum_{n'=0}^\infty
  \left(
  \prod_{i=1}^{n'}
  \frac{2C_F \alpha_s}{\pi^2}
  \int d^3\Ecal(k_i)\;
  \bar{P}(z^{\cal F}_i)\;
  \theta_{a_i>a_{i-1} \geq a_\Xi}\!\!
  \right)
  e^{-S_F}
  \theta_{\sum \beta_{i\in \cal F} < 1 -\beta_I}
  \right]
  \theta_{\sum\limits_{\cal I+F} \alpha_i<1}
  \delta_{x_{B}=1-\sum\limits_{{\cal I}+{\cal F}}\alpha_j}
  \Bigg\},
\end{split}
\end{equation}
\end{widetext}
where we have split the Sudakov form-factor 
into the ISR and FSR parts, 
$S=S_I+S_F$ and $ \beta_I= \sum \beta_{i\in \cal I}$.
In the above we have also neglected $\tilde{K}^2$
in $x_B=1-\sum_j \alpha_j -\tilde{K}^2$.
This is well justified at LO, but it turns out that it can
be done at NLO as well.
The alternative solution would be to make special effort in parametrizing
the phase space (part of the definition of the $\Pbbm'$ operator)
to ``protect'' $x_B$ as it was done for $\hat{x}=\hat{s}/s$
in the $W/Z$ production process.
We have decided that this is not worth the effort, as the dependence
on $x_B$ of the differential distributions is relatively mild.
We may come back to this idea if an additional justification is found.

Altogether, the final LO formula can be written
as a convolution of the PDF for ISR and 
the the resummed ``coefficient function'' 
$C_{\cal F}(z_F)$ for FSR:
\begin{widetext}
\begin{equation}
\label{eq:DIS-LO-compact}
\begin{split}
& \frac{d^2\sigma}{dt dx_{B}} = 
  \frac{2\pi Q_q^2 \alpha_{QED}^2}{t^2}\;
  W_0(y_B)
  \int dx_I dz_F\; \delta_{x_{B}=x_I z_F}\;
  D_{\cal I}(\Xi,x_I)\;
  C_{\cal F}(z_F),
\\&
D_{\cal I}(\Xi,x_I)=
  e^{-S_I}
  \sum_{n=0}^\infty
  \left(
  \prod_{i=1}^n
  \frac{2C_F \alpha_s}{\pi^2}
  \int d^3\Ecal(k_i)\;
  \bar{P}(z^{\cal I}_i)
  \theta_{a_\Xi \geq a_i >a_{i-1} }\!\!
  \right)
  \delta_{x_I=1-\sum_{j\in {\cal I}} \alpha_j},
\\&
C_{\cal F}(z_F)=
 e^{-S_F}\Bigg\{
 \delta_{1=z_F}
 +\sum_{n'=1}^\infty
  \left(
  \prod_{i=1}^{n'}
  \frac{2C_F \alpha_s}{\pi^2}
  \int d^3\Ecal(k_i)\;
  \bar{P}(z^{\cal F}_i)
  \theta_{a_i>a_{i-1} \geq a_\Xi}
  \right)
  \delta_{1-z_{F}= x_I^{-1} \sum_{j\in {\cal F}} \alpha_j }
  \Bigg\},
\\&
\int_0^1 dz_F\; C_{\cal F}(z_F) \equiv 1.
\end{split}
\end{equation}
\end{widetext}

The interesting pure FSR object $C_{\cal F}(x)$
is probing the FSR evolution variable,
instead of the FSR lightcone variable! 
In the LO version it is enough to keep only the trivial
$C_{\cal F}(x)=\delta(1-x)$ term, 
while for our NLO purpose it is enough to retain only
one more easily calculable term, $n'=1$, $\Xi=0$:
\begin{equation}
\label{eq:CFord1}
\begin{split}
&C_{\cal F}^{(1)}(x)=
\delta(1-x)
\\&~~~~~~+
  \frac{C_F \alpha_s}{\pi}
  \left(
  \int_0^1\int_0^1
  \frac{d\alpha_1 d\beta_1}{\alpha_1 \beta_1}\;
  \bar{P}(1-\beta_1)
  \theta_{\alpha_1<\beta_1}
  \delta_{1-x- \alpha_1}
  \right)_+
\\&
=\delta(1-x)+
\frac{C_F \alpha_s}{\pi}
\left( 
 -\frac{\ln(1-x)}{1-x} -\frac{3}{4}\frac{1}{1-x} +\frac{1}{4}(3+x)
\right)_+.
\end{split}
\end{equation}
NB. the above reproduces the bulk of the
coefficient function of eq.~(\ref{eq:C2angOrd}),
that is terms
like $\big(\frac{\ln(1-x)}{1-x}\big)_+$
and $\big(\frac{1}{1-x}\big)_+$.

The MC initial--state PDF obeys the LO DGLAP evolution
equation (limited to the non-singlet gluonstrahlung):
\begin{equation}
\begin{split}
2t\frac{\partial}{\partial t} D_I(t,x)
&=\frac{\partial}{\partial \Xi} D_I(t,x)
\\&
= \int dz dx\; \delta_{x_I=x z}\;
  \frac{C_F \alpha_s}{\pi}
  \left( \frac{\bar{P}(z)}{1-z} \right)_+ D_I(t,x),
\end{split}
\end{equation}
and the same is true for the structure function
$2F_1=C_{\cal F}\otimes D_I$:
\begin{equation}
\begin{split}
2t\frac{\partial}{\partial t} F_1(t,x)
&=\frac{\partial}{\partial \Xi} F_1(t,x)
\\&
= \int dz dx\; \delta_{x_I=x z}\;
  \frac{C_F \alpha_s}{\pi}
  \left( \frac{\bar{P}(z)}{1-z} \right)_+ F_1(t,x).
\end{split}
\end{equation}

\subsection{Exclusive ISR and FSR subtractions in DIS}

The two soft counterterms, for ISR and FSR,
can be identified in
the fully differential distribution
of the single-real gluon in the LO MC:
\begin{equation}
\label{eq:DISsingleLO}
\begin{split}
& d\sigma^{MCLO}_1 = 
  Q_q^2 \alpha_{QED}^2 dt d\varphi\;
   \frac{1}{t^2}\;
  \frac{d\psi}{2\pi}\;
  \delta_y\Big(\vec{k}_1\Big)\;
  \frac{C_F \alpha_s}{\pi}
  \frac{d\alpha_1 d\beta_1}{\alpha_1 \beta_1}\;
  \frac{d\phi_1}{2\pi}
\\&~~~~~~~~~~~~~~~~
  W_0(y_{B})
  \left\{
   \bar{P}(1-\alpha_1)
   \theta_{\beta_1<\alpha_1}
  +\bar{P}(1-\beta_1)
   \theta_{\beta_1>\alpha_1}
  \right\},
\end{split}
\end{equation}
where we define the $y_{B}=\frac{t}{s(1-\alpha_1)}$ variable and
the Born spin factor is $W_0(y) =1+(1-y)^2$.
On the other hand, the NLO-complete unsubtracted distribution is
\begin{equation}
\label{eq:DISsingleNLO}
\begin{split}
d\sigma^{NLO}_1 &= 
  Q_q^2 \alpha_{QED}^2 dt d\varphi\;
   \frac{1}{t^2}\;
  \frac{d\psi}{2\pi}\;
  \delta_y\Big(\vec{k}_1\Big)\;
\\&\times
  \frac{C_F \alpha_s}{\pi}
  \frac{d\alpha_1 d\beta_1}{\alpha_1 \beta_1}\;
  \frac{d\phi_1}{2\pi}
  W(\alpha_1,\beta_1,y_{B}),
\\
W(\alpha_1,\beta_1,y_{B})
  &\equiv \frac{s^2+u_1^2+s_1^2+u^2}{2s^2}.
\end{split}
\end{equation}
See eq.~(\ref{eq:DISmandelst}) for explicit Mandelstam invariants.

For the MC we shall use
the subtracted distribution with both the ISR and FSR counterterms:
\begin{equation}
\label{eq:DISbet}
\begin{split}
& d\sigma^{\Delta NLO}_1 =
  d\sigma^{NLO}_1 -d\sigma^{MCLO}_1=
\\&~~~~~~
 =Q_q^2 \alpha_{QED}^2 dt d\varphi\;
   \frac{1}{t^2}\;
  \frac{d\psi}{2\pi}\;
  \delta_y\Big(\vec{k}_1\Big)\;
  \frac{C_F \alpha_s}{\pi}
  \frac{d\alpha_1 d\beta_1}{\alpha_1 \beta_1}\;
  \frac{d\phi_1}{2\pi}\;\;
  \tbet_1(k),
\\&
  \tbet_1(k)=
     \tbet_I\theta_{\beta_1<\alpha_1}
    +\tbet_F\theta_{\beta_1>\alpha_1},
\\&
\tbet_I(\alpha_1,\beta_1,y_{B})
   =W(\alpha_1,\beta_1,y_{B})
   -W_0(y_{B}) \bar{P}(1-\alpha_1),
\\&
\tbet_F(\alpha_1,\beta_1,y_{B})
   =W(\alpha_1,\beta_1,y_{B})
   -W_0(y_{B}) \bar{P}(1-\beta_1),
\end{split}
\end{equation}
which defines (up to NLO) the following expression
\begin{equation}
\begin{split}
&\hat{C}_{\Delta NLO}(z,y_{B}/z)
 =
  \frac{C_F \alpha_s}{\pi}
  \int_0^1\int_0^1
  \frac{d\alpha_1 d\beta_1}{\alpha_1 \beta_1}\;
  \frac{d\psi}{2\pi}\;
\\&~~~~~~~~~~~~~~~~~~~~~~~~\times
 \left\{
     \tbet_I\theta_{\beta_1<\alpha_1}
    +\tbet_F\theta_{\beta_1>\alpha_1}\right\}
  \delta_{1-z=\alpha_1}
\\&
=
\frac{C_F \alpha_s}{\pi}
     W_0(y_{B})
     \left[\frac{1}{2} (1+z)\ln(1-z)+\frac{5}{4}z+\frac{1}{4}\right]
     -\, y_{B}^2\, z\,,
\end{split}
\end{equation}
to be used in the numerical tests of the MC implementations.

\subsection{Exclusive NLO correction to hard process in DIS MC}
In the following
we propose a MC weight which upgrades
the MC with the LO hard process and the LO evolution kernels
to the MC with the NLO hard process and the LO evolution kernels.
The distribution in the LO+NLO MC reads
\begin{equation}
\label{eq:DISmasterMCNLO}
\begin{split}
d\sigma^{NLO}_n &= 
  Q_q^2 \alpha_{QED}^2
   \frac{dt}{t^2}\; d\varphi\;
  \frac{d\psi}{2\pi}\;
  \delta_y\Big(\sum_j \vec{k}_j\Big)\;
\\&\times
  \left(
  \prod_{i=1}^n
  \frac{C_F \alpha_s}{\pi}
  \frac{d\alpha_i d\beta_i}{\alpha_i \beta_i}\;
  \frac{d\phi_i}{2\pi}
  \bar{P}(z^k_i)
  \theta_{a_i>a_{i-1}}\!\!
  \right)
\\&\times
  \theta_{\sum \alpha_i<1}
  \theta_{\sum \beta_i<1}\;
  e^{-S}\;
  W_0(y_B)\;\;
  w^{\Delta NLO}_{MC},
\end{split}
\end{equation}
where the key element is the following MC weight
\begin{equation}
\label{eq:WTmcNLO}
\begin{split}
w^{\Delta NLO}_{MC}
&=[1+\Delta_{S+V}]
 +\sum_{j\in {\cal I}} 
     \frac{\tbet_1(\alpha'_j,\beta'_j,y_{B})}%
          { W_0(y_B) \bar{P}(z_j)}
 +\sum_{j\in {\cal F}} 
     \frac{\tbet_1(\alpha''_j,\beta''_j,y_{B})}%
          {W_0(y_B) \bar{P}(z_j)}
\\
&=[1+\Delta_{S+V}]
 +\sum_{j\in {\cal I}} 
     \frac{\tbet_I(a_j, z^{\cal I}_j,y_{B})}%
          {W_0(y_B) \bar{P}(z^{\cal I}_j)}
 +\sum_{j\in {\cal F}} 
     \frac{\tbet_F(a_j, z^{\cal F}_j,y_{B})}%
          {W_0(y_B) \bar{P}(z^{\cal F}_j)},
\end{split}
\end{equation}
which adds the missing NLO correction of the real emission type
and  also includes $\Delta_{S+V}$,
representing the remaining NLO virtual+soft corrections.

The important point is the definition of the variables
$\alpha'_i,\beta'_i$ and $\alpha''_i,\beta''_i$
in terms of $a_i$ and $z_i$,
in the presence of many  ``spectator LO gluons''.
An extrapolation of the one-gluon
matrix element all over the multigluon phase space
is an inevitable feature of any scheme combining the fixed order
ME with the resummed ME,
and there is always certain freedom in doing that.
The above extrapolation is done in terms of $z_j$ and $a_j$.
In the ISR part of the sum, we proceed such that
first we define $\alpha'_j=1-z_j^{\cal I}$
and next from the evolution scale
$a^2_j/a^2_\Xi =\beta'_j/\alpha'_j$ we calculate $\beta'_j$
(apparently we proceed as if there were no spectator gluons).
In the FSR part we proceed similarly,
i.e. using $z_j^{\cal F}$ we define $\beta''_I=1-z_j^{\cal F}$.
Next, from the evolution scale
$a^2_j/a^2_\Xi = \beta''_j/\alpha''_j$ 
we calculate $\alpha''_j$.

\subsection{Analytical integration 
 of DIS MC distributions and determining $\Delta_{S+V}$}
A remarkable feature of the complicated multigluon
distribution defined within the exact phase space
(with the full energy-momentum conservation)
is that it can be integrated analytically!
This integration result will 
help us to determine the NLO soft+virtual
correction $\Delta_{S+V}$
and will also be used in the numerical cross-check of the MC code.

The result of the analytical phase space integration
for the DIS MC reads
\begin{equation}
\label{eq:DISMCanl}
\begin{split}
\frac{d^2\sigma^{NLO}_{MC}}{dt dx_{B}} &= 
  \frac{2\pi Q_q^2 \alpha_{QED}^2}{t^2}\;
  \int dx_I dz\; \delta_{x_{B}=x_I z}\;
  D_{\cal I}(t,x_I)\;
\\&\times
 \Big[
  W_0(y_B) (1+\Delta_{S+V})\delta_{1=z}
 +\bar{C}_I(z,y_{B})
\\&~~~~~~
 +\bar{C}_F(z,y_{B})
 +W_0(y_B) C^{[1]}_{\cal F}(z)_+
 \Big],
\end{split}
\end{equation}
where
\begin{equation}
\begin{split}
\bar{C}_I(z,y_{B})
&=\frac{2C_F \alpha_s}{\pi^2}\int\! d^3\Ecal(k)\;
 \tbet_I(\alpha,\beta,y_{B})
 \theta_{\beta<\alpha}\delta_{1-z=\alpha}
\\&
=\frac{2C_F \alpha_s}{\pi^2}\int\! d^3\Ecal(k)\;
\\&~~~\times
 [W(\alpha,\beta,y_{B}) -W_0(y_{B}) \bar{P}(1-\alpha)]
 \theta_{\beta<\alpha}\delta_{1-z=\alpha},
\\
\bar{C}_F(z,y_{B})
&=\frac{2C_F \alpha_s}{\pi^2}\int\! d^3\Ecal(k)\;
 \tbet_F(\alpha,\beta,y_{B})
 \theta_{\beta>\alpha}\delta_{1-z=\alpha}
\\&
=\frac{2C_F \alpha_s}{\pi^2}\int\! d^3\Ecal(k)\;
\\&~~~\times
 [W(\alpha,\beta,y_{B}) -W_0(y_{B}) \bar{P}(1-\beta)]
 \theta_{\beta>\alpha}\delta_{1-z=\alpha},
\end{split}
\end{equation}
and
\begin{equation}
\begin{split}
\label{eq86cogoniema}
&C^{[1]}_{\cal F}(z)_+
=\frac{2C_F \alpha_s}{\pi^2}
\int d^3\Ecal(k)\;
\bar{P}(1-\beta)
\theta_{\alpha<\beta}\delta_{1-z=\alpha} \theta_{\alpha>\delta}
\\&~~~~~~~~~~~~~~~~~
-\delta_{z=1} S_F(\delta) ,\quad
\\&
\int_0^1 dz\; C^{[1]}_{\cal F}(z)_+ =0,
\\&
S_F(\delta) = \frac{2C_F \alpha_s}{\pi^2}
\int d^3\Ecal(k)\; \bar{P}(1-\beta)
\theta_{\alpha<\beta} \theta_{\alpha>\delta}
\\&~~~~~~~~~
=\int_0^{1-\delta} dz\; C^{[1]}_{\cal F}(z).
\end{split}
\end{equation}
The plus prescription for $ C^{[1]}_{\cal F}(z)_+$ is provided
by the Sudakov form-factor of the MC.
$\bar{C}_I(z,y_{B})$ and $ \bar{C}_F(z,y_{B})$ are completely finite/regular,
without any $(...)_+$ parts.
The IR regulator $\delta$ will drop out at the end.

Let us now find out  $\Delta_{S+V}$ of the MC by means of
comparing/matching the 1st order
eq.~(\ref{eq:Altarelli97DIS}) and/or eq.~(\ref{eq:C2angOrd})
with eq.~(\ref{eq:DISMCanl}) truncated also to the 1st order.
Going back for a moment to $n=4+2\veps$ we find out
that the 1st order bare PDF of the LO MC is
\[
D_{\cal I}(t,x_I)|_{1st.~ord.}= \delta(1-x_I)+ K_I(x_I,\veps),\quad
\int_0^1 dz\; K_I(z,\veps)=0.
\]
Also, as anticipated, the contribution 
$C^{[1]}_{\cal F}(z)$ cancels the counterterm in $\bar{C}_F(z,y_B)$
\begin{equation}
\begin{split}
&W_0(y_B) C^{[1]}_{\cal F}(z) +\bar{C}_F(z,y_{B})
= \bar{D}_F(z,y_{B},\delta)
\\&~~~~~~~~~~~~~~~~~~~~~~~~~~~~~~~~~~~~~~~~~~~~
- W_0(y_B) S_F(\delta) \delta(1-z),
\\&
\bar{D}_F(z,y_B,\delta)=\frac{2C_F \alpha_s}{\pi^2}
 \int\! d^3\Ecal(k)\;
\delta_{1-z=\alpha} \theta_{1-z>\delta}
\\&~~~~~~~~~~~~~~~~~~~\times
  W(\alpha,\beta,y_{B}) 
  \theta_{\beta>\alpha}.
\end{split}
\end{equation}
Altogether the 1st order truncation
of eq.~(\ref{eq:DISMCanl}) reads
\begin{equation}
\label{eq:DISMCtrun}
\begin{split}
& \frac{d^2\sigma^{NLO}_{MC(1)}}{dt dx_{B}} = 
  \frac{2\pi Q_q^2 \alpha_{QED}^2}{t^2}
 \Big\{
   \delta_{1=x_B} W_0(y_B) [1  +\Delta_{S+V} -S_F(\delta)]
\\&~~~~~~~~~~~~~~~~~~~~
 +W_0(y_B) K_I(x_B,\veps)
 +\bar{C}_I(x_B,y_{B})
 +\bar{D}_F(x_B,y_{B},\delta)
 \Big\},
\\&
\bar{C}_I(z,y_{B})+\bar{D}_F(z,y_{B},\delta)
=\frac{2C_F \alpha_s}{\pi^2}\int\! d^3\Ecal(k)\;
\\&~~~~~~\times
\{
 W(\alpha,\beta,y_{B}) 
 -W_0(y_{B}) \bar{P}(1-\alpha) \theta_{\beta_1<\alpha_1}
\}
\delta_{1-z=\alpha}
\theta_{1-z>\delta}
\\&
= [ W_0(y_B) \theta_{1-z>\delta} \bar{C}_2^s(z) -y^2_B\; C_L(z)],
\end{split}
\end{equation}
where
\begin{equation}
\bar{C}_2^s(z)=
\frac{C_F \alpha_s}{\pi}
\left\{
 \frac{1+z^2}{2(1-z)}
 \ln\frac{1}{1-z}
 -\frac{3}{4}\frac{1}{1-z}
 +1+\frac{3}{2}z
\right\}.
\end{equation}
Remembering that (cf. eq.~(\ref{eq:C2angOrd}))
\[
\begin{split}
C_2^{MC}(z) &= (\bar{C}_2^s(z))_+
= \theta_{1-z>\delta} \bar{C}_2^s(z) -\delta_{z=1} T(\delta),
\\
T(\delta) &= \int_0^{1-\delta} dx\; \bar{C}_2^s(x),
\end{split}
\]
we finally get
\begin{equation}
\label{eq:DISMCtrun2}
\begin{split}
& \frac{d^2\sigma^{NLO}_{MC(1)}}{dt dx_{B}}
=  \frac{2\pi Q_q^2 \alpha_{QED}^2}{t^2}
\\&~~~~\times
 \Big\{
   \delta_{1=x_B} W_0(y_B) [1 
   +\Delta_{S+V}  -S_F(\delta) +T(\delta)]
\\&~~~~~~~~~~
 +W_0(y_B) K_I(x_B,\veps) 
 + W_0(y_B) C_2^{MC}(x_B) -y_B^2\; C_L(x_B)
 \Big\}.
\end{split}
\end{equation}
Comparing with the NLO-complete (real+virtual) calculation
(eg. eq.~(\ref{eq:SfunMSprim})),
we see that the matching with the above MC implementation dictates
the following relation (the Adler sum rule for $F_2$):
\begin{equation}
\label{eq:DeltaSVdis}
\Delta_{S+V} = S_F(\delta) -T(\delta)
=\int_0^1 dz\; [C^{[1]}_{\cal F}(z) - \bar{C}_2^s(z)].
\end{equation}
The above is finite in the $\delta \to 0$ limit.
This is not surprising, because 
$C^{[1]}_{\cal F}(z)$ integrates the FSR counterterm,
while $\bar{C}_2^s(x)$ comes from the ISR-subtracted exact ME 
-- they both coincide in the FSR collinear limit,
while the ISR collinear singularity is already removed from $\bar{C}_2^s(x)$.

Summarizing,
the complete analytical result for the structure
function from the DIS Monte Carlo (angular ordering)
defined in eq.~(\ref{eq:DISmasterMCNLO})
takes the following final form:
\begin{equation}
\label{eq:SfunDISMC}
\begin{split}
& \frac{d^2\sigma^{NLO}_{DIS}}{dt dx_B}
 = \frac{2\pi \alpha_{QED}^2  Q_q^2}{t^2} \;
\int dx dz\; \delta_{x_B=xz}\; D_I(t,x)
\\&~~~~~~~~~~~
\Big[ 
  W_0(y_B) (1+\Delta_{S+V})\delta_{1=z}
 +W_0(y_B) C_2^{MC}(z) -y_B^2 C_L(z) 
\Big],
 \\&
 C_2^{MC}(z)=
 \frac{C_F \alpha_s}{\pi}
 \left\{
 -\frac{1+z^2}{2(1-z)} \ln(1-z)
  -\frac{3}{4}\frac{1}{1-z}
  +1+ \frac{3}{2}z
 \right\}_+
 \\&
 C_L(z)=\frac{C_F \alpha_s}{\pi}z.
\end{split}
\end{equation}
The above formula is ``ready to go'' for numerical comparison with
the Monte Carlo.

The virtual+soft correction
$\Delta_{S+V}$ is given by eq.~(\ref{eq:DeltaSVdis}),
more precisely
\begin{equation}
\label{eq:DeltaSVdis2}
\begin{split}
\Delta_{S+V}
&=\frac{C_F \alpha_s}{\pi} \int_0^1 dz\;
 \bigg\{
  -\frac{\ln(1-z)}{1-z} -\frac{3}{4}\frac{1}{1-z} 
  +\frac{3}{4} +\frac{1}{4}z
\\&~~~~~~~~~~~~~~~~~
  +\frac{1+z^2}{2(1-z)} \ln(1-z)
  +\frac{3}{4}\frac{1}{1-z}
  -1- \frac{3}{2}z
 \bigg\}
\\&
=\frac{C_F \alpha_s}{\pi} \int_0^1 dz\;
 \bigg\{
   -\frac{1+z}{2} \ln(1-z)  
   -\frac{1}{4} -\frac{5}{4}z
 \bigg\}
=0.
\end{split}
\end{equation}
The above is just the result of the rigorous NLO calculation.

Notice also that the MC result features in a natural way
the exponentiation of the distributions like
\[
\begin{split}
f(z) &=
\delta(1-z)+\frac{C_F \alpha_s}{\pi} \left(\frac{\ln(1-z)}{1-z} \right)_+
\\&
\simeq
\frac{C_F \alpha_s}{\pi} \frac{\ln(1-z)}{(1-z)}
e^{ - \frac{C_F \alpha_s}{\pi} \frac{1}{2} \ln^2(1-z) },\quad
\\
\int_0^1 dz\; f(z) &= 1.
\end{split}
\]
Such an exponentiation can be included in the analytical formula.

Last but not least, let us write explicitly the difference between the
coefficient functions of the standard $\overline{MS}$ factorization scheme
of Eq.~(\ref{eq:bardeen78})
and the MC factorization scheme of Eq.~(\ref{eq:SfunDISMC})%
\footnote{Here, the usual $\overline{MS}$ 
assignment $t=\mu^2e^{-\omega_2}$ is done,
 see refs.~\cite{Mele:1990bq,Brock:1993sz}.}
\begin{equation}
\begin{split}
 \Delta C_2(z) &=
 C_2^s(z) - C_2^{MC}(z) =
\\&=
 \frac{C_F \alpha_s}{\pi}
 \left\{
 \frac{1+z^2}{2(1-z)} \ln\frac{(1-z)^2}{z}
 +\frac{1-z}{2}
 \right\}_+.
\end{split}
\end{equation}
The above function should be used to correct the existing
$\overline{MS}$ PDFs before using it to fix
input in our MC.
Alternatively, the coefficient function of Eq.~(\ref{eq:SfunDISMC})
should be used to fit DIS experimental data 
with the PDF function compatible with the presented MC%
\footnote{Similar corrections have to be also determined
 for the NLO inclusive kernels, once the NLO corrections
 are included in the ladder part of the MC,
 see the first incomplete results
 in ref.~\cite{Jadach:2011kc}.}.

\subsection{Factorization scheme independent relation
         between DY and DIS processes}
\label{sec:FSindep-relation}

In spite of the change of the factorization scheme in the MC,
the factorization scheme independent and the
regularization-independent relation of 
AEM (eq.~(91) in \cite{Altarelli:1979ub})
should be reproduced exactly, if we claim to protect the universality.
Let us verify it.
The original AEM relation reads%
\footnote{
Using again
$\int_0^1 dz\;  (1+z^2) \big( \frac{\ln(1-z)}{1-z} \big)_+ 
       = \frac{7}{4} $.
}
\begin{equation}
\label{DY-2DIS}
\begin{split}
\Delta^{\rm AEM}_q(z) &=
f_{q,DY}-2f_{q,2}
\\&
=\frac{C_F \alpha_s}{\pi}
\Bigg[
\delta_{z=1}\bigg( \frac{2}{3}\pi^2 + \frac{1}{2} \bigg)
+\frac{3}{2} \frac{1}{(1-z)_+}
\\&~~~~
+(-3-2z)
+(1+z^2)\bigg( \frac{\ln(1-z)}{1-z} \bigg)_+ \Bigg]
\\&
=\frac{C_F \alpha_s}{\pi}
\Bigg[
\delta_{z=1}
\bigg( \frac{2}{3}\pi^2 -\frac{7}{4} \bigg)
+\frac{3}{2} \frac{1}{(1-z)_+}
\\&~~~~
+(-3-2z)_+
+\bigg( (1+z^2)\frac{\ln(1-z)}{1-z} \bigg)_+
\Bigg].
\end{split}
\end{equation}
Using the result of the analytical integration 
of the DIS MC, eq.(\ref{eq:SfunDISMC}),
\[
 C_2^s(z)=
 \frac{C_F \alpha_s}{\pi}
 \left\{
 -\frac{1+z^2}{2(1-z)} \ln(1-z)
  -\frac{3}{4}\frac{1}{1-z}
  +1 +\frac{3}{2}z
 \right\}_+
\]
and the analogous analytical result for the DY MC of
eq.~(\ref{eq:DYfinalVS})
\[
C_{2}(z)=
\delta_{z=1}
\frac{C_F \alpha_s}{\pi}\;\left( \frac{2}{3}\pi^2 -\frac{7}{4} \right)
+\frac{C_F \alpha_s}{\pi}\; \left[ -(1-z)_+ \right],
\]
we obtain from our two MC implementations the same result
\begin{equation}
\label{eq:AEM78}
\begin{split}
&
C_{2}(z)-2C_2^s(z)
=\Delta^{\rm AEM}_q(z).
\end{split}
\end{equation}
In this way we have reproduced the AEM~\cite{Altarelli:1979ub} result
for the MC factorization scheme, confirming its universality.
The above agreement with  the AEM result
is easily traced back to the fact that
it holds already for the difference 
of the unsubtracted coefficient functions%
\footnote{The last term is of course absent for
 the usual assignment $\hat{s}=t=\mu^2$.}
\begin{equation}
f^{\rm AEM}_{DY}(z) -2 C^{\rm AEM}_{2,bare}(z)
=\Delta^{\rm AEM}_q(z)
+2\frac{C_F \alpha_s}{\pi} \ln\frac{\hat{s}}{t}
\end{equation}
(cf. eqs.~(\ref{eq:AEM79}) and (\ref{eq:Altarelli97DIS}))
and because 
in both MC formulae the same ISR counterterm
of eqs.~(\ref{eq:KFz}) and (\ref{eq:KIdim}) is subtracted%
\footnote{The ISR counterterm is defined in the DIS and DY processes
 at the exclusive level, involving $\Pbbm'$ and kinematic mapping,
 so the statement that ``it is the same'' is
 more non-trivial than in the case of the CFP inclusive counterterms.}.
In particular, terms due to the $\veps \frac{1-z}{2}$
component in the $\gamma$-trace present in
the DY and DIS coefficient functions necessarily cancel out.

It is fair to mention that in ref.~\cite{Altarelli:1979ub}
the relation of eq.~(\ref{DY-2DIS}) is treated as the pQCD result
for the coefficient function of the DY process in the DIS factorization
scheme.
On the other hand, this relation can be turned into an experimentally testable
relation between the structure functions of the DY and DIS processes,
testing the important principle of universality (process independence)
of collinear singularities in pQCD predictions,
independently of any particular choice of the factorization scheme and the PDFs.

\section{Summary and outlook}
\label{sec:summary}

We have presented a complete method for implementing
NLO corrections to the hard process in the LO MC for DY and DIS
processes.
This method was originally developed for introducing NLO corrections
in the ladder MC~\cite{Jadach:2010ew,Jadach:2010aa},
therefore it is well suited to be extended to
include NLO corrections in both hard process and ladder parts.

The presented method is based on a new factorization
scheme~\cite{Jadach:2010ew,Jadach:2011kc}
extending the collinear factorization
theorems~\cite{Ellis:1978ty,Curci:1980uw} to the
fully exclusive (unintegrated) form,
which can serve as a base for the MC distributions.
All differences between the $\overline{MS}$ and this new MC scheme
are kept under strict control,
and we are elaborating on that in a quite some detail.
In particular, we make a powerfull cross-check
of the whole MC factorization scheme by showing (analytically)
that the NLO MC results are reproducing
the factorization scheme independent relation of
Altarelli--Ellis--Martinelli~\cite{Altarelli:1979ub}
between the Drell--Yan and DIS
processes, see eq.~(\ref{eq:AEM78}).

The main practical results of this work are the multiparton distributions
of eq.~(\ref{eq:NLODYMCmaster}) and eq.~(\ref{eq:DISmasterMCNLO}),
for the EW boson production in hadron--hadron collision and
electron--hadron deep-inelastic scattering, respectively,
which are ready for Monte Carlo implementation.
These distributions feature the NLO corrections in the hard process part
and the LO pQCD evolution in two multiparton ladder parts.
The NLO corrections to the hard process
are introduced by means of a single MC weight
on top of the LO distributions -- it is, therefore, critical
that the LO MC covers the multiparton phase space without any gaps or overlaps.
This is achieved by means of using the angular ordering, which
is also essential for good control of the soft
gluon behaviour beyond LO, already in the LO MC.
The correct soft limit also assures good behaviour of the MC weight;
weights are positive and small (peaked near 1).
For the weights distributions and other numerical cross-checks
of the presented method we refer reader to ref.~\cite{Jadach:2012vs}.

In our opinion this work solves the main obstacles on the way to the NLO MC,
based rigorously on the new MC factorization scheme.
There are still many less important problems to be solved on the
way to the practical level, i.e. construction of the MC program
applicable in the LHC data analysis.
Let us signal some of these problems and their solutions.
(i) For simplicity in our formulas
we have omitted the initial PDF of the quark in hadron at low
factorization scale $Q\sim 1$~GeV. This can be easily included
in the MC.
(ii) If we are aiming in a fully NLO MC the ladder parts
have to be upgraded to the NLO level, and this work is already well
advanced~\cite{Skrzypek:2011zw,Jadach:2010ew,Jadach:2010aa}.
(iii) In the presented MC scheme the QCD coupling was constant,
non-running. It is quite trivial to make it running within the LO MC.
It will be less trivial, but also profitable, to disentangle the
running-coupling 
effect from the NLO corrections in the MC implementation of the NLO ladder.
This problem is under study and will be treated in a separate publication.
(iv) All the MC distributions presented in this work are defined
for quarks and gluons, hence in the practical level MC code they will be
subject to a hadronization procedure, using one of 
the existing MC tools, such as 
HERWIG~\cite{Webber:1984if} or PYTHIA~\cite{Sjostrand:1985xi}.

Obviously, the proposed scheme of implementing
the NLO corrections to the hard process combined
with the MC parton showers (ladders) is different from the existing ones.
In section \ref{sec:MCatNLO} we comment on the differences between
our scheme and that of the MC@NLO~\cite{Frixione:2002ik} and POWHEG~\cite{Nason:2004rx}.
More systematic comparisons with these and other
schemes~\cite{Tanaka:2011ig,Tanaka:2007mi} will be done separately,
at the time of the numerical MC implementation.

The presented method of implementing
the NLO corrections to the hard process does not have any principal
limitations -- it can be extended to more diagrams and other processes.
However, at the practical level its application requires that LO parton
shower provides for the full coverage (no gaps nor dead zones) of
the hard process phase space relevant at NLO level.
This requirement is typically not fulfilled by the classic MC parton showers
like HERWIG or PYTHIA. It is not excluded that modernized
version of these MCs will provide for better phase space coverage,
notably using tools developed for the MC@NLO and POWHEG implementations.
Otherwise, LO parton shower has to be reconstructed,
for instance using scheme proposed in the present work.
(This may turn out to be mandatory for implementing NLO corrections
in the ladder parts of the parton shower MC.)

Summarizing, this work represents an important step into
a new area in the pQCD calculations for hadron colliders in the MC form,
in which the NLO corrections are implemented both in the hard process
and the ladder parts in a completely exclusive (unintegrated) way,
in full compatibility with the redefined, fully exclusive pQCD
factorization.


\section*{Acknowledgment}
This work is partly supported by 
  the Polish Ministry of Science and Higher Education grant 
  No.\ 1289/B/H03/2009/37,
  Foundation for Polish Science grant Homing Plus 2010-2/6,
  by DOE grant DE-FG02-09ER41600,
 the Polish National Science Centre grant DEC-2011/03/B/ST2/02632,
  by the U.S.\ Department of Energy
  under grant DE-FG02-04ER41299 and the Lightner-Sams Foundation.
One of the authors (S.J.) is grateful for the partial support
and warm hospitality of the TH Unit of the CERN PH Division,
and of Max Planck Institut f\"ur Physik, M\"unchen,
while completing this work.

\appendix
\numberwithin{equation}{section}

\section{Kinematics of EW boson production process}
\label{app:A}

Let us consider the case of a single real
(not necessarily collinear) gluon emission,
relevant to the NLO level  description of
the hard process
\begin{equation}
q(p_{0F})+\bar{q}(p_{0B})\rightarrow l^-(q_1)+l^+(q_2) +g(k),
\end{equation}
which is the classic EW vector boson production process
in the annihilation of the quark--antiquark pair
(Drell--Yan process in case of $\gamma^*$)
decaying into a lepton pair.
Note that in the definition below we omit
the distribution of the quark (antiquark) in the proton.
This can be always added easily in the MC. 

The following kinematical variables
\begin{equation}
\begin{split}
&
x
=\frac{\hat{s}}{s}
=\frac{(P-k)^2}{P^2}
=1-\alpha-\beta,\quad
\\&
P=p_{0F}+p_{0B},\quad
Q=\hat{P}=p_{0F}+p_{0B}-k,\quad
\\&
s=P^2,\quad
\hat{s}=\hat{P}^2=Q^2=(P-k)^2=P^2-2k\cdot P,\quad
\\&
\alpha=\frac{2k\cdot p_{0B}}{ P^2},\quad
\beta =\frac{2k\cdot p_{0F}}{ P^2},\quad
\end{split}
\end{equation}
are used in this work.
The most important of them is the invariant mass squared $\hat{s}$
of the produced colourless boson.

For the emitted gluons
we are using dimensionless {\em eikonal phase space} parametrized
in terms of various variables: 
\begin{equation}
\begin{split}
\label{eq:EikonalPHSP}
d^3\Ecal(k) &=
\frac{d^3 k}{2k^0}\;
\frac{1}{\bk^2}
= \frac{1}{2}
\frac{dk^+}{k^+}
\frac{d^2 \bk}{\bk^2}
= \frac{\pi}{2}
\frac{d\phi}{2\pi}
\frac{d\alpha}{\alpha}
\frac{d \ba^2}{\ba^2}
\\&
= \frac{\pi}{2}
\frac{d\phi}{2\pi}
\frac{d\alpha}{\alpha}
\frac{d\beta}{\beta}
= \pi
\frac{d\phi}{2\pi}
\frac{d\alpha}{\alpha}
d \eta
= \pi
\frac{d\phi}{2\pi}
\frac{d\beta}{\beta}
d \eta,
\end{split}
\end{equation}
where $\bk=(k^1,k^2)$ is a transverse Cartesian 2-vector
($k_T^2=|\bk|^2= s\alpha\beta$),
the Sudakov (lightcone) variables are
\[
 k^\pm = k^0\pm k^3,\quad
 \alpha= \frac{2k^+}{\sqrt{s}},\quad
 \beta = \frac{2k^-}{\sqrt{s}}.
\]
Moreover, we introduce variable $\ba \equiv \bk/\alpha$, and
the conventional rapidity variable $\eta$ is defined as
\[
  \eta=\frac{1}{2}\ln\frac{\alpha}{\beta}
      =\frac{1}{2}\ln\frac{k^+}{k^-}
      =-\ln\frac{|\ba|}{\sqrt{s}},\quad
 a=|\ba|=e^{-\eta} \sqrt{s}.
\]

Multiparticle phase space is defined as
\begin{equation}
d\tau_n (P;p_1,p_2,...,p_n)=
\delta^{(4)}
\Big(P-\sum_{i=1}^n p_i \Big)\;
\prod_{i=1}^n \frac{d^3 p_i}{2p_i^0}.
\end{equation}
The two-dimensional phase space for massless particles is then
\begin{equation}
d\tau_2 (Q;q_1,q_2)= \frac{1}{2} d\Omega.
\end{equation}

\begin{widetext}
\section{1-real gluon NLO correction, analytical integration}
\label{app:B}

We are going to integrate analytically 1-real gluon NLO correction
as defined in eq.~(\ref{eq:DYbeta1F}).
The contribution from the F hemisphere is easily calculable:
\begin{equation}
\label{eq:C2FDY}
\begin{split}
\frac{1}{2}C_{2r}(x) &=
 \frac{C_F \alpha_s}{\pi}
\int\limits_0^\infty\!\! d\alpha\!\!
 \int\limits_0^\infty\!\! d\beta \;
\Big[
 \frac{\bar{P}(x)-\alpha\beta}{\alpha\beta}\;
 \theta_{\alpha+\beta<1}
 \theta_{\beta<\alpha}
 \delta_{x=1-\alpha-\beta}
-\frac{\bar{P}(x)}{\alpha\beta}
 \theta_{\beta<\alpha}
 \delta_{x=1-\alpha}
\Big]
\\&
=\frac{C_F \alpha_s}{\pi}
\Bigg[
 \bar{P}(x)
 \int\limits_\Delta^{(1-x)/2}\!\!\! d\beta \;
 \frac{1}{(1-x-\beta)\beta}\;
-\int\limits_0^{(1-x)/2}\!\!\! d\beta \;
-\bar{P}(x)
 \int\limits_\Delta^{(1-x)}\!\! d\beta \;
 \frac{1}{(1-x)\beta}
\Bigg]
\\&
=\frac{C_F \alpha_s}{\pi}
 \frac{\bar{P}(x)}{(1-x)}
\Bigg[
 \int\limits_0^{(1-x)/2}\!\!
 \frac{d\beta}{1-x-\beta}\;
+\int\limits_\Delta^{(1-x)/2}\!\!
 \frac{d\beta}{\beta}\;
-\int\limits_\Delta^{(1-x)}\!\!
 \frac{d\beta}{\beta}
\Bigg]
-\frac{C_F \alpha_s}{\pi} \frac{1-x}{2}
=-\frac{C_F \alpha_s}{\pi} \frac{1-x}{2}.
\end{split}
\end{equation}

\section{Inclusive NLO factorization formula for DY MC}
\label{app:C}

We are going to prove the formula of eq.~(\ref{eq:DYanxch}),
representing the MC with two LO ladders and the NLO-corrected hard process,
by means of reorganizing the phase space integration
of eq.~(\ref{eq:NLODYMCmaster}).
Let us consider the part of the
total cross section of eq.~(\ref{eq:NLODYMCmaster})
proportional to the term $j\in F$ in
the MC weight of eq.~(\ref{eq:NLODYMCwt}).
The summation and integration over the ``spectator'' LO gluons
in the B part of the phase space can be easily folded into the LO PDF.
What remains to be considered is the following sum of integrals:
\begin{equation}
\label{eq:C2rEval}
\begin{split}
\sigma_I^{NLO} &=
\int dx_F\; dx_B\;
\sum_{n_1=1}^\infty\;
e^{-S_{_F}}
\int_{\Xi<\eta_{n_1}}
\bigg(
\prod_{i=1}^{n_1}
 d^3\Ecal(\bar{k}_i)
 \theta_{\eta_i<\eta_{i-1}}
 \frac{2C_F\alpha_s}{\pi^2} \bar{P}(z_{Fi})
\bigg)\;
\sum_{j\in F} 
\frac{\tbet_1(\hat{s},\hat{p}_F,\hat{p}_B;a_j, z_{Fj})}%
      {\bar{P}(z_{Fj})}\;
G_B(\Xi,x_B)\;
\delta_{x_F =\prod_i z_{Fi}}
\\&
=\int dx_F\; dx_B\;
\sum_{n_1=1}^\infty\;
e^{-S_{_F}}
\sum_{j=1}^{n_1}\;
\int
\bigg(
\prod_{i=1,i\neq j}^{n_1}
 d^3\Ecal(\bar{k}_i)
 \theta_{\eta_i<\eta_{i-1}}
 \frac{2C_F\alpha_s}{\pi^2} \bar{P}(z_{Fi})
\bigg)\;
\\&~~~~~~~~~~\times
\int
d^3\Ecal(\bar{k}_j)\;
\theta_{ \eta_{j+1}< \eta_j < \eta_{j-1} }
\tbet_1(\hat{s},\hat{p}_F,\hat{p}_B;a_j, z_{Fj})\;
G_B(\Xi,x_B)\;
\delta_{x_F =\prod_i z_{Fi}}.
\end{split}
\end{equation}
The essential step in transforming
each $j$-th term
is relabelling the gluons $i\to i'$ such that
$i'=i$ for $i=1,2,...,j-1$ and
$i'=i-1$ for $i=j+1,...,n_1$,
hence $i'=1,2,...,n_1-1$ without any gap, 
and finally $i=j$ is relabelled as $j'=0$.
Using the symmetry of the integrand, integrals over $k_{i'}$
can be pulled out, and the sum over adjacent integration
ranges of $k_{j'}=k_0$ is factorized off:
\begin{equation}
\begin{split}
\sigma_I^{NLO} &= \int dx_F\; dx_B\;
\delta_{x=x_F x_B}
\sum_{n_1=1}^\infty\;
e^{-S_{_F}}
\int
\bigg(
\prod_{i'=1}^{n_1-1}
 d^3\Ecal(\bar{k}_{i'})
 \theta_{\eta_{i'}<\eta_{i'-1}}
 \frac{2C_F\alpha_s}{\pi^2} \bar{P}(z_{Fi'})
\bigg)\;
\\&\times
\sum_{i'=1}^{n_1-1}\;
\int
d^3\Ecal(\bar{k}_{0})\;
\theta_{ \eta_{i'}< \eta_{0} \leq \eta_{i'-1} }
\tbet_1(\hat{s},\hat{p}_F,\hat{p}_B;a_0, z_{F0})\;
G_B(\Xi,x_B)\;
\delta(x_F - z_{F0} \prod\limits_{i'=1}^{n_1-1} z_{Fi'}).
\end{split}
\end{equation}
The sum over the adjacent integration intervals
is combined into a single integral 
\[
 \int\limits_{0}^{a_1}   \tbet_1 da_0
+\int\limits_{a_1}^{a_2} \tbet_1 da_0
+\int\limits_{a_2}^{a_3} \tbet_1 da_0
\dots
+\int\limits_{a_{n_1-2}}^{a_{n_1-1}} \tbet_1 da_0
=\int\limits_{0}^{a_{n_1-1}}   \tbet_1 da_0
\]
and factorized off,
while the remaining integrals over the spectator gluons 
$i'=1,2,...,n_{1}-1$ give rise to the LO PDF:
\begin{equation}
\begin{split}
\sigma_I^{NLO}
&=\int dx_F dx'_F dx_B\;
\bigg\{
\sum_{n_1=1}^\infty\;
e^{-S_{_F}}
\int
\bigg(
\prod_{i'=1}^{n_1-1}
 d^3\Ecal(\bar{k}_{i'})
 \theta_{\eta_{i'}<\eta_{i'-1}}
 \frac{2C_F\alpha_s}{\pi^2} \bar{P}(z_{Fi'})
\bigg)
\delta_{x'_F=\prod_{i'} z_{Fi'}}
\bigg\}
\\&~~~~~~~~~\times
\int
d^3\Ecal(\bar{k}_{0})\;
\tbet_1(\hat{s},\hat{p}_F,\hat{p}_B;a_0, z_{F0})\;
G_B(\Xi,x_B)\;
\delta_{x_F = z_{F0} x'_F }
\delta_{x=x_F x_B}
\\&
=\int dx_B\; dx'_F\; dz_{F0}\;
G_F(\Xi,x'_F)\;
G_B(\Xi,x_B)\;
\frac{1}{2}C_{2r}( z_{F0})\; \sigma_B(sx)
\delta_{x=x_{B} x'_F z_{F0}},
\end{split}
\end{equation}
where we have replaced the integration variable $x_F$ with $z_{F0}=x_F/x'_F$.
In the last step we were able to use the integral defined in
eq.~(\ref{eq:DYbeta1F}) 
and evaluated in Appendix \ref{app:B}, eq.~(\ref{eq:C2FDY}).

The other part of the
total cross section of eq.~(\ref{eq:NLODYMCmaster})
proportional to the term $j\in B$ in
the MC weight of eq.~(\ref{eq:NLODYMCwt})
gives the same result.
For the LO part times $(1+\delta_{S+V})$ 
we use eq.~(\ref{eq:LOfactDY2LO}).

As already noted, the key part of the above algebra is reminiscent
of that in ref.~\cite{yfs:1961}, except
that here the resummed singularities are in the angle
while in ref.~\cite{yfs:1961} they are in the energy variable.
\end{widetext}

\bibliographystyle{apsrev}
\bibliography{radcor}

\end{document}